\documentclass[a4paper,11pt]{article}
\pdfoutput=1 

\usepackage{jcappub} 

\usepackage[T1]{fontenc} 
\usepackage{hyperref}
\usepackage{cleveref}
\usepackage{hypcap}
\usepackage[font=small,labelfont=bf,labelsep=period]{caption}
\usepackage{graphicx}
\usepackage[dvipsnames,table]{xcolor}
\usepackage{mathrsfs}
\usepackage{subfig}
\usepackage{comment}
\usepackage{amsfonts}
\usepackage{float,subfloat}
\usepackage{array}
\usepackage{mathrsfs}
\usepackage{slashed}
\usepackage{latexsym}
\usepackage{amssymb,amsmath,tabularx}
\usepackage{braket}
\usepackage{bm}
\usepackage{adjustbox}
\usepackage{placeins}
\usepackage[normalem]{ulem}
\hypersetup{allcolors=[rgb]{0.0 0.0 0.6},linkcolor=[rgb]{0.75 0.05 0.05}}

\usepackage{tikzfeynman}
\usepackage{tikz}
\usepackage{tkz-euclide}
\usetikzlibrary{shapes.misc}
\usetikzlibrary{decorations.pathreplacing,calligraphy,patterns}
\usetikzlibrary{decorations.pathmorphing} 
\tikzset{
	v/.style={decorate, decoration={snake, segment length=3mm, amplitude=0.75mm}, draw},
	f/.style={draw=black, postaction={decorate},
		decoration={markings,mark=at position .6 with {\arrow[very thick]{latex}}}},
	fb/.style={draw=black, postaction={decorate},
		decoration={markings,mark=at position .4 with {\arrowreversed[very thick]{latex}}}},
	fnar/.style={draw=blue,thick},
	a/.style={double distance=2.2pt,draw=red,line width =1pt
        },
	g/.style={decorate, draw=black,
		decoration={coil,amplitude=3pt, segment length=3.5pt}},
	s/.style={dashed,draw=black, postaction={decorate},
		decoration={markings,mark=at position .55 with {\arrow[very thick]{latex}}}},
	sb/.style={dashed,draw=black, postaction={decorate},
		decoration={markings,mark=at position .55 with {\arrowreversed[draw=black,very thick]{latex}}}},
	snar/.style={draw=black,line width =1.5pt},
	cross/.style={cross out, draw=black, minimum size=2*(#1-\pgflinewidth), inner sep=0pt, outer sep=0pt},
	cross/.default={3pt},
	none/.style={draw=white, postaction={decorate},
		decoration={markings,mark=at position .6 with {\arrow[very thick]{latex}}}}
		}

\definecolor{darkblue}{rgb}{0,0.1,0.5}
\definecolor{darkgreen}{rgb}{0,0.5,0.2}
\definecolor{darkred}{RGB}{153,26,0}
\definecolor{seablue}{rgb}{0,0.2,0.6}
\definecolor{light}{rgb}{0,0.2,0}
\definecolor{viola}{RGB}{134,41,198}
\definecolor{myGreen}{RGB}{20,140,0}

\newcommand{\be}{\begin{equation}}
\newcommand{\ee}{\end{equation}}
\newcommand{\bes}{\begin{equation*}}
\newcommand{\ees}{\end{equation*}}
\newcommand{\Eq}[1]{eq.~\eqref{#1}}
\newcommand{\Eqs}[2]{eq.~\eqref{#1} and~\eqref{#2}}

\newcommand{\Sec}[1]{section~\ref{#1}}
\newcommand{\App}[1]{appendix~\ref{#1}}
\newcommand{\Fig}[1]{figure~\ref{#1}}
\newcommand{\Figs}[2]{figures~\ref{#1} and~\ref{#2}}

\newcommand{\ztwo}{$\mathbb{Z}_2$}
\newcommand{\ZT}{$\mathbb{Z}_3$}

\newcommand{\Mpl}{M_{\rm Pl}}

\newcommand{\mdm}{m_S}
\newcommand{\xdm}{x_S}
\newcommand{\epsdm}{\epsilon_S}
\newcommand{\epsphi}{\epsilon_\phi}

\newcommand{\vmol}{v_\textup{\tiny{Møl}}}
\newcommand{\algn}[1]{\begin{aligned} #1\end{aligned}}
\newcommand{\obar}[1]{\mkern 1.5mu\overline{\mkern-1.5mu#1\mkern-1.5mu}\mkern 1.5mu}

\newcommand{\GeV}{\mathrm{GeV}}

\newcommand{\TeV}{\mathrm{TeV}}

\newcommand{\cms}{\mathrm{cm^3/s}}

\title{\boldmath Cosmic-Ray Signatures of Annihilating and Semi-Annihilating Dark Matter via One-Step Cascades}

\author[a,b]{Francesco D'Eramo,}
\author[c]{Silvia Manconi,}
\author[a,b]{Tommaso Sassi}

\affiliation[a]{Dipartimento di Fisica e Astronomia, Università degli Studi di Padova, \\
Via Marzolo 8, 35131 Padova, Italy}
\affiliation[b]{Istituto Nazionale di Fisica Nucleare (INFN), Sezione di Padova, \\
Via Marzolo 8, 35131 Padova, Italy}
\affiliation[c]{Sorbonne Université \& Laboratoire de Physique Théorique et Hautes Énergies (LPTHE), CNRS, 4 Place Jussieu, Paris, France}

\emailAdd{francesco.deramo@pd.infn.it}
\emailAdd{manconi@lpthe.jussieu.fr}
\emailAdd{tommaso.sassi@phd.unipd.it}

\abstract{We present a framework in which three classes of dark matter number-changing processes can affect both the relic abundance via thermal freeze-out in the early universe and the generation of indirect cosmic-ray signals today. These processes are: (i) direct annihilations into Standard Model final states; (ii) annihilations into metastable on-shell mediators that subsequently decay into Standard Model particles; (iii) semi-annihilation processes featuring a dark matter particle in the final state, accompanied by a metastable mediator. A central element of our analysis is the systematic inclusion of semi-annihilation alongside the more commonly considered channels. This setup is largely model-independent, as we only assume the presence of one or more of these processes with unsuppressed $s-$wave contributions. We analyze representative benchmarks for the dominant decay modes of the mediator and show how the resulting injection spectra for $\gamma$ rays, neutrinos, and cosmic-ray antimatter vary with the relative importance of the three classes of processes. As an application, we evaluate the observable $\gamma$-ray fluxes from dwarf spheroidal galaxies in the GeV–TeV window. Finally, we provide explicit model realizations in which multiple processes coexist, and discuss how their interplay shapes indirect detection signatures. Our results provide a consistent connection between early-universe dynamics and present-day observables, revealing distinctive features that arise when multiple dark matter processes contribute simultaneously.}

\begin{document}
\maketitle
\flushbottom

\section{Introduction}
\label{sec:intro}

Astrophysical and cosmological observations across a wide range of length scales~\cite{Bertone:2004pz,Feng:2010gw,Cirelli:2024ssz,Bozorgnia:2024pwk} have firmly established the existence of dark matter (DM), an unknown component of the universe that is far more abundant than baryonic matter and accounts for approximately $27\%$ of its total energy budget. Beyond the fact that it interacts gravitationally, our knowledge of DM is limited to a few essential properties that any viable particle candidate must satisfy (see \cite{Cirelli:2024ssz} and references therein): it must be cosmologically stable, effectively electrically neutral, non-relativistic at the time of matter-radiation equality so as not to spoil structure formation, and essentially collisionless. Consequently, the microscopic nature of DM remains elusive, and uncovering its particle identity constitutes one of the most compelling open problems in physics beyond the Standard Model (SM).

The thermal freeze-out paradigm~\cite{Lee:1977ua,Gondolo:1990dk} provides an appealing explanation for the origin of the present DM relic abundance. In this framework, DM interacts with the visible sector at a rate that allows DM particles in the primordial plasma to remain in thermal equilibrium. However, the DM number density is continuously diluted by the Hubble expansion until the interaction rate drops below the expansion rate, causing the DM particles to decouple from the thermal bath. Representative realizations include scenarios in which the DM particle is charged under the SM electroweak gauge group, as in the minimal DM framework~\cite{Cirelli:2005uq}, as well as models in which it is a SM singlet coupled to the visible sector through new portal interactions. In all such cases, DM interactions maintain thermal equilibrium between the dark and visible sectors in the early universe. If the processes mediating these interactions have cross sections of the typical size of SM weak interactions, the subsequent departure from thermal equilibrium naturally yields a relic abundance of the order of the observed value $\Omega_{\rm DM} h^2 = 0.120 \pm 0.001$~\cite{Planck:2018vyg}. Remarkably, well-motivated solutions to the SM hierarchy problem predict the existence of weakly interacting massive particles (WIMPs) with precisely these properties~\cite{Goldberg:1983nd,Ellis:1983ew,Scherrer:1985zt,Srednicki:1988ce}, giving rise to the so-called \textit{WIMP miracle}. The ability to reproduce the observed relic density, together with this theoretical motivation, has driven an extensive experimental program to search for WIMPs. Indeed, DM-SM interactions are expected to induce observable signals in direct detection experiments, at particle accelerators and colliders, and through indirect detection (ID) searches in the fluxes of cosmic ray (CR) particles.

In this work, we focus on indirect searches that aim to detect anomalous signals on top of astrophysical backgrounds potentially originating from DM. These experiments target regions with large expected DM densities, such as the Galactic Centre (GC) and dwarf spheroidal satellite galaxies (dSphs), and search for stable SM final states, including $\gamma$ rays, neutrinos, and antimatter. Within a vanilla WIMP framework, annihilations of DM with its antiparticle\footnote{Or, in the case of a self-conjugate particle such as a real scalar or a Majorana fermion, with itself.} govern both the relic abundance obtained through thermal freeze-out and the present-day CR fluxes. As a result, imposing the relic density requirement together with current ID limits excludes sizable portions of the parameter space. Complementary constraints from direct and collider probes may further restrict the remaining region~\cite{Cirelli:2024ssz}. 

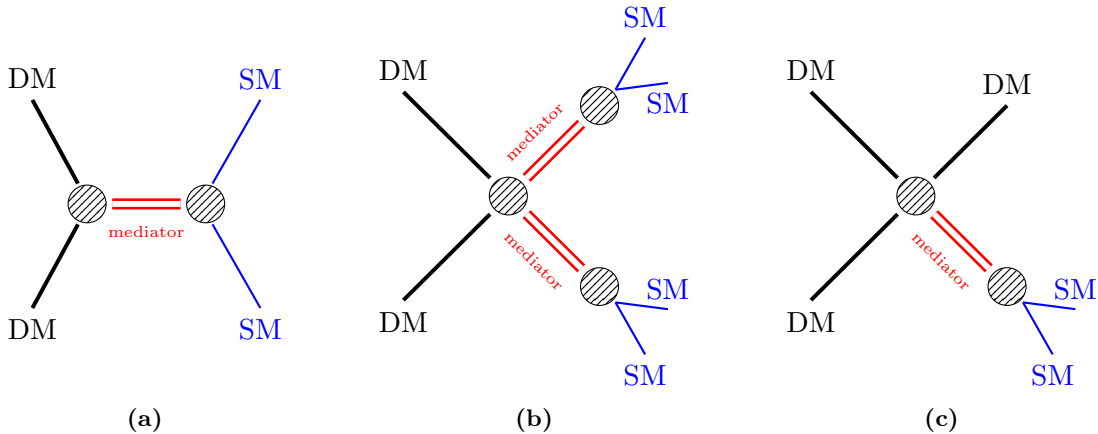
\begin{figure}[t]
\centering
\subfloat[\label{fig:annihilation}]{
\begin{tikzpicture}[scale=1.2]
    \draw[snar] (-0.75,1.15) node[above] {$\mathrm{DM}$} -- (-0.25,0.225);
    \draw[snar] (-0.75,-1.15) node[below] {$\mathrm{DM}$} -- (-0.25,-0.225);
    \node[circle, draw, minimum size=0.5cm, pattern=north east lines] at (-0.15,0) {};
    \draw[a] (0.125,0)--node[midway,below=1.5mm]{\color{red}{\tiny{mediator}}}(0.875,0);
    \node[fill=gray,circle, draw, minimum size=0.5cm, pattern=north east lines] at (1.15,0) {};
    \draw[fnar] (1.225,0.225) -- (1.75,1.15) node[above] {\color{blue}{$\mathrm{SM}$}};
    \draw[fnar] (1.225,-0.225) -- (1.75,-1.15) node[below] {\color{blue}{$\mathrm{SM}$}};
    \path (0,-2) node[opacity=0] {}; 
\end{tikzpicture}
}\qquad
\subfloat[\label{fig:one-step-annihilation}]{
\begin{tikzpicture}[scale=1.2]
 	\draw[snar] (-1.15,1.15) node[above] {$\mathrm{DM}$} -- (-0.2,0.2);
    \draw[snar] (-1.15,-1.15) node[below] {$\mathrm{DM}$} -- (-0.2,-0.2);
    \node[circle, draw, minimum size=0.5cm, pattern=north east lines] at (0,0) {};
    \draw[a] (0.2,0.2) --node[midway, above=1.5mm, sloped] {\color{red}{\tiny{mediator}}} (0.8,0.8);
    \draw[a] (0.2,-0.2) -- node[midway, below=1.5mm, sloped] {\color{red}{\tiny{mediator}}}(0.8,-0.8) ;
    \node[circle, draw, minimum size=0.5cm, pattern=north east lines] at (1,1.) {};
    \node[circle, draw, minimum size=0.5cm, pattern=north east lines] at (1,-1.) {};
    \draw[fnar] (1.175,1.175) -- (1.5,1.75) node[above] {\color{blue}{$\mathrm{SM}$}};
    \draw[fnar] (1.175,1.175) -- (1.75,1.25) node[below] {\color{blue}{$\mathrm{SM}$}};
    \draw[fnar] (1.175,-1.175) -- (1.5,-1.75) node[below] {\color{blue}{$\mathrm{SM}$}};
    \draw[fnar] (1.175,-1.175) -- (1.75,-1.25) node[above] {\color{blue}{$\mathrm{SM}$}};
\end{tikzpicture}
}\qquad
\subfloat[\label{fig:one-step-semiannihilation}]{
\begin{tikzpicture}[scale=1.2]
 	\draw[snar] (-1.15,1.15) node[above] {$\mathrm{DM}$} -- (-0.2,0.2);
    \draw[snar] (-1.15,-1.15) node[below] {$\mathrm{DM}$} -- (-0.2,-0.2);
    \node[circle, draw, minimum size=0.5cm, pattern=north east lines] at (0,0) {};
    \draw[snar] (0.2,0.2) -- (1,1)node[above] {$\mathrm{DM}$};
    \draw[a] (0.2,-0.2) -- node[midway, below=1.5mm, sloped] {\color{red}{\tiny{mediator}}}(0.8,-0.8) ;
    \node[circle, draw, minimum size=0.5cm, pattern=north east lines] at (1,-1.) {};
    \draw[fnar] (1.175,-1.175) -- (1.5,-1.75) node[below] {\color{blue}{$\mathrm{SM}$}};
    \draw[fnar] (1.175,-1.175) -- (1.75,-1.25) node[above] {\color{blue}{$\mathrm{SM}$}};
\end{tikzpicture}
}
\caption{Schematic representation of the three distinct classes of processes responsible for CR production considered in this study: annihilations (\ref{fig:annihilation}), one-step annihilations (\ref{fig:one-step-annihilation}), and one-step semi-annihilations (\ref{fig:one-step-semiannihilation}). All initial states involve two DM (anti-)particles (solid black lines). A mediator that couples to both the dark and visible sectors is exchanged or produced in each process (double red lines). In the annihilation case, the DM pair connects to the SM through an off-shell mediator. The other two processes proceed as cascade reactions: an initial $2 \rightarrow 2$ interaction produces one or more on-shell metastable mediators, which subsequently decay into SM primary products (blue legs).}
\label{fig:diagrams}
\end{figure}

Moving beyond the standard vanilla WIMP picture, in which ID signals arise exclusively from DM annihilations into SM states, we consider a more general and richer framework in which multiple processes contribute to an exotic component of the observed CR fluxes. In this setup, a mediator field lighter than the DM particle couples to both DM and SM states. We assume that the underlying symmetries of the model allow for the three classes of processes schematically illustrated in \Fig{fig:diagrams} and described below. \vspace{-0.2cm} 
\begin{itemize}
    \item \textbf{Annihilation:} The conventional process in which DM particles annihilate into pairs of SM states through an off-shell mediator. \vspace{-0.2cm}
    \item \textbf{One-step cascade annihilation:} DM particles annihilate into a pair of on-shell mediators, which subsequently decay into SM states. Such processes have been extensively studied over the past two decades, both in model-dependent~\cite{DiMauro:2025jsb,Hooper:2019xss} and model-independent settings~\cite{Mardon:2009rc,Fortin:2009rq}, and have been suggested to explain several ID anomalies---such as the GC $\gamma$-ray excess~\cite{Elor:2015tva,Tempel:2012ey,Bai:2012qy}, the positron excess~\cite{Bergstrom:2008ag,Cholis:2008wq}, and the antiproton excess~\cite{Cholis:2019ejx,Hooper:2019xss}---while evading stringent direct detection constraints and reproducing the observed relic abundance. Although conventional astrophysical sources may account for the intensity and morphology of some of these anomalies, a DM contribution has not yet been conclusively excluded.  \vspace{-0.2cm}
    \item \textbf{One-Step Cascade Semi-annihilation:} DM particles annihilate into a final state containing one mediator and one DM particle~\cite{Hambye:2008bq,Hambye:2009fg,Arina:2009uq,DEramo:2010keq}. The mediator subsequently decays into SM states, as in the previous case, generating an exotic CR flux. Since this process reduces the total number of DM particles by only one unit, it was dubbed \textit{semi-annihilation} in Ref.~\cite{DEramo:2010keq}. Processes of this kind arise when DM is stabilized by symmetries more general than an Abelian $\mathbb{Z}_2$. As pointed out in Ref.~\cite{DEramo:2010keq}, such reactions can have a significant impact on the relic density, especially in scenarios in which the dark sector contains more than one stable component. Moreover, the resulting ID spectra are typically richer than in the standard case due to the different kinematics~\cite{DEramo:2010keq,DEramo:2012fou}, which produces different spectral features. More recently, secluded dark sectors with semi-annihilation processes have been explored as viable alternatives for generating DM-induced ID signals in different CR channels, including $\gamma$ rays~\cite{Queiroz_2019,Guo:2023kqt,Guo:2024jdj} and antiprotons~\cite{Arcadi:2017vis}. 
\end{itemize} \vspace{-0.2cm}
In what follows, we refer to the latter two collectively as \textit{one-step cascade processes}.

We consider a setup in which all three classes of processes illustrated in \Fig{fig:diagrams} are simultaneously present, allowing for a unified treatment of their combined contributions. Standard DM annihilation into SM particles has been extensively studied, while annihilation into metastable on-shell mediators has also received considerable attention in the recent literature. In this context, the systematic inclusion of one-step semi-annihilation processes constitutes the main novelty of this work.

The goal of this paper is to characterize the CR \textit{injection spectra}, which encode the energy distribution of messenger particles produced at the source in DM-induced collisions. This enables a direct connection between relic density calculations and the source terms of CR fluxes, for both neutral and charged species. Our framework extends previous studies, which typically focus on a single interaction channel at a time, by accounting for the interplay of multiple processes within a common setup. Importantly, the approach is largely model independent, as it relies only on the existence of the processes in \Fig{fig:diagrams} with unsuppressed $s-$wave contributions. It can be applied to any underlying particle physics realization by specifying the relevant masses, cross sections, and branching ratios. As a result, it provides a versatile tool to explore DM scenarios that yield richer and more general CR injection spectra than those obtained in vanilla WIMP frameworks. Setting aside the effect of CR evolution in the interstellar medium, such as solar modulation, energy losses, secondary scattering, magnetic fields impact, this work serves as a global template study of several CR messenger injection spectra. We describe the tools necessary to compute the CR injection spectra from one-step cascade chain reactions, systematically accounting for the presence of semi-annihilations and simultaneous presence of multiple DM collision processes. The study, performed under the simplified assumption of mediator coupling dominantly to a unique SM field, provides an insight of the spectral features that characterize the CR injection spectrum via each production mechanism prior to propagation. This is the fundamental input to compute the observable CR flux once this latter ingredient is accounted for, allowing for a comparison with existing data end the extrapolation of new bounds upon a tailored dedicated data-analysis.

The paper is organized as follows. In \Sec{sec:rates}, we introduce the general framework, briefly review the thermal freeze-out determination of the DM relic abundance, and derive the corresponding CR injection spectra arising from the combined contributions of the different processes, providing the analytical expressions that underpin our analysis. In \cref{sec:gammas,sec:neutrinos,sec:chargedCR}, we analyze the injection spectra of both neutral and charged messengers, first comparing the spectral features of the individual processes and then exploring their interplay. As a concrete application, \Sec{sec:gamma_flux} combines these injection spectra with the relevant astrophysical ingredients to compute the $\gamma$-ray flux. Finally, \Sec{sec:models} discusses representative DM models that realize the phenomenological scenarios introduced in this work. We conclude in \Sec{sec:conclusions}, and relegate technical and computational details to the appendices.

\section{Relic abundance and cosmic ray production}
\label{sec:rates}

The CRs studied in this work originate from the three classes of processes shown in \Fig{fig:diagrams}. The corresponding interaction rates per unit time and volume determine both the DM relic abundance and the present-day production of SM primaries, which subsequently shower and hadronize, yielding the observable CR flux.

In this section, we introduce the interaction rates for each process and establish the notation and conventions adopted throughout this work. These rates involve a thermal average over all possible kinematical configurations of the DM (anti-)particles in the initial state. During freeze-out, this average is taken over a thermal equilibrium distribution. The corresponding averaging procedure for ID depends on the astrophysical environment. However, the non-relativistic velocities of DM in all such systems imply that only the $s$-wave contribution is phenomenologically relevant. Focusing on this leading term simplifies the thermal averaging, and its formal definition is provided in the next subsection.

We will apply these results to both freeze-out and CR production. For relic density calculations, only the total interaction rate is required, obtained as a specific linear combination of the processes shown in \Fig{fig:diagrams}. The same overall rate also determines the total number of DM interactions producing CRs today, while the relative weights of the different channels shape the resulting spectra.

\subsection{Notation, conventions, and thermal averages}
\label{sec:notation}

The scenario in which each of the processes in \Fig{fig:diagrams} contributes to CR production is rather general, as it does not rely on specific assumptions about the nature of the DM particle or the mediator. It can be realized for a variety of spin assignments, as well as for different choices of SM fields in the final state, as expanded in \Sec{sec:models}. However, not all combinations are allowed due to constraints such as angular momentum conservation and SM gauge invariance. Assuming the DM field to be neutral under the SM gauge group, semi-annihilation processes with three DM fields on external legs imply that the mediator cannot carry any gauge charge either. Moreover, semi-annihilation requires the mediator to share the same spin statistics as the DM particle. Once these conditions are satisfied, the DM can consistently be either bosonic or fermionic.

In this work, we adopt an explicit notation inspired by analogies with scenarios commonly discussed in the literature. Semi-annihilation processes are often realized in models in which DM is a complex scalar stabilized by an Abelian discrete symmetry (e.g.~$\mathbb{Z}_3$), and we denote it by $S$. Likewise, the mediator is typically a real scalar that is neutral under the SM gauge group, and we denote it by $\phi$. We further assume that the mediator decays either into SM fermion-antifermion pairs or into gauge bosons; both possibilities are included in our analysis, provided that SM gauge invariance is preserved. For notational simplicity, we collectively denote the final-state SM field by $\psi_{\mathrm{SM}}$, with the understanding that it does not necessarily correspond to a fermion.

With the notation established, the processes can be explicitly expressed as follows.
\begin{subequations} 
\begin{align} 
\label{eq:annSM} \text{Annihilation to SM pairs:} & \, \qquad  \qquad S S^\star \, \rightarrow \, \psi_{\rm SM} \overline{\psi}_{\rm SM} \ , \\ 
\label{eq:annmed} \text{Annihilation to $\phi$ pairs:} & \, \qquad  \qquad S S^\star \, \rightarrow \, \phi \phi \ , \\ 
\label{eq:semiA} \text{Semi-annihilation to single $\phi$:} & \, \qquad  \qquad S S \, \rightarrow \, S^\star \phi  \ , \\
\label{eq:semiB} & \, \qquad \qquad S^\star S^\star \, \rightarrow \, S \phi \ . 
\end{align} 
\label{eq:processes}
\end{subequations}
Whenever a mediator $\phi$ appears in the above equations, it is always assumed to be produced on-shell and to decay into SM states via $\phi \rightarrow \psi_{\mathrm{SM}} \overline{\psi}_{\mathrm{SM}}$ only after its production. Crucially, we consider scenarios in which the \textit{primary} SM states $\psi_{\rm SM}$ produced by $\phi$ decays are the same as those identified as final states of the standard DM annihilations. Namely, we work under the assumption of a dark sector uniquely coupled to a single SM field.\footnote{Strictly speaking, it does not have to be uniquely coupled to a single SM field. Our analysis holds as long as the mediator couples predominantly to a single SM field, see \Sec{sec:injection_spectra} for further discussion. The less minimal scenario in which the mediator $\phi$ couples to multiple SM states $\psi_{\rm SM}$ requires a generalization that is nonetheless straightforward.} This is convenient, as it allows us to directly compare and combine the resulting CR spectra, since they are generated via the same $\phi$ decay channel and $\psi_{\rm SM}$ evolution.

Notice that we include two distinct processes for the semi-annihilations in \Eqs{eq:semiA}{eq:semiB}, depending on whether the initial state contains two DM particles or two antiparticles. This reflects our assumption that DM is not a self-conjugate field and has distinct antiparticles (e.g.~the DM field is a complex scalar or a Dirac fermion). Both processes produce a mediator $\phi$ in the final state and therefore contribute to the CR flux.

We summarize the general procedure to evaluate collision rates and collect the relevant equations in \App{app:theaver}. Here, we limit ourselves to recalling that, for each process under consideration, the relevant quantity to be thermally averaged is the product of the total cross section $\sigma$ and the M{\o}ller velocity $\vmol$ defined in \Eq{eq:Mvel}. For both relic density and ID rate calculations, it is useful to take the non-relativistic limit via the partial-wave expansion
\be
\sigma \vmol = \sigma_0 + \sigma_1 \vmol^2 + \ldots \ .
\label{eq:partialwaves}
\ee
The thermal average of the constant $s-$wave contribution $\sigma_0$ is straightforward. The p-wave term, proportional to $\sigma_1$, scales with the squared velocity and, up to order-one factors, has a thermal average proportional to the ratio between the DM temperature and the DM mass, $\langle \vmol^2 \rangle \propto T_\chi / m_\chi$.\footnote{By temperature we refer either to the DM temperature in the early universe, assuming equilibrium distribution, or to the appropriately normalized width of the DM velocity distribution in astrophysical environments.} Higher partial waves are encoded in the ellipsis.

In our model-independent analysis, we focus on scenarios in which the leading term in the expansion in \Eq{eq:partialwaves} is non-vanishing, $\sigma_0 \neq 0$. Otherwise, the annihilation process would be velocity-suppressed and the corresponding ID signals strongly reduced. In this work, we retain only the $s-$wave contribution and neglect higher partial waves. This choice is particularly convenient, as each process in \Eq{eq:processes} is then characterized by a single parameter $\sigma_0$, and it provides an excellent approximation for ID given the typically low DM velocities in astrophysical environments. For thermal freeze-out, neglecting higher partial waves is less accurate, since after decoupling DM velocities remain mildly relativistic.\footnote{Once an explicit model is specified, higher partial waves can be straightforwardly included using standard methods (see, e.g., Refs.~\cite{Bernstein:1988bw,Gondolo:1990dk,Jungman:1995df,DEramo:2017ecx}). Restricting to the $s-$wave for freeze-out induces only minimal corrections of order $T_\chi/m_\chi \sim \mathcal{O}(1/25)$, leading to percent-level effects on the relic abundance while enabling a direct comparison between early- and late-universe observables.}

The produced mediator $\phi$ subsequently decays into SM particles via $\phi \to \psi_{\rm SM}\,\overline{\psi}_{\rm SM}$. These decays do not affect the computation of DM freeze-out, since only DM number-changing processes are relevant. As long as the mediator remains in thermal equilibrium with the thermal bath (see, e.g., the freeze-out analysis in Ref.~\cite{DEramo:2025xef}), its decays can be safely neglected when computing the DM relic abundance. They do not affect the overall ID rates either, but are crucial in shaping the CR spectra, as the produced primary particles undergo different hadronization and evolution, resulting in distinct spectra of stable SM particles.

\subsection{Dark matter relic density}
\label{sec:freezeout}

The interaction rates relevant for the production of detectable CRs are such that DM reaches thermal equilibrium in the early universe, placing us within the standard thermal freeze-out framework. We assume that the DM particle is distinct from its own antiparticle, and that there is no matter-antimatter asymmetry in the dark sector, such that the number densities satisfy $n_S = n_{S^\star}$. We track the evolution of the DM particle $S$ abundance; the total DM number density then differs only by an overall factor of $2$. The fundamental tool to describe the evolution of the DM number density is the Boltzmann equation, which takes the general form
(see, e.g., Refs.~\cite{Bernstein:1988bw,DEramo:2020gpr,DEramo:2025xef} for a complete derivation)
\be
\frac{dn_S}{dt} + 3 H n_S = - \Sigma_{\rm ann} \left(n_S^2 - n_S^{\rm eq\, 2} \right) - \Sigma_{\rm semi} \left(n_S^2 - n_S n_S^{\rm eq} \right) \ .
\label{eq:BEgen}
\ee

The left-hand side of the Boltzmann equation encodes the effect of the expansion of the universe through the Hubble parameter $H$. We assume that DM production takes place during the radiation-dominated epoch, in which the Hubble rate scales with the temperature of the radiation bath $T$ as
\be
H(T)=\frac{\pi\sqrt{g_\star(T)}}{3 \sqrt{10}}\frac{T^2}{\Mpl} \ ,
\ee
where $g_\star(T)$ denotes the number of effective relativistic degrees of freedom in the primordial plasma. The right-hand side of \Eq{eq:BEgen} describes the underlying dynamics and contains the DM equilibrium distribution $n_S^{\rm eq}$ given in \Eq{eq:nSeq}. It accounts for all DM number-changing processes, which can be broadly classified into annihilations and semi-annihilations. The former involve $S S^\star$ in the initial state and lead to final states consisting of either SM particle pairs or metastable mediators. The latter involve $S S$ or $S^\star S^\star$ in the initial state and produce a DM particle or antiparticle in the final state, accompanied by a single metastable mediator.

Our next step is to derive how the quantities $\Sigma_{\rm ann}$ and $\Sigma_{\rm semi}$ depend on the thermal averages of the cross sections of the processes in \Eq{eq:processes}. Here, potential subtleties arise in dealing with factors of $2$, so we clarify our results and conventions by spelling out all the details. The general definition of the thermal average for a given process is provided in \Eq{eq:thavdef}. These quantities are constant in the $s-$wave limit of interest, but the expressions presented in this paragraph are valid in full generality. Whenever identical particles are present in the initial and/or final states, this is accounted for in the thermal average through the statistical factor defined in \Eq{eq:Sabcd}. With these conventions, the presence of two identical particles in the initial state ($S S$ or $S^\star S^\star$) for semi-annihilations introduces no additional factors of $2$ on the right-hand side of the Boltzmann equation.  However, there is another source of factors of $2$ that is not accounted for by the thermal average: the fact that the process with $S S$ in the initial state destroys \textit{two} DM particles. This is the reason why the associated collision operator carries an additional factor of $2$. This factor is not present for the process with $S^\star S^\star$ in the initial state, since it creates only \textit{one} DM particle.\footnote{It is worth recalling that we are constructing the Boltzmann equation for the number of DM particles $n_S$ and not for the number of DM antiparticles $n_{S^\star}$.} Finally, CP conservation in the dark sector implies that the squared matrix elements for the semi-annihilation processes in \Eqs{eq:semiA}{eq:semiB} are identical.  Putting all this together, we obtain
\begin{subequations} 
\begin{align} 
\label{eq:SigmaANN} \Sigma_{\rm ann} \equiv & \, \langle \sigma_{S S^\star \, \rightarrow \, \psi_{\rm SM} \overline{\psi}_{\rm SM}} \vmol \rangle + \langle \sigma_{S S^\star \, \rightarrow \, \phi \phi } \vmol \rangle \ , \\
\label{eq:SigmaSEMI} \Sigma_{\rm semi} \equiv & \, 2  \langle \sigma_{S S \, \rightarrow \,  S^\star \phi } \vmol \rangle - \langle \sigma_{S^\star S^\star \, \rightarrow \,  S \phi } \vmol \rangle = \langle \sigma_{S S \, \rightarrow \,  S^\star \phi } \vmol \rangle \ .
\end{align} 
\label{eq:SigmaDef}
\end{subequations}

It is convenient to rewrite the Boltzmann equation in terms of dimensionless comoving variables. We introduce the DM yield $Y_S \equiv n_S/s$, with the entropy density defined as
\be
s=\frac{2\pi^2}{45}g_{\star s}(T)T^3\,.
\ee
Here, $g_{\star s}(T)$ denotes the temperature-dependent number of entropic relativistic degrees of freedom in the thermal bath. In this language, we define the equilibrium comoving number density of DM as $Y_S^{\rm eq} \equiv n_S^{\rm eq}/s$. We also trade the time variable for the inverse temperature $x \equiv \mdm/T$. The Boltzmann equation then takes the form
\be
\frac{dY_S}{dx}=-\left( 1 - \frac13\frac{d \ln g_{\star s}}{d \ln x} \right)\frac{s(x)}{xH(x)}
\left[ \Sigma_{\rm ann} \left(Y_S^2 - Y_S^{{\rm eq}\,2}\right)+ \Sigma_{\rm semi} \left(Y_S^2 - Y_S Y_S^{\rm eq} \right) \right]\, ,
\label{eq:BE_Y}
\ee
which can be solved to obtain the asymptotic DM yield $Y_S^\infty \equiv Y_S(x\rightarrow\infty)$. The present DM abundance is then given by
\be
\Omega_S h^2 = 2 \times \frac{\mdm \, Y_S^\infty \, s_0}{\rho_c/h^2}\,,
\ee
where the overall factor of $2$ accounts for the presence of both DM particles and antiparticles. Here, $s_0 = 2891.2 \,\mathrm{cm}^{-3}$ is the present entropy density and $\rho_c/h^2 \simeq 1.054\times10^{-5} \,\GeV/\mathrm{cm}^3$ the reduced critical energy density of the universe~\cite{ParticleDataGroup:2024cfk}.

The Boltzmann equation in \Eq{eq:BE_Y} is a special case of a Riccati equation with vanishing linear term and, as such, it does not admit closed-form solutions. Its asymptotic value for the DM yield must therefore be computed numerically. Semi-analytical approximations are nevertheless desirable because they provide useful guidance for understanding the dependence of the relic density on the key quantities. Ref.~\cite{DEramo:2010keq} showed that semi-annihilations behave very similarly to annihilations: they lead to a freeze-out temperature that depends logarithmically on the DM mass and interaction strength, and they produce a relic abundance that scales as the inverse of the semi-annihilation cross section. In particular, DM particles decouple when the bath temperature $T_{\rm FO}$ satisfies the condition
\be
x_{\rm FO} \equiv \frac{m_S}{T_{\rm FO}} = 
\ln\!\left[\frac{0.038\, M_{\rm Pl}\, m_S \, \Sigma_{\rm eff}} {\sqrt{g_\star(x_{\rm FO})\, x_{\rm FO}}} \right] \ ,
\ee
which leads to the DM relic density
\be
\Omega_S h^2 = 2 \times \frac{1.07 \times 10^9 \, {\rm GeV^{-1}} \, x_{\rm FO}}{\sqrt{g_\star(x_{\rm FO})} M_{\text{Pl}} \, \Sigma_{\rm eff}} \ .
\label{eq:Omegah2}
\ee
We introduce two quantities to parametrize the strength of the DM number-changing processes: the effective total thermally averaged cross section $\Sigma_{\rm eff}$ and the relative semi-annihilation contribution $\eta$, defined as
\begin{subequations}
    \begin{align}
        \label{eq:Sigmaeff}
        \Sigma_{\rm eff}&=\Sigma_\textup{ann}+\Sigma_\textup{semi} \ ,\\
        \label{eq:eta}
        \eta&=\frac{\Sigma_{\rm semi}}{\Sigma_{\rm eff}} \ .
    \end{align}
\end{subequations}
Notice that both the freeze-out temperature and the relic abundance are uniquely determined by the total thermally averaged effective cross section $\Sigma_{\rm eff}$. Therefore, the relic density constraint fixes only its overall magnitude, but does not determine the relative size of the cross sections of the different processes, encoded in the weight $\eta$.

\begin{figure}
    \centering
    \includegraphics[width=0.7\linewidth]{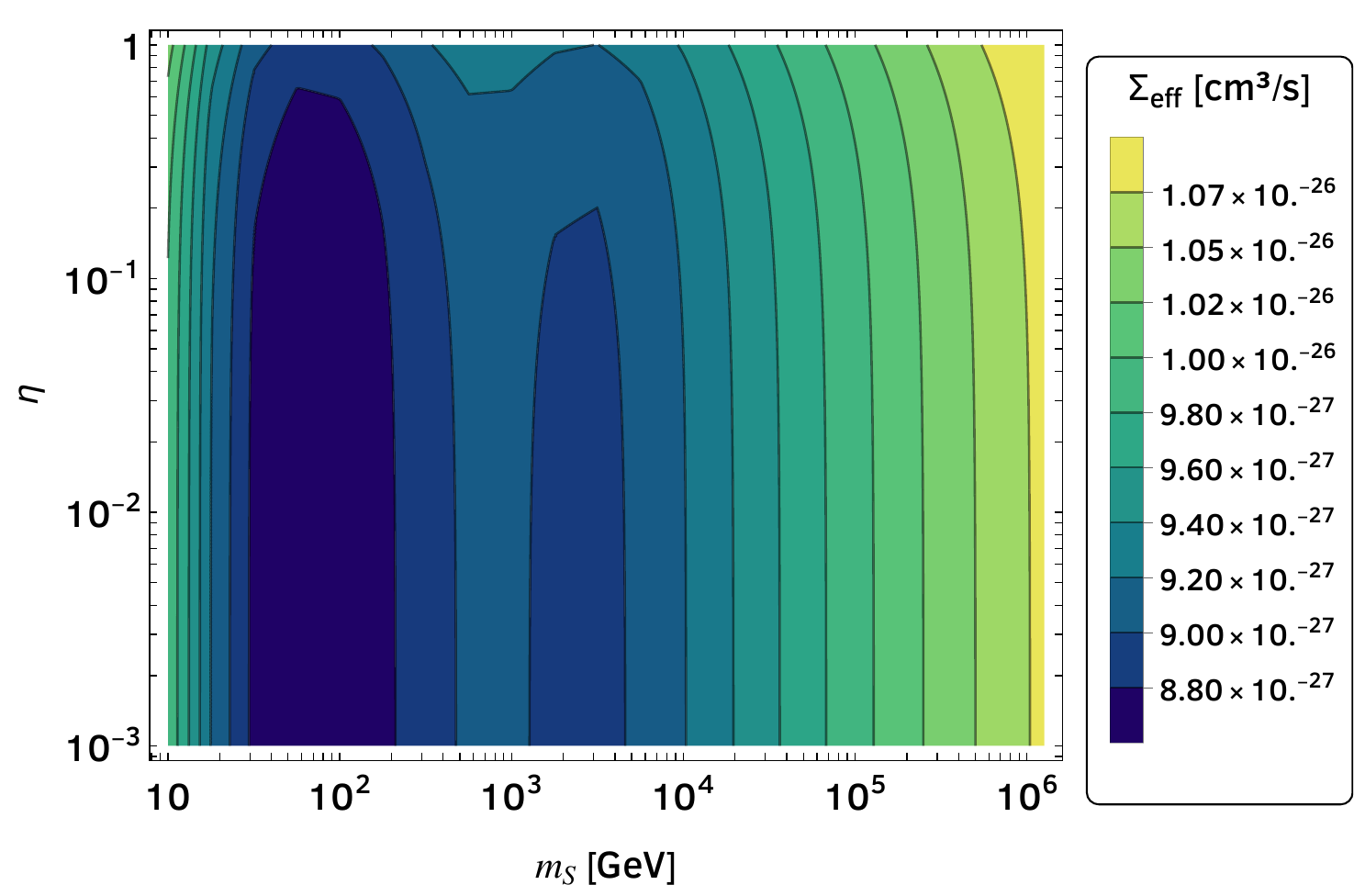}
    \caption{Relative semi-annihilation contribution $\eta$ (defined in \Eq{eq:eta}) as a function of the mass of the DM particle $m_S$ for constant values of the total, effective averaged cross section $\Sigma_{\rm eff}$. All the parameter combinations reproduce the DM abundance observed today. }
    \label{fig:relic}
\end{figure}

This property, identified by the semi-analytical solution, is confirmed to high accuracy by the dedicated numerical analysis illustrated in \Fig{fig:relic}. In the $(m_S, \eta)$ plane, we identify isocontours of $\Sigma_{\rm eff}$ by requiring that the observed DM abundance is correctly reproduced. One observes only an extremely mild dependence on the value of the relative weight $\eta$. In other words, for any choice of $\eta$, the relic density is reproduced for essentially the same value of the total effective cross section $\Sigma_{\rm eff}$. For a fixed DM mass, as long as $\Sigma_{\rm eff}$ matches the typical thermal value targeted by ID experiments $(\sim 10^{-26}\,\cms)$, the relative contributions of the different processes do not alter the freeze-out abundance. This implies that our model-independent formalism captures a broad spectrum of scenarios, in which the individual scattering cross sections of each process are generally unconstrained, provided that their sum reproduces the observed DM relic density.

Before moving to the discussion of the CR spectra, we summarize here the assumptions underlying the relic density calculation adopted throughout this work. We assume that the DM abundance is determined by unsuppressed $s$-wave processes, that no dark sector asymmetry is present, and that the mediator is kept in thermal equilibrium with the SM bath. Furthermore, we neglect the effects of coannihilations, Sommerfeld enhancement, bound-state formation, resonant effects, and late mediator dynamics. Within this framework, higher partial waves are expected to provide only subleading corrections.

\subsection{Cosmic ray fluxes}
\label{sec:injection_spectra}

Having clarified the freeze-out production at early times, we now turn to late-time ID signals. The key observables in this context are the fluxes of stable SM particles produced via DM (semi-)annihilations. The theoretical prediction of detectable CR spectra involves two main steps: the determination of the production rate per unit volume of stable SM particles at the source, and a detailed description of their propagation in the interstellar or intergalactic medium. The former is set by a combination of the astrophysical DM distribution, the underlying particle-physics framework for DM interactions, and SM hadronization/showering, while the latter depends on the specific messenger. A detailed study of propagation effects on the final CR flux for all messengers lies beyond the scope of this work. Instead, we focus on the injection spectra and explicitly compute exemplary CR fluxes only for $\gamma$ rays in \Sec{sec:gamma_flux}. Nevertheless, we present in what follows a complete formalism applicable to all messengers, including charged ones, to facilitate future use of the framework introduced in this paper.

In what follows, we denote by $X$ the specific messenger under consideration, corresponding to the final state of the secondary showering of primary SM states $\psi_{\rm SM}$ produced by the (cascade) processes. Explicit examples of $X$, which will be addressed in the next sections, include photons, neutrinos, stable SM charged particles, and anti-nuclei. The processes in \Eq{eq:processes} yield a detectable $X$ flux through the following chains
\begin{subequations}\label{eq:reaction_chain} 
\begin{align} 
\text{Annihilation to SM pairs:} & \, \qquad  \qquad S S^\star \, \rightarrow \, \psi_{\rm SM} \overline{\psi}_{\rm SM} \, \rightarrow \, X + \ldots \\ 
\text{Annihilation to $\phi$ pairs:} & \, \qquad  \qquad S S^\star \, \rightarrow \, \phi \phi \, \rightarrow \, 2\times\psi_{\rm SM} \overline{\psi}_{\rm SM} \, \rightarrow \, X + \ldots \\ 
\text{Semi-annihilation to single $\phi$:} & \, \qquad  \qquad S S \, \rightarrow \, S^\star \phi \, \rightarrow \, 
S^\star \psi_{\rm SM} \overline{\psi}_{\rm SM} \, \rightarrow \, S^\star \, X + \ldots .  \\
& \, \qquad  \qquad S^\star S^\star \, \rightarrow \, S \phi  \, \rightarrow \, 
S \psi_{\rm SM} \overline{\psi}_{\rm SM} \, \rightarrow \, S \, X + \ldots .  
\end{align} 
\end{subequations}
Notice that semi-annihilation processes also produce a boosted DM particle in the final state, which can be searched for using other methods~\cite{Agashe:2014yua,Berger:2014sqa,Necib:2016aez,Giudice:2017zke}.

The expected number of produced cosmic messengers $X$ per unit time, volume, and energy bin, prior to propagation, can be evaluated from three fundamental ingredients: the DM number density at the source, the DM collision rate, and the differential energy distribution of $X$ particles produced per single DM collision. This latter quantity, known as the differential \textit{injection spectrum}, is the main focus of this work.

We start from neutral messengers, i.e.~high-energy photons and neutrinos directly created by DM collisions and prior to secondary CR scattering. We focus here on dSphs as the primary sources, which will be used as a showcase in \Sec{sec:gamma_flux}. For a given process of type $a$, the number of collisions per unit time and volume is obtained by multiplying the local squared DM number density at the source by the thermally averaged scattering cross section of DM particles $\braket{\sigma_a \vmol}$.\footnote{The DM number density is given by its energy density divided by its mass.} The quadratic dependence reflects the number of DM particles in the initial state participating in the process, as we consider only ID signals sourced by binary collisions. The resulting differential flux of prompt $\gamma$ rays (a completely analogous expression holds for neutrinos), produced in a dSph and integrated over a solid angle $\Delta\Omega$, is given by~\cite{Cirelli:2010xx,Cirelli:2024ssz}:
\be\label{eq:gamma_flux_standard}
\frac{d\Phi_\gamma}{dE_\gamma}= \sum_a \frac{\kappa_a}{2} \,\frac{\braket{\sigma v}_a}{4 \pi\mdm^2} \times\,\underbrace{\frac{dN^a_\gamma}{dE_\gamma}}_{\substack{\textit{injection}\\ \textit{spectrum}}} \,\times\,J(\Delta\Omega)\,\,.
\ee
This equation is valid under the assumption of velocity-independent DM cross sections. If thermal averages were dominated by higher multipoles in the partial wave expansion, \Eq{eq:gamma_flux_standard} should be modified to account for a velocity-weighed astrophysical factor. This expression is thus consistent with the assumptions taken so far.
The sum over $a$ runs over all processes with two DM (anti-)particles in the initial state. The combinatorial factor $\kappa_a$ accounts for the number of colliding DM pairs; we will provide its explicit value for each process below. The highlighted injection spectrum $dN^{a}_{\gamma}/dE_{\gamma}$ is the differential energy distribution of photons produced in a single process of type $a$. The $J$ factor encodes the astrophysical input and corresponds to the line-of-sight integral of the squared DM energy density from the source to the Earth~\cite{Cirelli:2010xx,Cirelli:2024ssz}. Accounting for the integration of the signal over a solid angle $\Delta\Omega$ corresponding to the angular extent of the dSph, the $J$ factor is defined as
\be\label{eq:Jfactorgen}
J(\Delta\Omega)=\int_{\Delta\Omega}\int_{\text{l.o.s.}}\rho_S^2(\ell,\,\Omega)\,d\ell\,d\Omega\,\,,
\ee
where $\ell$ is the radial distance along the line of sight from the Earth to the source, and $\rho_S$ is the DM density profile of the dSph.

It is worth stressing that the differential flux in \Eq{eq:gamma_flux_standard} is derived under the assumption of a mediator coupled to a single SM field $\psi_{\rm SM}$. This assumption simplifies the expression, as it avoids the need to account for the superposition of CR injection spectra arising from different primary production channels for a given process $a$. The extension to a more general scenario, in which the mediator $\phi$ couples to multiple SM fields, requires a straightforward modification to include: (i) a sum over the different primary production channels; and (ii) the corresponding branching ratios of the mediator decays into different SM.\footnote{The resulting generalization of \Eq{eq:gamma_flux_standard} for the total $\gamma-$ray injection spectrum reads \be
\frac{d\Phi_\gamma}{dE_\gamma}= \sum_a \frac{\kappa_a}{2} \,\frac{\braket{\sigma v}_a}{4 \pi\mdm^2} \times\,\sum_{\psi_{\rm SM}}{\rm BR}(\phi\to\overline{\psi}_{\rm SM}\psi_{\rm SM}) \ \frac{dN_\gamma^{a(\psi_{\rm SM})}}{dE_\gamma} \,\times\,J(\Delta\Omega) \ .
\label{eq:dphigammaBR}
\ee
} 

Propagation effects through the (inter)galactic medium are highly non-trivial for charged CR messengers and do not allow for an expression analogous to \Eq{eq:gamma_flux_standard} to be derived in these cases. Nonetheless, the propagation equation contains a fundamental ingredient related to the injection spectrum: the \textit{source term} $\mathcal{Q}$. Considering DM binary collisions occurring in the Galactic halo, this quantity can be defined as the product of the collision rate at a given location within the source and the CR injection spectrum:
\be\label{eq:chargedX_source_gen}
\mathcal{Q}_X(r,E_X)=\sum_a\frac{\kappa_a}{2}\braket{\sigma v}_a\left(\frac{\rho_S(r)}{\mdm}\right)^2\frac{dN_X^{a}}{dE_X} \ . 
\ee
The combinatorial factor $\kappa_a$ is the same as in \Eq{eq:gamma_flux_standard}, as it accounts for the number of possible initial states. The fundamental difference with respect to the case of neutral species lies in the astrophysical input, which is now the DM energy density $\rho_S(r)$ evaluated at position $r$ in Galactic coordinates. This reflects the fact that DM acts as a diffuse source throughout the halo, leading to a spatial modulation of the signal intensity with the local DM density. The expression for the source term of charged CRs can be generalized to include the effect of the existence of multiple mediator coupling to different SM particles analogously to \Eq{eq:dphigammaBR}.

Looking back at \Eqs{eq:gamma_flux_standard}{eq:chargedX_source_gen}, we note that the thermally averaged cross sections and the DM mass are key particle physics inputs. These quantities must combine in such a way that the observed DM relic abundance is correctly reproduced. As shown in \Sec{sec:freezeout}, this requirement places only mild constraints on the DM mass $m_S$. Another essential particle physics ingredient shaping the resulting CR fluxes is the differential injection spectrum of $X$ messengers (either neutral or charged) for a single event among those listed in \Eq{eq:processes}. While the spectra for DM direct annihilation (or decay) are typically tabulated~\cite{Cirelli:2010xx,Arina:2023eic,Arina:2025ner}, their derivation is more involved in the case of one-step processes. The next subsection presents suitable expressions for the injection spectra from one-step cascade processes.

\subsection{Injection spectra for one-step cascade processes}

DM (semi-)annihilations take place approximately in the Galactic Frame (GF), given the tiny kinetic energies involved. For one-step processes, primary SM states are produced by in-flight decays of on-shell metastable mediators. As a consequence, the physical spectra of stable $X$ messengers computed in the GF are obtained by combining the CR spectra at production, computed in the $\phi$ rest frame, with the Lorentz boost to the center-of-mass frame of DM collisions. 

The injection spectrum boosted to the GF is derived in \App{app:kinematics}. Under the assumption of isotropic decays of the mediator in its rest frame,\footnote{This condition is certainly fulfilled when the mediator is a spin-0 bosonic field. A dependence on the emission angle would arise, for instance, for spin-1 vector mediators as discussed in Ref.~\cite{Elor:2015tva}.} the boosted injection spectrum of a given messenger $X$ is given by
\be\label{eq:dNdE}
\frac{dN_X}{dE_X}=\frac{\ell}{2}\times\frac{1}{\gamma_L\beta_L}\int_{E_X^{'\,\rm min}}^{E_X^{'\,\rm max}} \frac{dE_X'}{\sqrt{E_X'^{\,2}-m_X^2}} \frac{dN_X}{dE_X'}\,\,.
\ee
The factor $\ell$ counts the number of on-shell $\phi$ particles produced in the collision process: it takes the value $1$ for DM semi-annihilation and $2$ for one-step annihilations (see \Eq{eq:reaction_chain}). The Lorentz factors $\gamma_L$ and $\beta_L$ characterize the boost between the $\phi$ rest frame and the DM center-of-mass frame. They affect not only the normalization of the one-step injection spectrum, but also the integration interval: the integration bounds $E_X^{'\,\rm min/max}$, corresponding to the minimum and maximum energies of $X$ particles in the $\phi$ rest frame, are determined by kinematics and depend on the Lorentz boost, see \App{app:kinematics}. Finally, the integrand features what we dub the \textit{zero-step} injection spectrum $dN_X/dE_X'$, which corresponds to the spectrum of $X$ particles in the $\phi$ rest frame.

The main effect of the boost is to suppress the signal intensity and shift its maximum to lower energies with respect to the direct annihilation case. This reasoning applies, although to different extents, to both one-step cascade annihilation and semi-annihilation spectra. To better appreciate both qualitatively and quantitatively the differences in the spectral shapes associated with each process, it is convenient to consider the ratio of injection spectra
\be\label{eq:ratio}
R_{ab}(E_X) = \frac{dN^a_X / dE_X}{dN^b_X / dE_X} \ , \qquad\qquad a \neq b\,\,,
\ee
where $a,b$ denote direct annihilation or one-step (semi-)annihilation. Notice that, as discussed in detail in \App{app:kinematics}, one-step semi-annihilation spectra generally have a smaller support than those arising from (one-step) annihilation processes.

\subsection{Injection spectra for arbitrary linear combinations of processes}

It is realistic to expect that DM models may naturally accommodate the coexistence of more than one type of collision process. We defer to \Sec{sec:models} the description of minimal Lagrangians featuring DM candidates with two or more of the processes illustrated in \Fig{fig:diagrams}. Here, we present a framework to evaluate CR fluxes in terms of effective dimensionless ratios that quantify the relative weights of the different CR production channels. 

Whether we consider the production of neutral or charged CRs, with differential fluxes and source terms given respectively by \Eqs{eq:gamma_flux_standard}{eq:chargedX_source_gen}, we need to specify the coefficient $\kappa_a$, which counts the number of binary collisions of type $a$. Following standard conventions in the literature, these expressions are normalized such that for standard annihilation of Majorana DM we have $\kappa_a = 1$. Indeed, when DM particles coincide with their own antiparticles, the number of possible annihilations per unit volume is $n_{\rm DM}^2/2 = \rho_{\rm DM}^2 / (2 m_{\rm DM}^2)$, from which the explicit factor of $1/2$ arises. 

However, we perform our analysis under the assumption that DM is not a self-conjugate particle. Assuming no matter-antimatter asymmetry in the dark sector, the total DM density is $\rho_{\rm DM}=\rho_S+\rho_{S^\star}=2\rho_S$. Within this framework, the number of DM annihilations per unit volume is $n_S n_{S^\star} = \rho_S^2 / (4 m_S^2)$, and therefore $\kappa_{\rm ann}=1/2$. For semi-annihilations $SS\to S^\star\phi$, we consider a process involving two identical DM particles. The number of possible initial states is $n_S^2/2 = \rho_S^2 / (8 m_S^2)$, which implies $\kappa_{\rm semi}=1/4$. 

We put everything together by performing the explicit sum over the different processes. We present an explicit derivation for the $\gamma$-ray flux; the one for the source term of charged CRs is identical. The sum in \Eq{eq:gamma_flux_standard} can be written explicitly as
\be
\begin{split}
  \frac{d\Phi_\gamma}{dE_\gamma} = \frac{J(\Delta\Omega)}{8 \pi m_S^2} 
  & \left[ \frac{1}{2} \braket{\sigma_{\scriptscriptstyle SS^\star\to \psi_{\rm SM}\bar\psi_{\rm SM}}\, v} \frac{dN^{\scriptscriptstyle SS^\star\to \psi_{\rm SM}\bar\psi_{\rm SM}}_\gamma}{dE_\gamma} +  \frac{1}{2} \braket{\sigma_{\scriptscriptstyle SS^\star\to \phi\phi}\, v} \frac{dN^{\scriptscriptstyle SS^\star\to \phi\phi}_\gamma}{dE_\gamma} \right. 
  \\ & \left. + \frac{1}{4}  \braket{\sigma_{\scriptscriptstyle SS\to S^\star\phi}\, v} \frac{dN^{\scriptscriptstyle SS\to S^\star\phi}_\gamma}{dE_\gamma} + \frac{1}{4} \braket{\sigma_{\scriptscriptstyle S^\star S^\star\to S\phi}\, v} \frac{dN^{\scriptscriptstyle S^\star S^\star\to S\phi}_\gamma}{dE_\gamma}
  \right]   \ .
\end{split}
\ee
The second line contains the contributions from semi-annihilation, including the conjugate process. As mentioned before, we work under the assumption of CP conservation that implies identical rates for both semi-annihilation processes $S S \to S^\star\phi$ and $S^\star S^\star\to S\phi$. Furthermore, the CR injection spectra are also identical, as they arise from the decays of the same mediator $\phi$. Consequently, for all practical purposes, the overall contribution from semi-annihilation to the combinatorial factor is equivalent to that from annihilation.

We can rewrite the result obtained above in such a way that it effectively takes the form of a single process producing $\gamma$ rays
\be
\frac{d\Phi_\gamma}{dE_\gamma} = \frac{J(\Delta\Omega)}{16 \pi m_S^2} \, \Sigma_{\rm eff} \,  \frac{dN^{\scriptscriptstyle \rm eff}_\gamma}{dE_\gamma} \ ,
\label{eq:dPhigammagen}
\ee
where the effective cross section $\Sigma_{\rm eff}$ has already been defined in \Eq{eq:Sigmaeff}. The effective injection spectrum is defined as
\be
\frac{dN^{\scriptscriptstyle \rm eff}_\gamma}{dE_\gamma} =  (1 - \alpha - \beta) \frac{dN^{\scriptscriptstyle SS^\star\to \psi_{\rm SM}\bar\psi_{\rm SM}}_\gamma}{dE_\gamma} +  \alpha \frac{dN^{\scriptscriptstyle SS^\star\to \phi\phi}_\gamma}{dE_\gamma} + \beta \frac{dN^{\scriptscriptstyle SS\to S^\star\phi}_\gamma}{dE_\gamma}  \ ,
\label{eq:phigammagen}
\ee
with relative weights expressed in terms of ratios of cross sections,
\begin{align}\label{eq:alphabeta}
\alpha&\equiv\frac{\braket{\sigma v}_{SS^\star\to \phi\phi}}{\Sigma_{\rm eff}}\,\,,& \beta&\equiv\frac{\braket{\sigma v}_{SS\to S^\star\phi}}{\Sigma_{\rm eff}} = \eta\,\, .
\end{align}
From \Eq{eq:dPhigammagen}, we observe that the differential $\gamma$-ray flux has an overall normalization set by the same quantity $\Sigma_{\rm eff}$ that determines the DM relic density. In other words, even in the most general case where a generic linear combination of processes is present, there remains a direct correspondence between the relic density requirement and the $\gamma$ ray flux normalization. The spectral shape, instead, is determined by the relative weights $\alpha$ and $\beta$; in particular, $\beta$ coincides with the parameter $\eta$ defined in \Eq{eq:eta}. 

While the expression in \Eq{eq:phigammagen} has been derived for the specific case of $\gamma$ rays, its generalization to arbitrary messengers is straightforward. An analogous expression holds for the neutrino flux. For charged messengers, the generalization of \Eq{eq:chargedX_source_gen} reads explicitly
\be\label{eq:chargedX_mixing}
\mathcal{Q}_X(r,E_X) = \frac{1}{4} \left(\frac{\rho_S(r)}{\mdm}\right)^2 \, \Sigma_{\rm eff} \, \frac{dN^{\scriptscriptstyle \rm eff}_X}{dE_X} \ .
\ee
The overall rate is again proportional to the effective cross section $\Sigma_{\rm eff}$, while the effective injection spectrum is obtained from \Eq{eq:phigammagen} by the replacement $\gamma \rightarrow X$.  

From this point on, we treat $(\alpha,\,\beta)$ as independent variables to characterize the spectral shape. The configurations $(\alpha,\beta)=(0,0), (1,0), (0,1)$ correspond to benchmark scenarios in which DM phenomenology is entirely determined by a single process, namely direct annihilation, one-step annihilation, and one-step semi-annihilation, respectively. We consider both these limiting cases and more general combinations, in which the three injection spectra are mixed, each contributing its own characteristic features.  In particular, we consider the following combinations: $(\alpha,\,\beta)=(0.5,0)$, referred to as the \textit{global $U(1)_S$ scenario}, since this symmetry forbids the semi-annihilation channel; $(\alpha,\,\beta)=(0.5,0.5)$, labeled as the \textit{cascades scenario}, since direct annihilations are absent in this case; and $(\alpha,\,\beta)=(0,0.5)$, corresponding to the \textit{direct+semi scenario}, named after the processes that are active. Finally, allowing for a generic mixture of all three contributions leads to what we refer to as an \textit{anarchic scenario}. For definiteness, we fix $(\alpha,\,\beta)=(1/3,1/3)$ to illustrate the case in which the three collision processes contribute with equal weight to both the relic density and ID phenomenology. The resulting signal therefore reflects the interplay among the different components, which may be enhanced or suppressed depending on their relative weights. In the following, for all computational purposes, we will use the most updated CR injection spectra provided by Ref.~\cite{Arina:2023eic} for the direct DM annihilation  case. The same functions are fed as the zero-step injection spectra from $\phi$ decays into \Eq{eq:dNdE} to compute the differential distribution of $X$ particles generated by one-step cascade processes.


\section{Gamma-ray injection spectra}
\label{sec:gammas}

Photons, and $\gamma$ rays in particular, are an interesting and convenient type of indirect messenger of DM. 
A variety of ID facilities including the \textit{Fermi}-Large Area Telescope (LAT) \cite{2009ApJ...697.1071A,McDaniel:2023bju}, H.E.S.S. \cite{Hinton:2004eu,HESS:2016mib}, and forthcoming experiments like the Southern Wide-field Gamma-ray Observatory (SWGO) \cite{SWGO:2025taj} and the Cherenkov Telescope Array Observatory (CTAO) \cite{CTAConsortium:2021cvs}, aim at exploring the $\gamma-$ray sky searching for signatures tracing the expected spectral (and spatial) characteristics of DM interactions, the former shaped by the injection spectra as detailed in the previous section.

We start illustrating ${{dN^a_\gamma}/{dE_\gamma}}$, the  photon differential injection spectra per unit energy per \textit{single} DM collision, as produced in a scenario in which the phenomenology is determined  by a single scattering process among the three alternatives (labeled by $a$, see \Eq{eq:reaction_chain}) discussed in this paper, or by their combination. Since we are interested in the $\gamma-$ray spectra produced per single DM collision event, we do not make any distinction between the two conjugate processes that participate to semi-annihilations, and keep only one of them since they yield the same photon distribution. On the other hand, when looking at the superposition of the injection spectra of different scattering processes (see below), we  account for both contributions as far as semi-annihilations are involved. This is required to properly quantify  the total contribution of semi-annihilation to $\gamma-$ray yields.

For illustrative purposes, we fix the benchmark values $(\mdm=1\,\TeV,\,m_\phi=200\,\GeV)$ in all shown figures, unless otherwise stated. However, we remind that our approach to compute CR injection spectra is valid as long as cascades and mediator decays are both kinematically allowed by the conditions $m_S > m_\phi > 2 m_{\psi_{\rm SM}}$. Within this region, the derivation is valid regardless of the mass hierarchy between the particles involved in the processes. On top of this condition, the variation of spectral features with the DM and mediator masses (for fixed $\psi_{\rm SM}$ and $X$ states) would be better appreciated by parametrizing CR injection spectra as functions of the energy fraction $\xdm=E_X/\mdm$. In terms of this variable, CR spectra computed for different values of $\mdm$ are rescaled into the same domain and factor out $\mdm$ effects. This leaves to $m_\phi$ the role of imprinting different features onto spectral functions. The DM mass $\mdm$ first determines the amount of hadronization/EW showering, hence shaping the resulting spectrum. This role is left to the mediator mass $m_\phi$ in case of cascade processes. In the latter scenario, the ratio $m_\phi/\mdm$ crucially defines the Lorentz boost of final CR particles, affecting both the normalization and the shape of the injection spectrum. While we defer to \App{app:kinematics} for a detailed derivation of the expression of CR injection spectra in terms of $\xdm$ and a thorough discussion of finite mass effects, we will touch upon the effect of varying the mediator-to-DM mass ratio later in this section. In what follows, we will always present the injection spectra in terms of the physical energy scale $E_X$ to make a direct comparison with experimental signals. This choice has the additional benefit of keeping track of the physical mass scales in the process and identifying what final states are kinematically allowed.

\begin{figure}[t]
    \centering
    \includegraphics[width=0.5\linewidth]{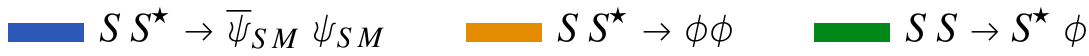}
    
    \medskip
    \subfloat[]{%
        \includegraphics[width=.48\linewidth]{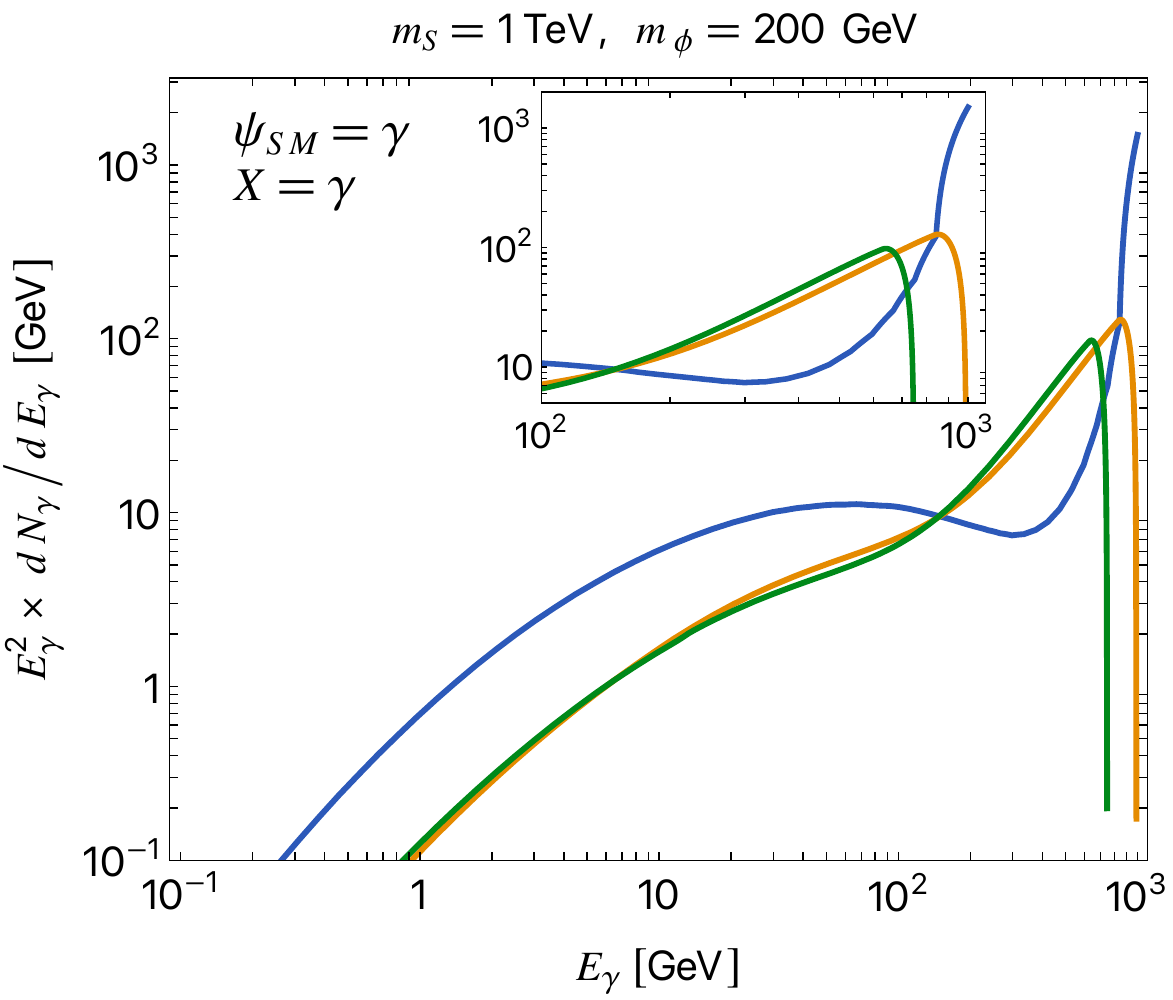}%
        \label{fig:cfr_gamma_gamma}%
    }\quad
    \subfloat[]{%
        \includegraphics[width=.48\linewidth]{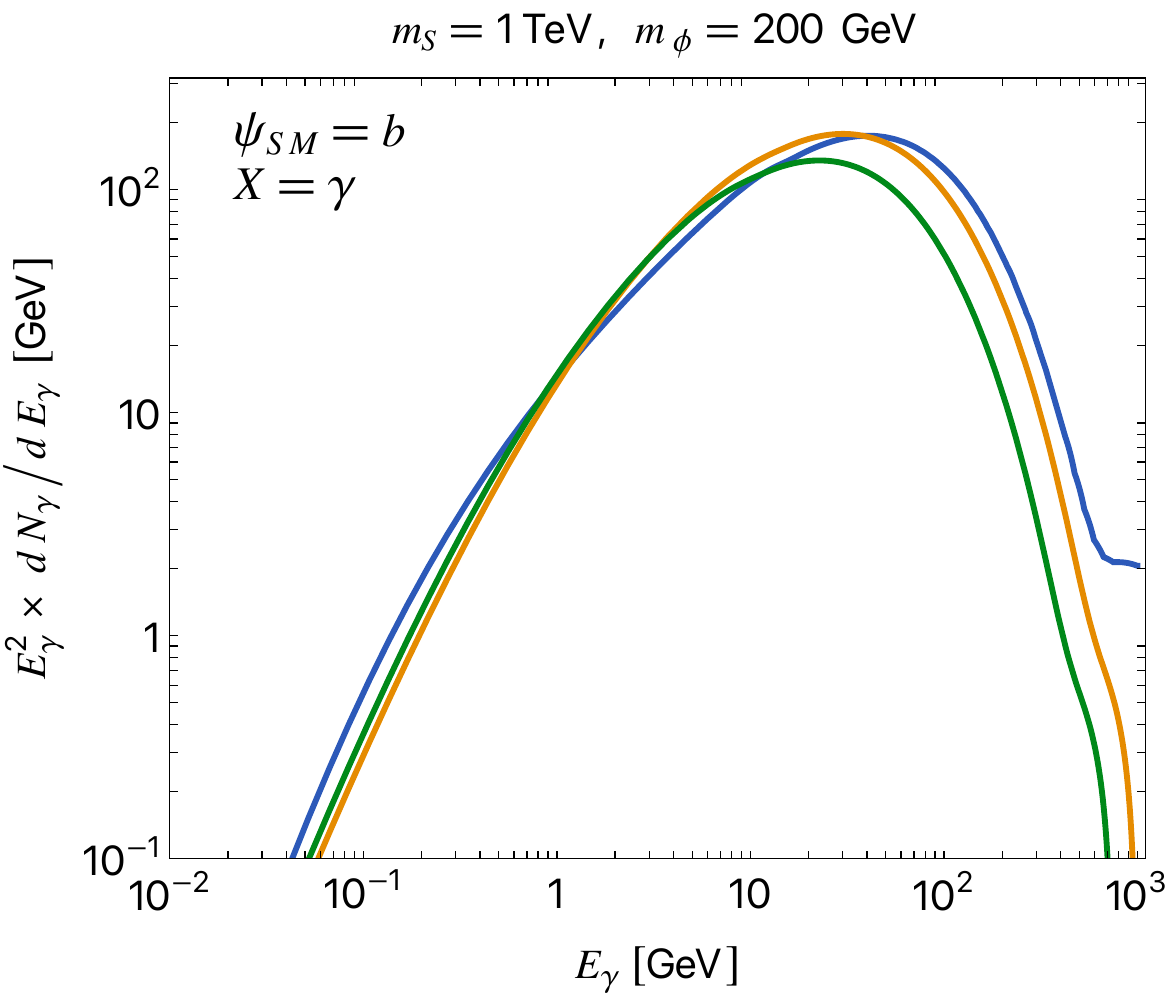}%
        \label{fig:cfr_b_gamma}%
    }\\
    \subfloat[]{%
        \includegraphics[width=.48\linewidth]{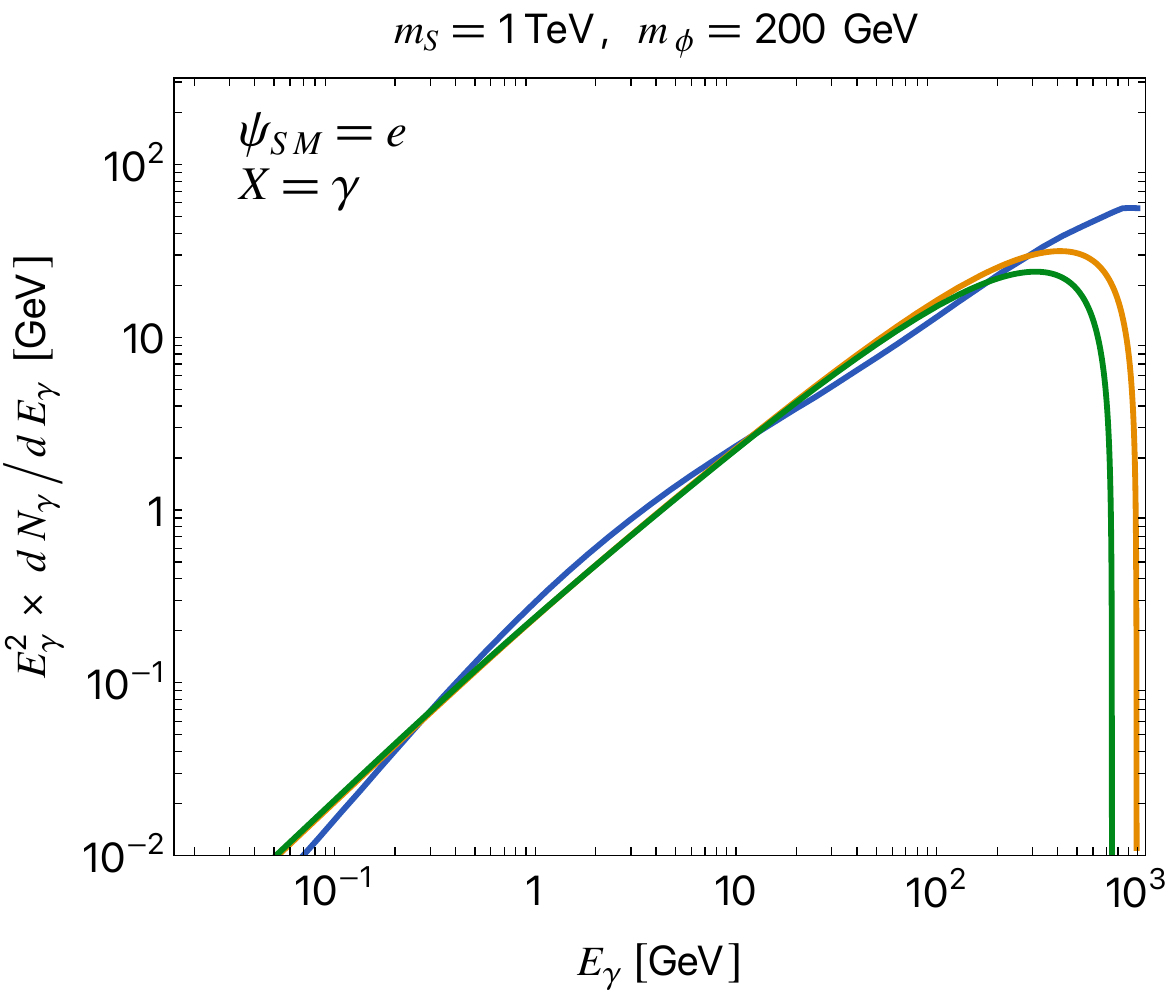}%
        \label{fig:cfr_e_gamma}%
    }\quad
    \subfloat[]{%
        \includegraphics[width=.475\linewidth]{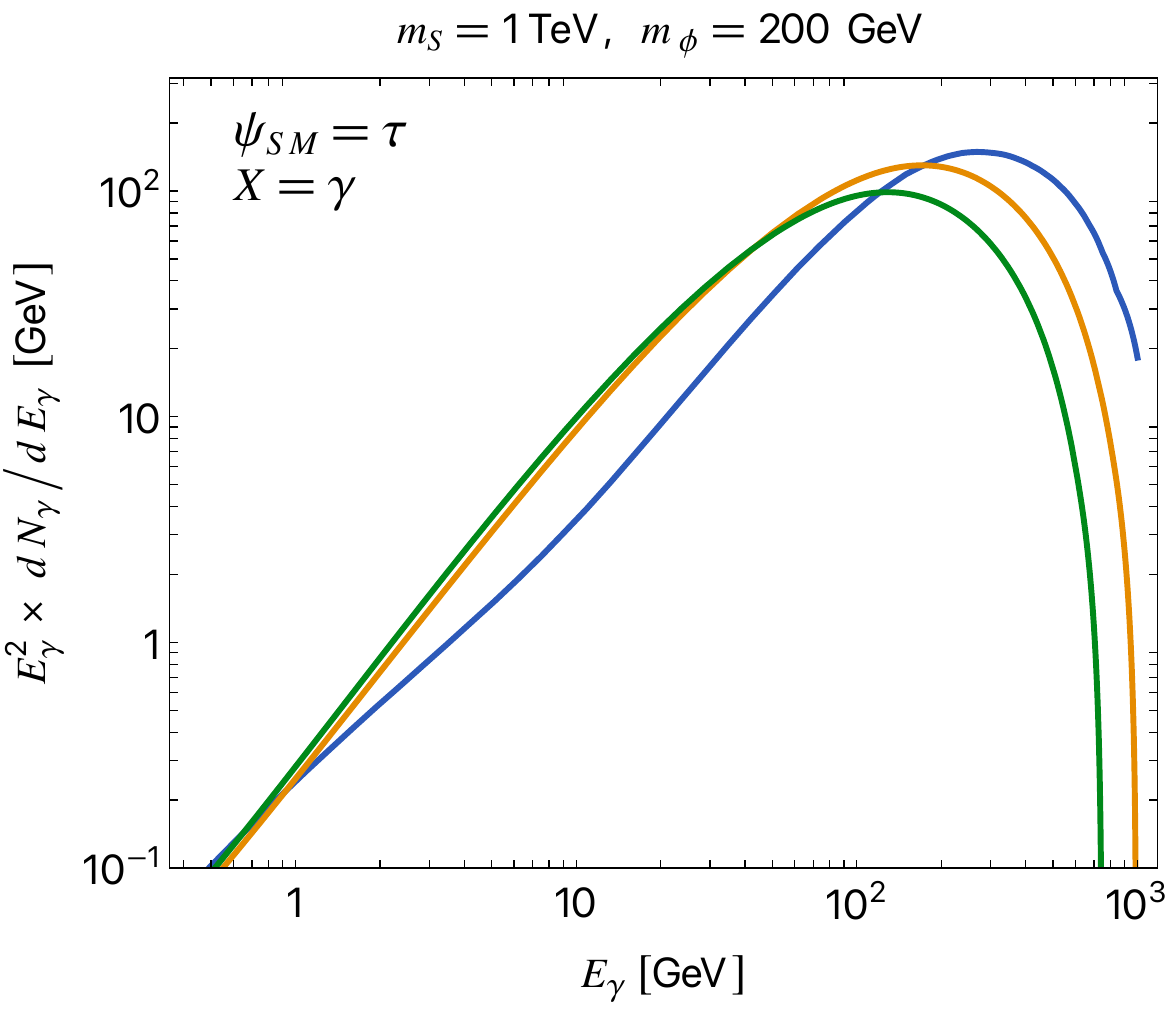}%
        \label{fig:cfr_tau_gamma}%
    } 
    \caption{\textit{Single process injection spectra: $\gamma$ rays.}
    Differential injection spectra of $\gamma$ rays per single DM collision process when the signal is determined exclusively by: direct annihilation (\textit{blue}), one-step annihilation (\textit{orange}) or one-step semi-annihilation (\textit{green}) processes. Each panel (a,b,c,d) refers to a different final state particle $\psi_{\rm SM}=(\gamma,\,b,\,\tau,\, e)$, going clockwise starting from the top-left panel.}
    
    \label{fig:cfr_gammas}
\end{figure}

The four panels of \Fig{fig:cfr_gammas} illustrate the differential injection spectra of $\gamma$ rays per single DM collision process, each for a different final state particle $\psi_{\rm SM}=(\gamma,\,b,\,\tau,\,e)$: photons, bottom quarks, tau leptons and electrons and positrons. In each panel, the three lines correspond to a different process: direct annihilation (blue), one-step annihilation (orange) or one-step semi-annihilation (green). We note that for all three cases, the total energy budget available for the collision is set by the DM mass $\mdm$. Indeed, for cold relics, and within the $s-$wave approximation of the thermally averaged cross section, velocity effects are safely negligible.
Inspecting the features that characterize each $\gamma-$ray spectrum, the first important difference is observed for processes ending into primary photons (\Fig{fig:cfr_gamma_gamma}). 
The direct annihilation signal is given, in first approximation, by the monochromatic emission of pairs of prompt high energetic photons with $E_\gamma=\mdm$. The spectrum then decays smoothly at lower energies, even presenting a small bump (still suppressed by orders of magnitude with respect to the main peak), due to secondary radiation emission from QED or EW showering \cite{Cirelli:2010xx,Arina:2023eic}. 
This pronounced line-like feature is instead not present for $\gamma-$ray spectra produced by one-step cascade processes, and leaves space to a 
``blunt box'' shape, as magnified in the zoom-in panel.
As discussed in \Sec{sec:injection_spectra}, these collisions proceed by first producing intermediate on-shell $\phi$ states, which later decay in-flight into pairs of primary states (photons in this case). This requires the application of a Lorentz boost to obtain the injection spectra in the center of mass frame of the colliding DM particles.
In the case of primary photon emission, the naive prediction is that this would create a box-shaped spectrum \cite{Ibarra:2012dw,DiazSaez:2021pmg}, which sharp edges gets attenuated in reality by the secondary EW/QED emission. Notice how the ``blunt box'' for the semi-annihilation scenario is shifted to lower energies and ends at $\sim3\mdm/4$. 
We note that this is expected for cases where $m_S\gg m_\phi$, i.e. in a large mass hierarchy where the mediator is much lighter than the DM particle. Instead, when the mediator mass is not negligible, this behavior gets corrected by factors of the order of $m_\phi^2/\mdm$, see \Eq{eq:semi-ann kinematics} in \App{app:kinematics}. 

\begin{figure}[t]
    \centering
    \includegraphics[width=0.5\linewidth]{Figures/legenda_spectra.pdf}
    \\
    \medskip
    \subfloat[\label{fig:cfr_eps_onestep}]{\includegraphics[width=.48\linewidth]{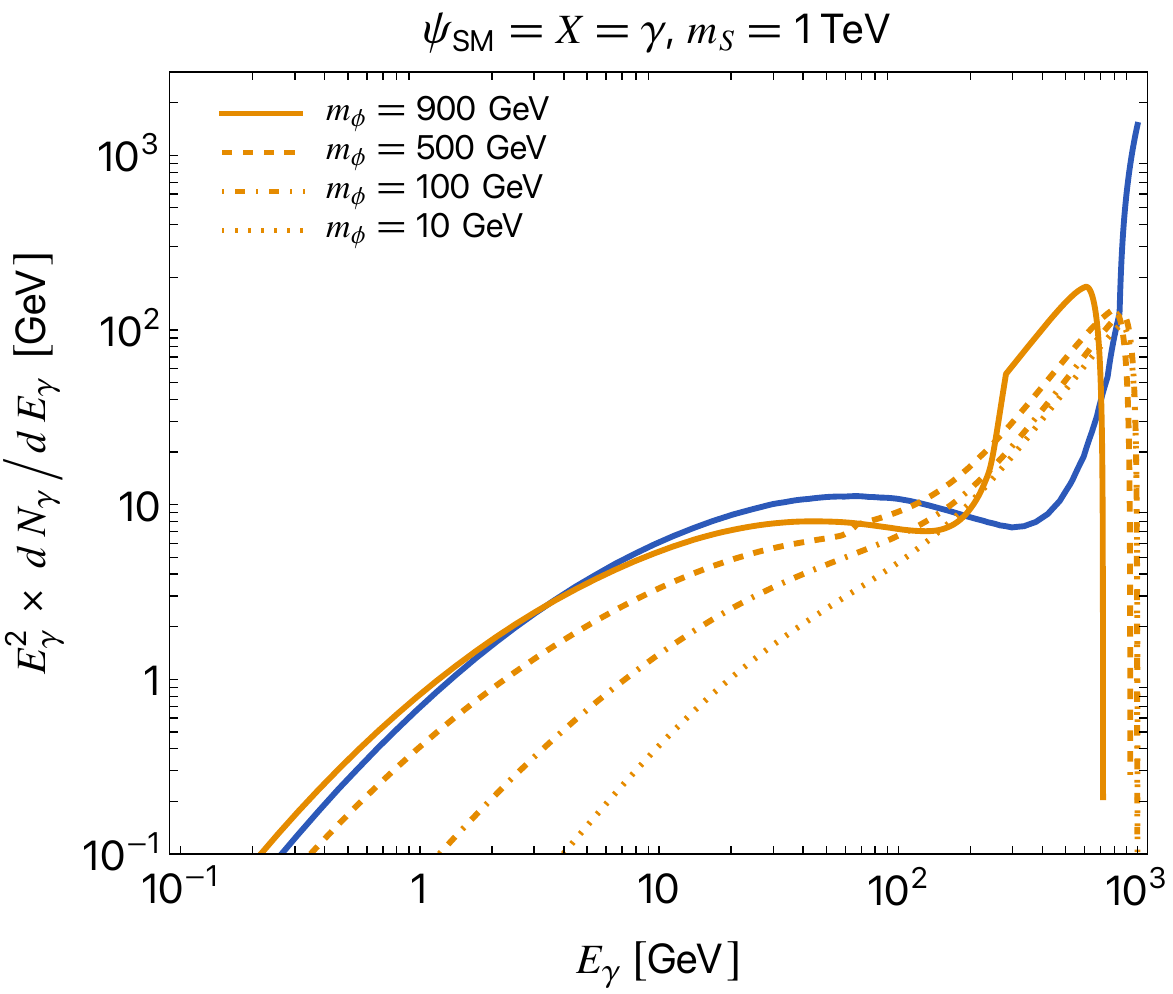}}\quad
    \subfloat[\label{fig:cfr_eps_semi}]{\includegraphics[width=.48\linewidth]{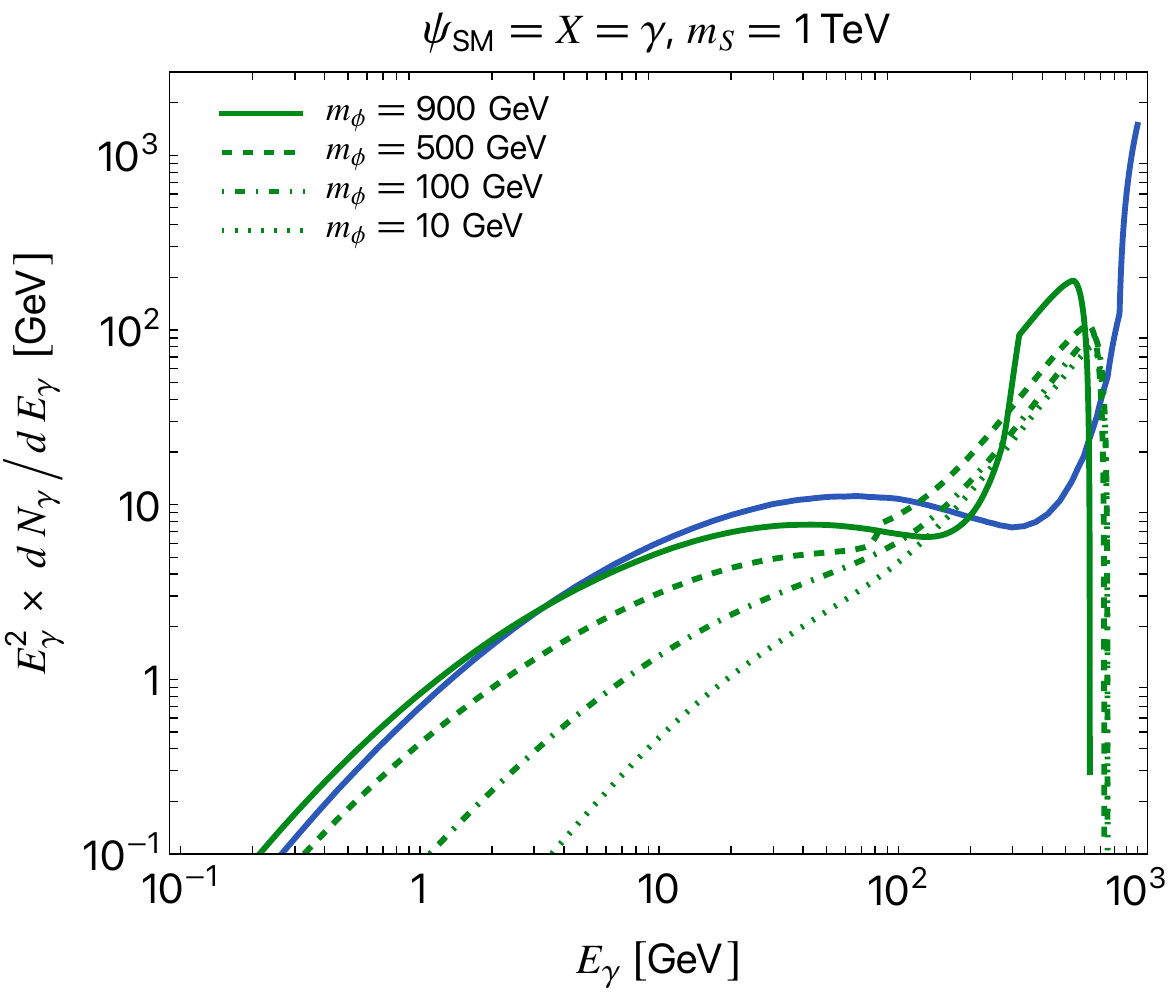}}\\
    \caption{We study the effect of varying the mediator-to-DM mass ratio $m_\phi/\mdm$  onto the spectral features of $\gamma$ rays. We fix the DM mass to $\mdm=1\,\TeV$ and vary $m_\phi=(900,500,100,10)\,\GeV$. The injection spectrum generated by direct DM annihilation is shown in \textit{blue} for reference, while the spectra yielded by one-step annihilations and one-step semi-annihilations are shown in \textit{orange} and \textit{green} in the \textit{left} and \textit{right} panels, respectively.}
    \label{fig:cfr_massratio_gamma}
\end{figure}

We find it convenient to pick this specific benchmark case of study $(\psi_{\rm SM}=X=\gamma)$ to shed light on the role that the mass ratio $m_\phi/\mdm$ plays in shaping the $\gamma-$ray injection spectra generated by one-step cascade reactions. Indeed, the peculiar blunt-box shaped spectra provides a good benchmark to discern the effect of a heavier/lighter mediator in the spectrum of mass eigenstates (for fixed DM mass, SM primaries and observable CR) as this will affect both the normalization and the edges of the spectra. The results can be appreciated in \Fig{fig:cfr_massratio_gamma} where we compare the one-step annihilation and one-step semi-annihilation $\gamma-$ray injection spectra obtained by varying the mediator mass from heavy towards light values with respect to the DM mass scale which, for reference, we keep to be $\mdm=1\,\TeV$.

Looking at the one-step annihilation spectrum in \Fig{fig:cfr_eps_onestep}, heavier mediators generate spectra with narrower and more peaked box shapes, approaching the line-like spectrum yielded by standard DM annihilations with $\mdm=m_\phi/2$. This is easily understood in terms of kinematics: when the mediator and DM particles are nearly degenerate in mass, the on-shell metastable state is generated nearly at rest and the resulting CR spectrum is approximately the same of a decaying DM particle in it rest frame, namely a monochromatic photon emission peaked at half the DM (mediator) mass (up to the low-energy tail arising from secondary EW showering). On the other hand, when light mediator states are involved in the cascade reaction, the blunt-box shape becomes shallower and gets smoothed out since the kinematics resembles the emission of massless particles.

Completely analogous features are observed in the injection spectrum of $\gamma$ rays generated by one-step semi-annihilations, as show in \Fig{fig:cfr_eps_semi}. Indeed, the different kinematics which distinguishes this process from one-step cascade annihilations becomes more evident in the heavy mediator regime as the edges of the box-shaped spectrum are clearly modified. On the other hand, for lighter mediators, it becomes more difficult to differentiate among the two processes.

On the other hand, the  injection spectra from DM annihilation into other primary states do not show pronounced spectral features. 
However, $\gamma$ ray production through one-step cascade processes yields distinguishable spectral changes with respect to standard annihilation. 
From figures~\ref{fig:cfr_b_gamma} to \ref{fig:cfr_tau_gamma},  we observe that the peaks of the spectra for one-step annihilation (orange) and semi-annihilation (green) are always shifted to lower energies and suppressed, with the exception of one-step annihilation going into primary $b$ quarks showing smaller effects. 

\begin{figure}[t]
    \centering
    \includegraphics[width=0.9\linewidth]{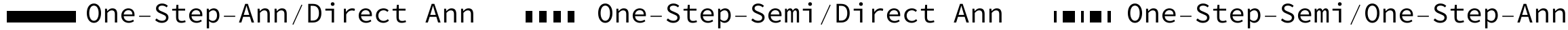}
    \\
    \medskip
    \subfloat[\label{fig:r_gamma_gamma}]{\includegraphics[width=.465\linewidth]{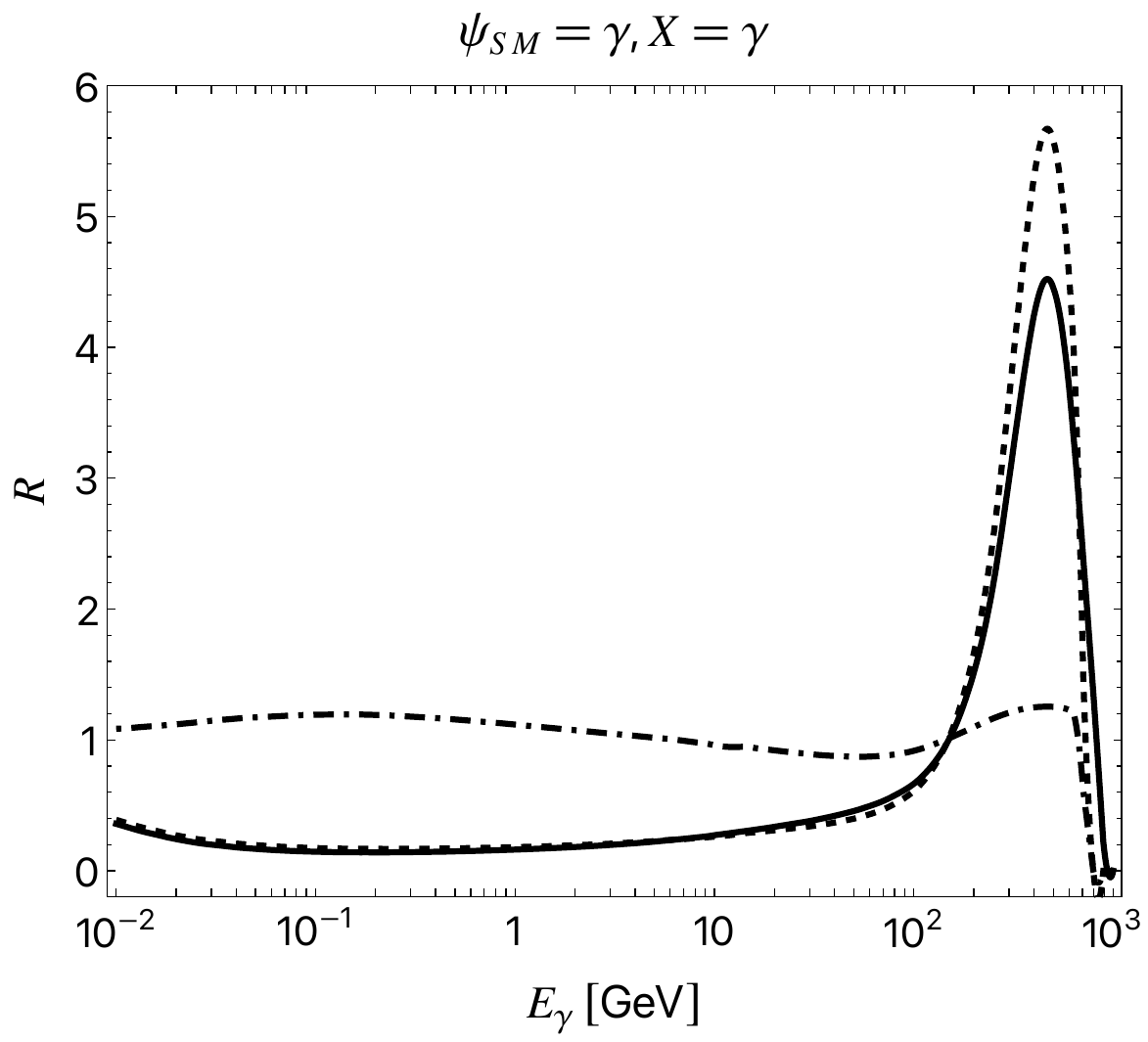}}\quad
    \subfloat[\label{fig:r_b_gamma}]{\includegraphics[width=.48\linewidth]{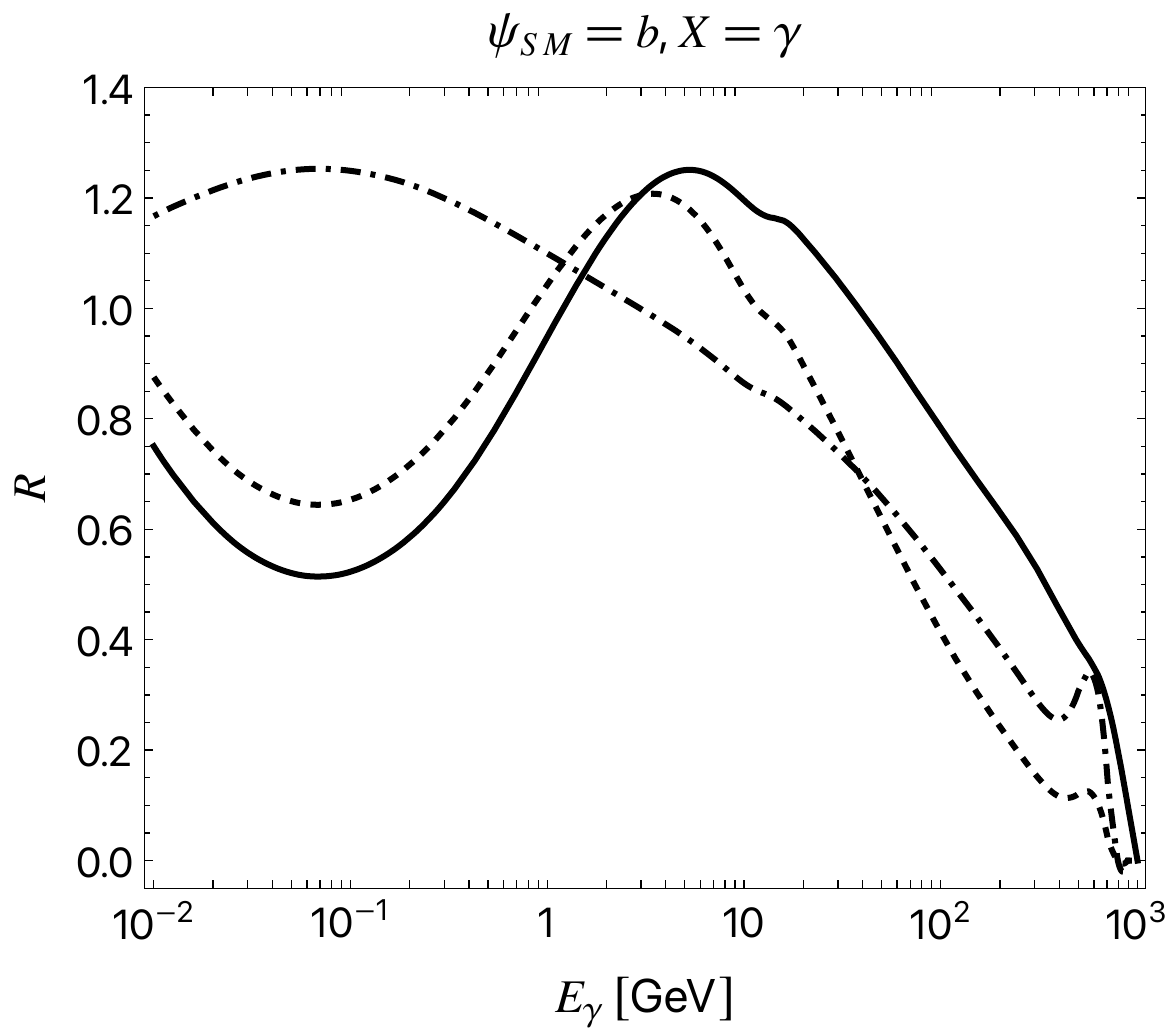}}\\
    \subfloat[\label{fig:r_e_gamma}]{\includegraphics[width=.48\linewidth]{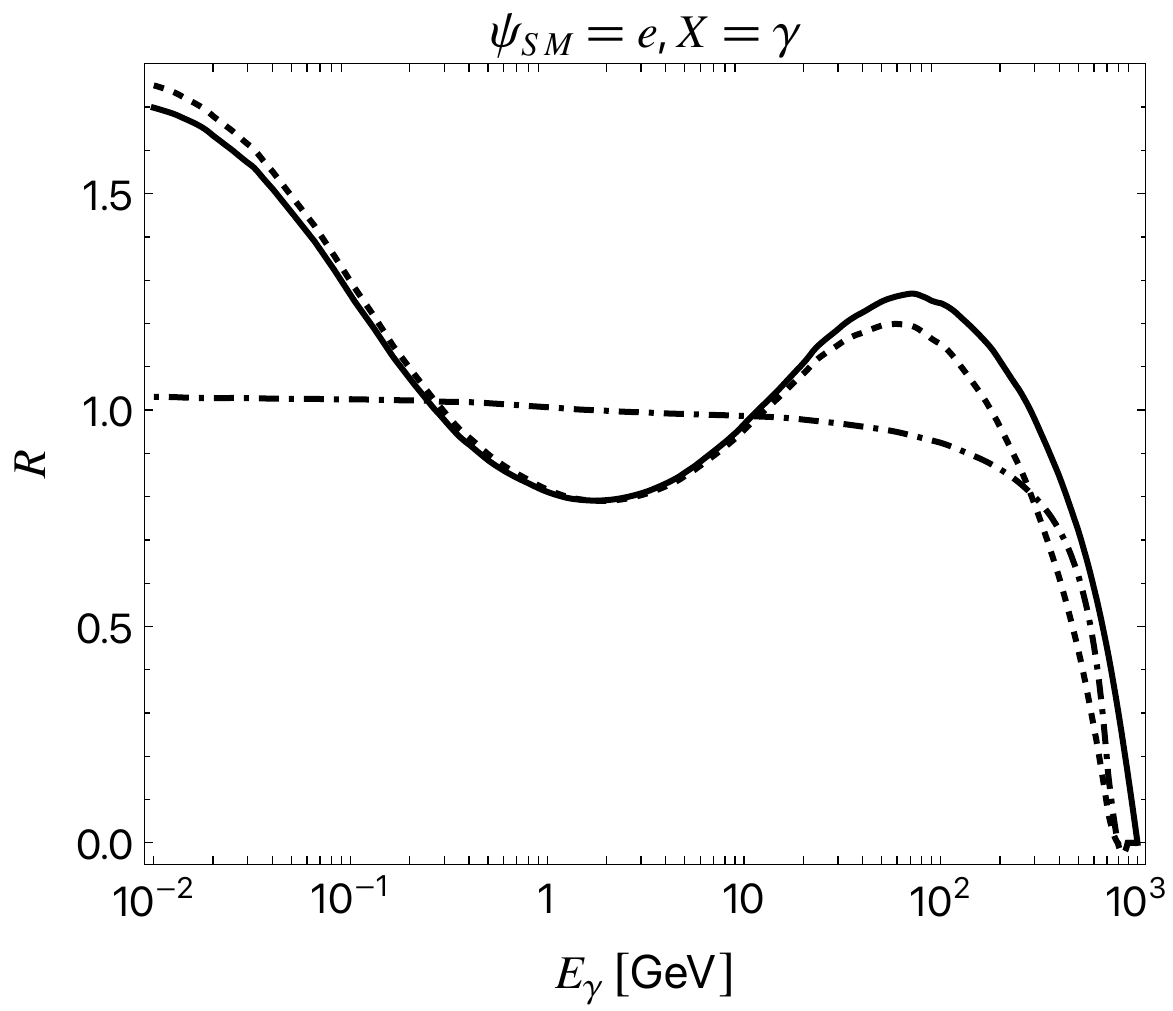}}\quad
    \subfloat[\label{fig:r_tau_gamma}]{\includegraphics[width=.48\linewidth]{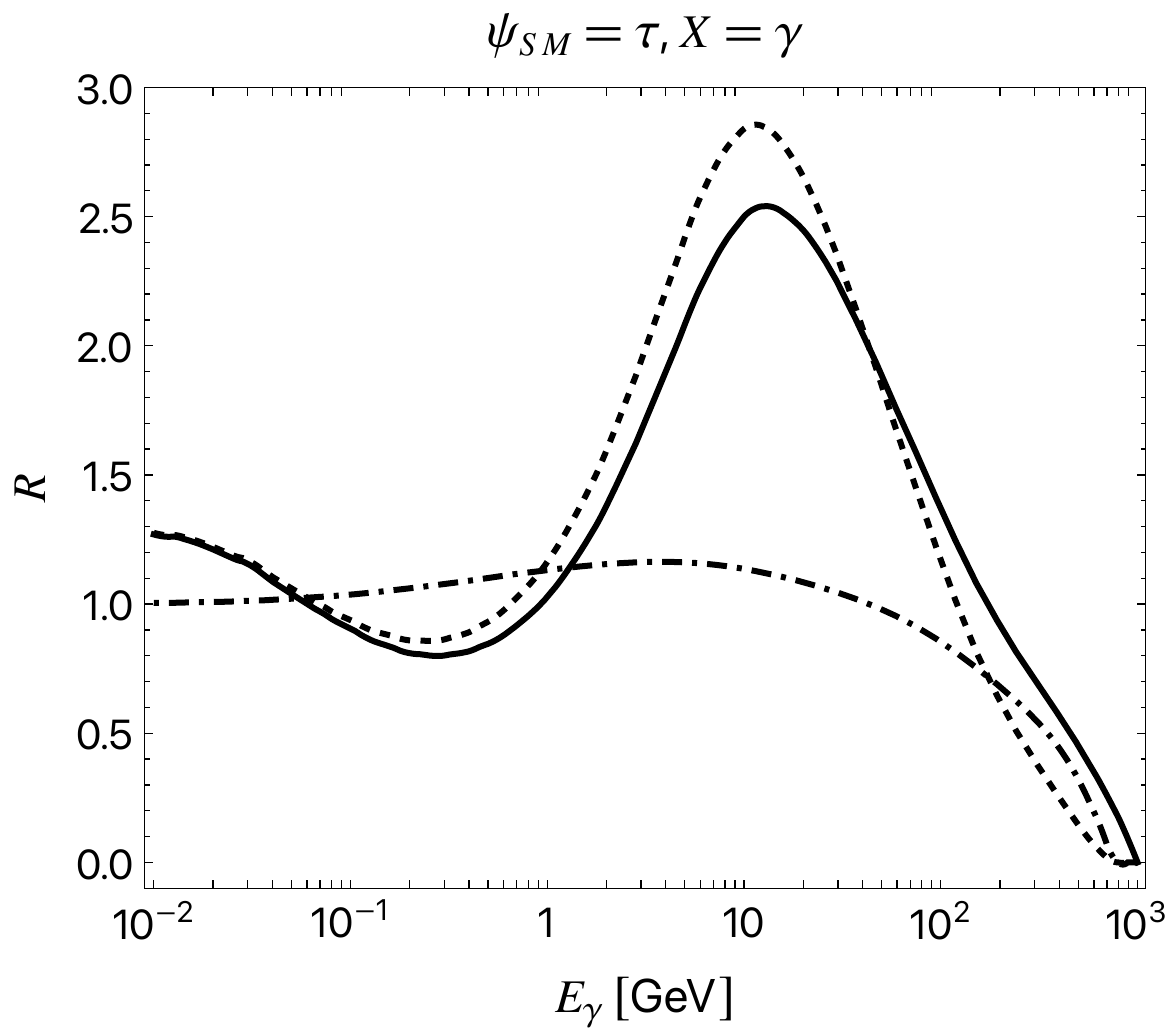}}
    \caption{ \textit{Ratio of $\gamma-$ray spectra from different single process.}
    Ratio $R$ (see \Eq{eq:ratio}) of the differential injection spectra of $\gamma$ rays per single DM collision process in scenarios where the ID signal is uniquely determined by a single process among the ones illustrated in \Fig{fig:diagrams}. Each panel refers to a different final state particle $\psi_{\mathrm{SM}} = (\gamma, b, \tau, e)$, going clockwise starting from the top-left panel.}
\label{fig:ratios_gammas}
\end{figure}

The size of these effects is quantified in \Fig{fig:ratios_gammas} where the ratio of pair of spectra (defined according to \Eq{eq:ratio}) is plotted, again for the same SM primary state particles. 
The ratio $R$ among the one-step annihilation and the direct annihilation is reported with solid lines, while dotted  (dashed-dotted) lines refer to the ratio among one-step semi-annihilation and direct (one-step) annihilation. 
We observe that the $\gamma-$ray emission coming from the one-step processes   dominates over the standard $SS^\star\to\gamma\gamma$ at the high energy end of the spectrum (apart for the line-feature) by a factor of $\sim5$, but, on the other side, is actually suppressed by a factor of $\sim2$ going at lower energies. The ratio among one-step cascade-generated photon spectra (dot-dashed lines) appears to be relatively less affected by the different kinematics, since $R$ remains $\sim O(1)$ for energies up to about 100~GeV, where, instead, the different kinematics among one-step annihilation and semi-annihilation becomes relevant.  
The only exception is shown by the $b$-channel in \Fig{fig:r_b_gamma}, where the ratio between semi-annihilation and one-step annihilation is almost a monotonically decreasing function towards higher energies. Quite interesting is also the difference among $\gamma-$rays injection spectra produced via primary $\psi_{\rm SM}=\tau$ emission (\Fig{fig:r_tau_gamma}), where the signals for one-step processes are enhanced, up to a factor of three, over an energy interval centered around about 10~GeV, with respect to the direct annihilation case (solid and dashed lines).
Overall, we observe modifications in the injection spectra given by one-step annihilation and semi-annihilation that change the overall normalization and spectral shape of the expected ID signal with respect to the standard annihilation by maximal factors that range from about 6 ($\psi_{\rm SM}=\gamma$) to tens of percent ($\psi_{\rm SM}=b$), which are not (only) overall rescaling factors, but introduce a rich family of spectral features throughout the whole energy range. These effects induce spectral modifications that are potentially relevant for forthcoming $\gamma$-ray searches, in particular for direct annihilation into photons. Given the distortions produced by one-step (semi-)annihilation processes with respect to the standard annihilation scenarios investigated, e.g.~in \cite{DeLaTorreLuque:2023fyg,CTAO:2024wvb}, a straightforward recasting of existing results is not possible. Dedicated searches, exclusion limits, and sensitivity projections based on \textit{Fermi}-LAT data and CTAO simulations are therefore required to quantitatively assess the sensitivity to such processes. 

\begin{figure}[t]
    \centering
    \includegraphics[width=0.8\linewidth]{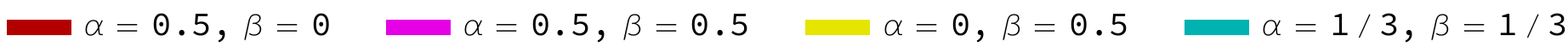}
    \\
    \medskip
    \subfloat[\label{fig:mix_gamma_gamma}]{\includegraphics[width=.48\linewidth]{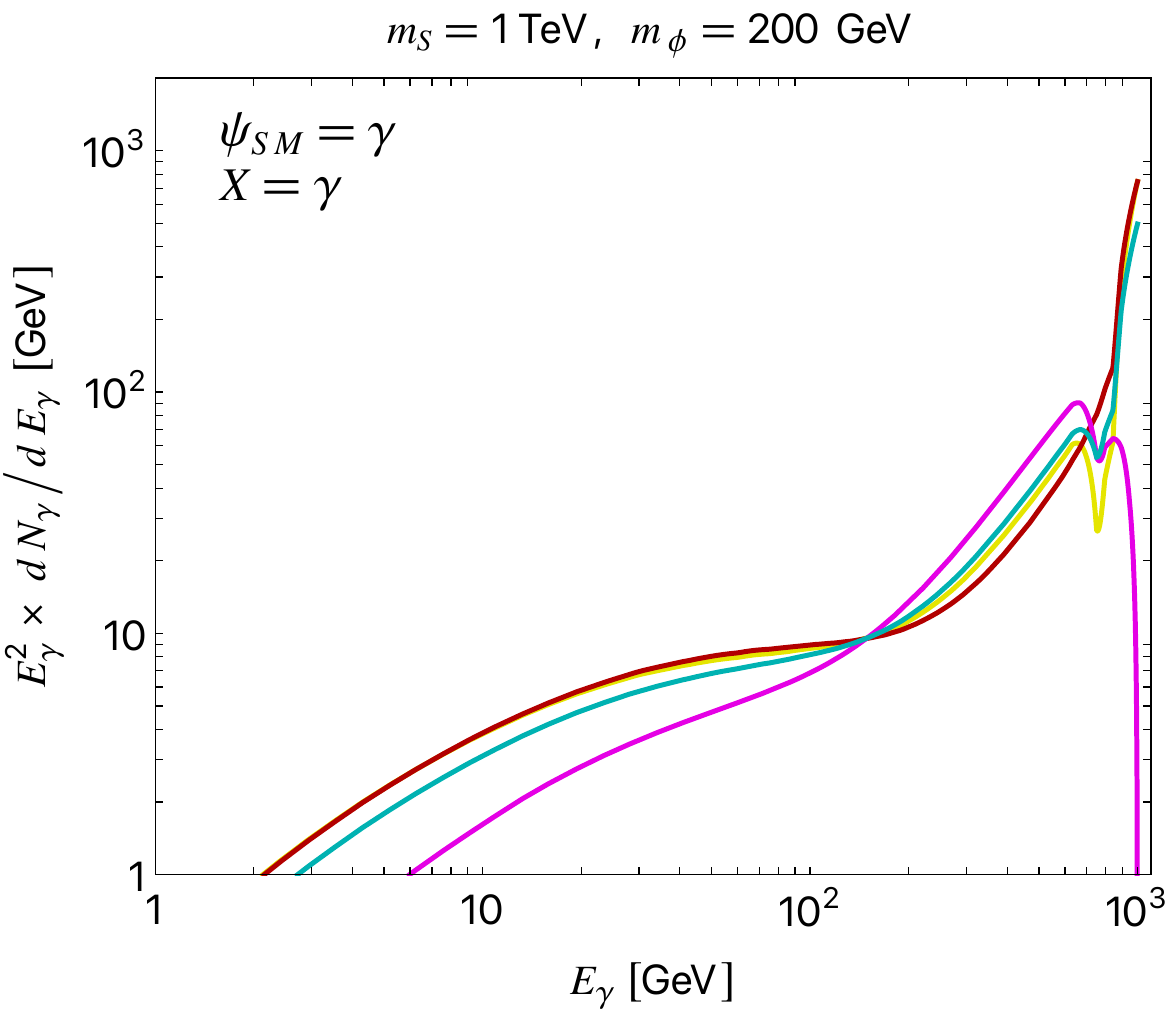}}\quad
    \subfloat[\label{fig:mix_b_gamma}]{\includegraphics[width=.475\linewidth]{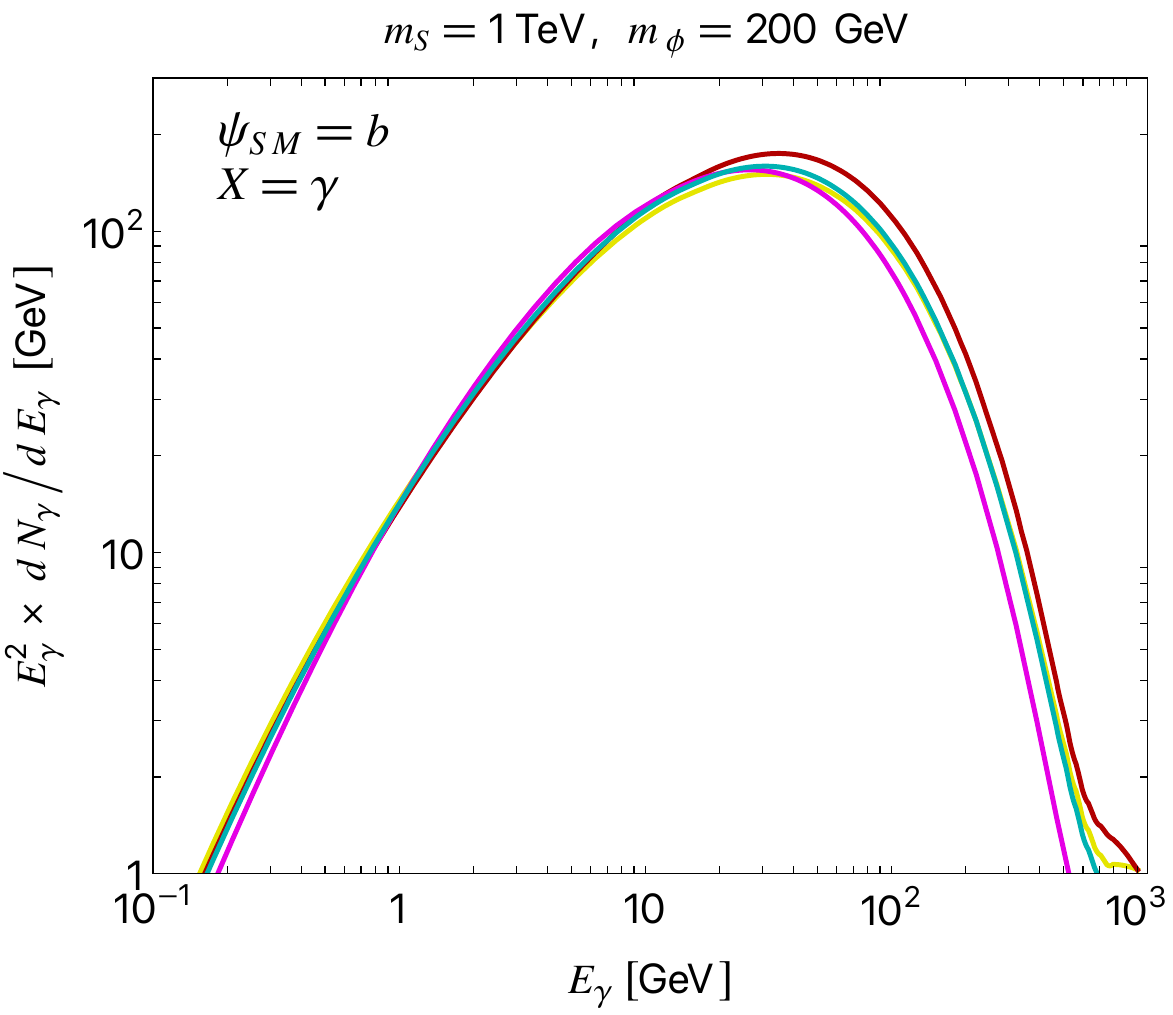}}\\
    \subfloat[\label{fig:mix_e_gamma}]{\includegraphics[width=.48\linewidth]{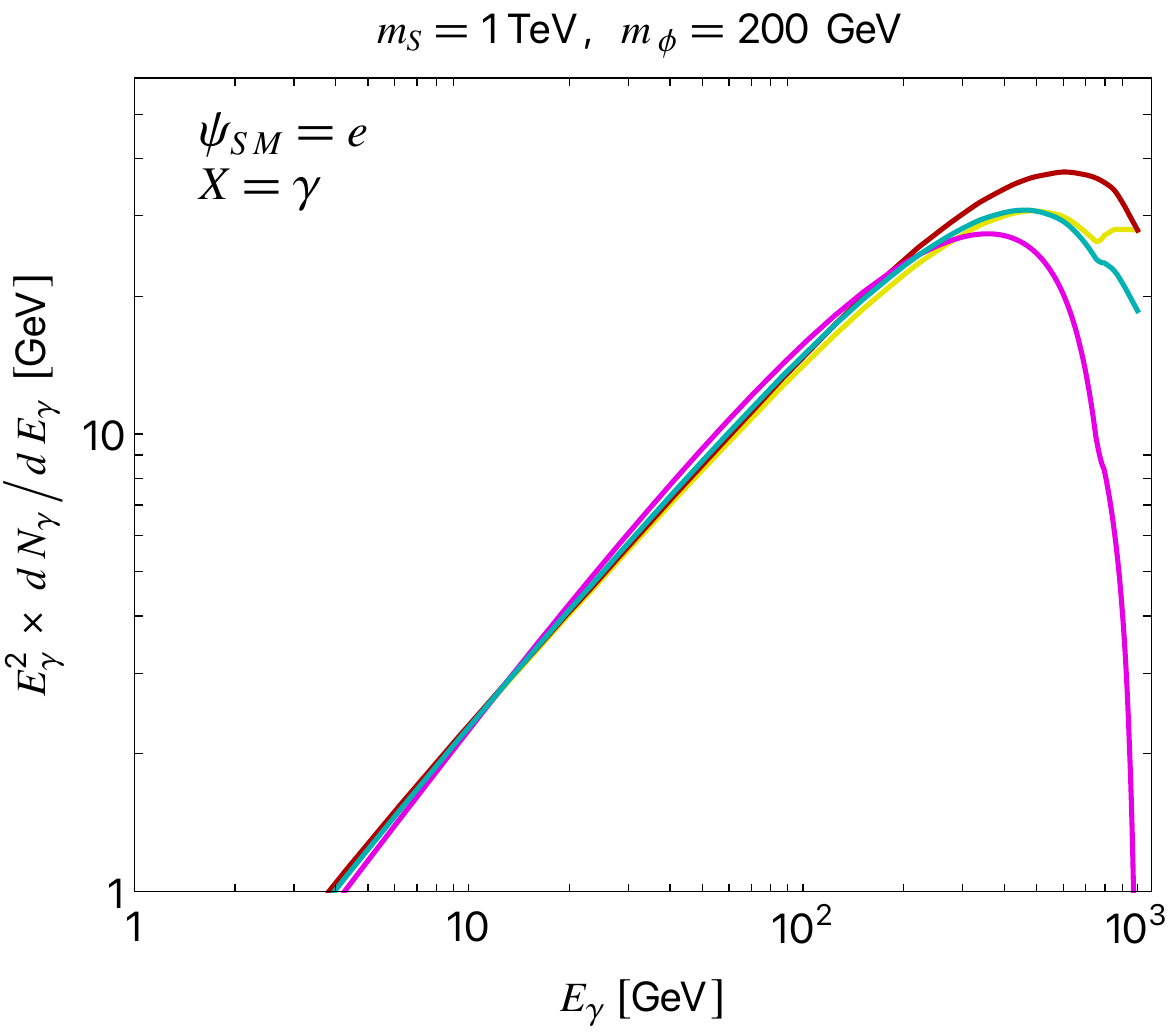}}\quad
    \subfloat[\label{fig:mix_tau_gamma}]{\includegraphics[width=.48\linewidth]{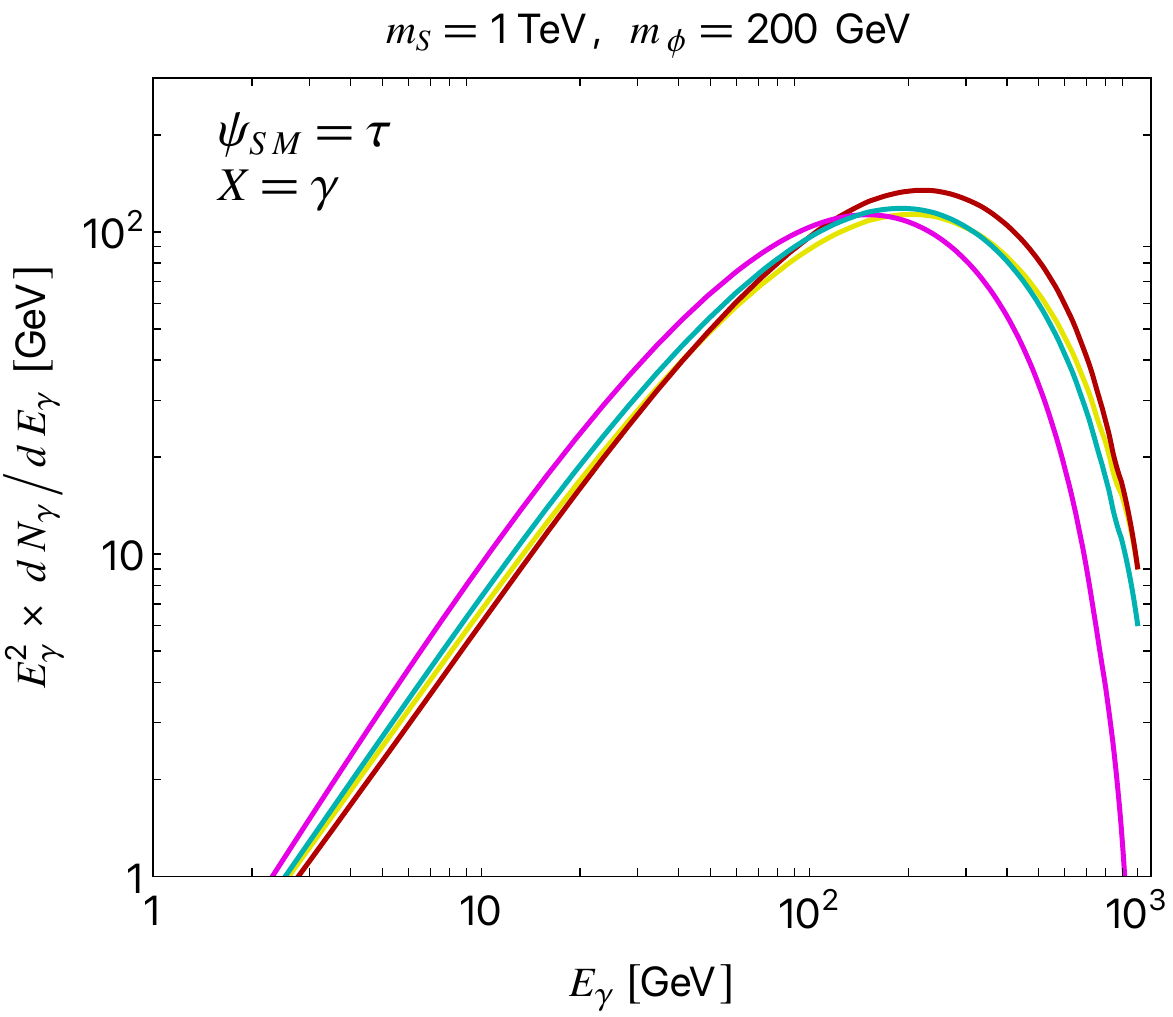}}
\caption{\textit{Injection spectra from a superposition of processes: $\gamma$ rays.}
Differential injection spectra of $\gamma$ rays per single DM collision process in scenarios where the DM  signal is generated by the superpositions of the contributions coming by a mixture of  two (yellow, red, magenta) all of the three processes (cyan) under consideration. Each panel refers to a different final state particle $\psi_{\rm SM}=(\gamma,\,b,\,\tau,\,e)$, going clockwise starting from the top-left panel.}
\label{fig:mixing_gammas}
\end{figure}

Following the discussion of \Sec{sec:injection_spectra}, we now extend the phenomenology covered by previous works by investigating the spectra generated by the superposition of signals produced by a linear combination of the collision processes under consideration, as defined in \Eq{eq:alphabeta}. The results are presented in \Fig{fig:mixing_gammas}. We focus on the following benchmark cases of study: the scenario when only two out of three production mechanisms are active, namely when $\alpha$ and(or) $\beta$ equal $1/2$, and a scenario where all three mechanisms are present, for an evenly weighted sum $\alpha=\beta=1/3$. As expected, if strong features are present in the spectra at production, i.e. annihilation spectra, they will withstand the superposition of differently shaped one-step spectra, unless heavily suppressed $(\alpha+\beta\simeq1)$. Interestingly, for $(\alpha,\,\beta)=(0,\,0.5)$, the photon spectrum presents both a distinguishable peak followed, at lower energies, by the blunt box shape typical of direct annihilations and semi-annihilations, respectively. When only one-step processes are active $(\alpha+\beta=1)$ we generally obtain the most peculiar spectral shapes, showing all the characteristics described before for single production mechanisms via cascade reactions. Namely, for $\alpha=\beta=1/2$ the overall signal is generally suppressed and/or softened. On the other hand, all other combinations would still lead to the presence of a more pronounced peak for primary $\gamma$ emission, while no line-like feature is relevant in the full one-step spectrum as it even appears to be further suppressed. Besides mild suppression and softening towards the endpoint of the spectrum, no particular spectral feature is observed in the benchmark scenarios for the other SM primary channels. 

We remind that the phenomenology  of ID signals of multi-step annihilating (or decaying) DM as source of $\gamma$ rays  has been studied also in previous literature \cite{Mardon:2009rc,Elor:2015tva,Elor:2015bho,Fortin:2009rq,Tempel:2012ey}, for example characterizing the ID phenomenology of secluded dark sectors \cite{Profumo:2017obk,NFortes:2022dkj}. 
On the other hand, the only piece of work we are aware of exploring $\gamma-$ray signals from DM semi-annihilations in a model-indipendent fashion is provided by \cite{Queiroz_2019}, while in \cite{Guo:2023kqt,Guo:2024jdj,Arcadi:2017vis} specific model realizations are investigated.  
While developing our framework, we have compared the CR spectra we obtained applying the Lorentz boost to the zero-step spectra via \Eq{eq:dNdE} with a number of previous references. As discussed further in details in \App{app:kinematics}, our results match those of Ref.~\cite{Elor:2015bho} for one-step annihilations  under some constraining assumption on the scale separation between the masses of particles. Similarly, our results have been validated by the comparison with Ref.~\cite{Queiroz_2019}.
In the present work, we move forward investigating the effects of an extended DM phenomenology in which a superposition of signals produced by a linear combination of the collision processes under consideration, as defined in \Eq{eq:phigammagen}, is present, and could be e.g.~realized in the specific models presented in \Sec{sec:models}. Moreover, we generalize this treatment to neutrinos and charged CRs, which are discussed in the next sections. 

\section{Neutrinos injection spectra}
\label{sec:neutrinos}

DM collisions primarily ending into ``invisible'' states like neutrinos could explain the lack of signals at indirect searches so far \cite{Arguelles:2019ouk}, and have been adopted to constrain other non-standard scenarios beyond thermally produced DM \cite{Kanemura:2025byi}. 
Facilities like ANTARES \cite{ANTARES:2016xuh,ANTARES:2019svn,ANTARES:2025exu}, IceCube \cite{IceCube:2021kuw,IceCube:2021sog,IceCube:2021xzo}, Super \cite{Frankiewicz:2015zma} and Hyper-Kamiokande \cite{Bell:2020rkw} have been looking for decades for neutrino signals produced by DM annihilations in different supposedly DM-rich environments, like the Sun or towards the GC, providing so far stringent constraints, which are usually evaluated for standard annihilation \cite{Arguelles:2019ouk}.  
Neutrino experiments are typically capable of flavor tagging in incoming fluxes, and/or are sensitive to the interaction of a single neutrino flavor (most likely the muon neutrino).   However, DM collisions are usually assumed to produce all three neutrino flavors in an equal amount, unless a specific hierarchical flavor coupling structure is specified by a well motivated model. For this reason, we will discuss the total injection spectrum of neutrinos summing over all flavors.

\begin{figure}[t]
    \centering
    \includegraphics[width=0.5\linewidth]{Figures/legenda_spectra.pdf}
    \\
    \medskip
    \subfloat[\label{fig:cfr_gamma_nu}]{\includegraphics[width=.48\linewidth]{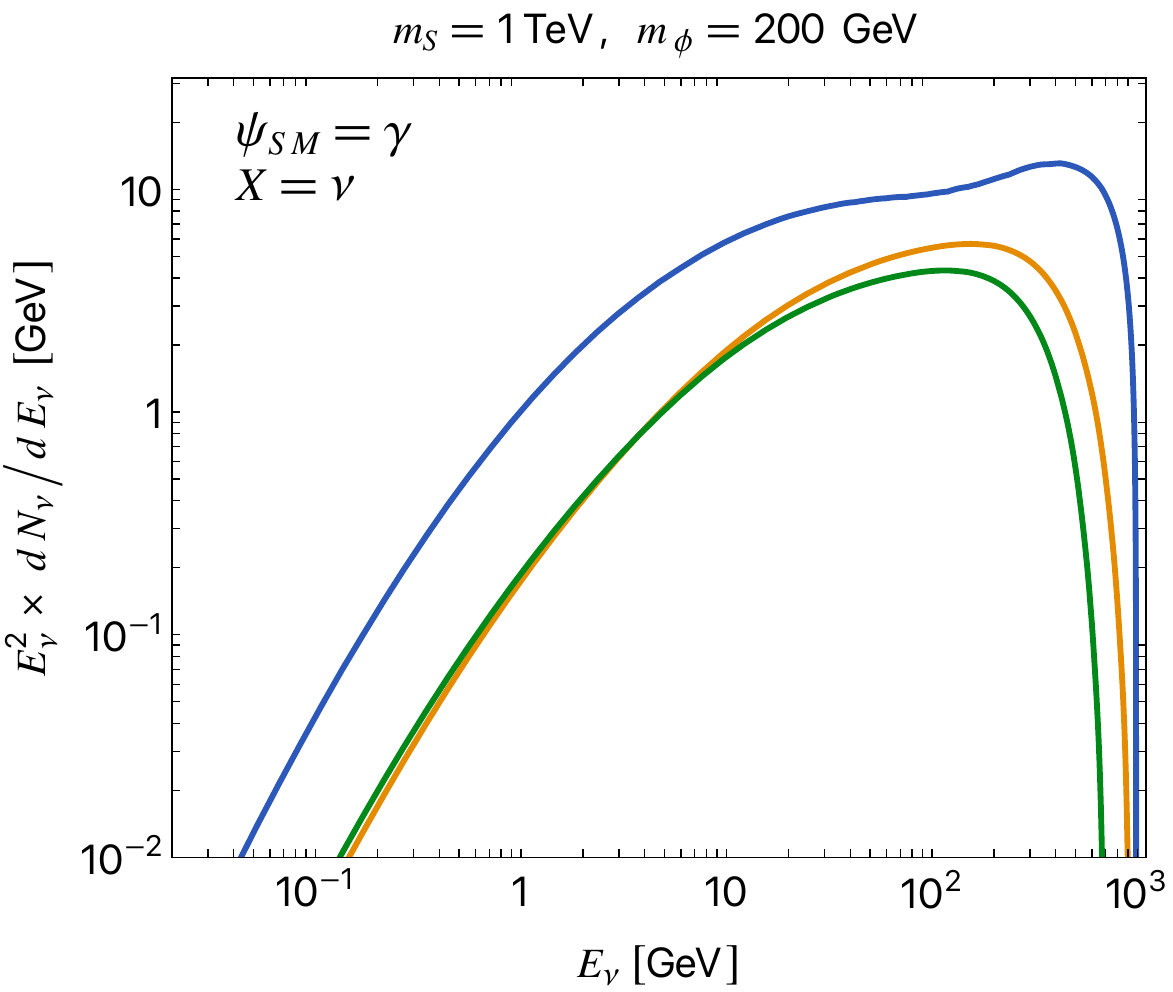}}\quad
    \subfloat[\label{fig:cfr_b_nu}]{\includegraphics[width=.48\linewidth]{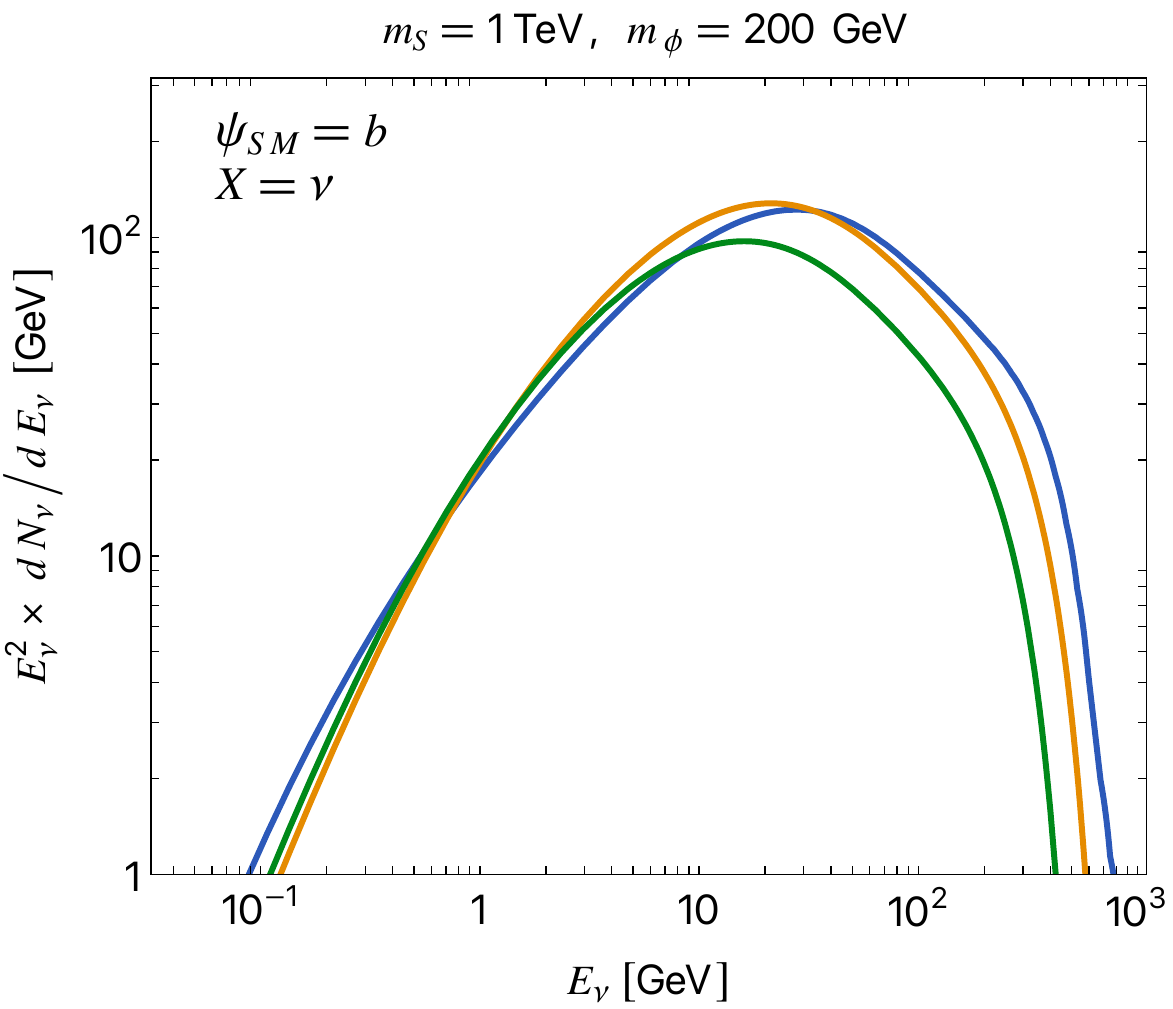}}\\
    \subfloat[\label{fig:cfr_e_nu}]{\includegraphics[width=.48\linewidth]{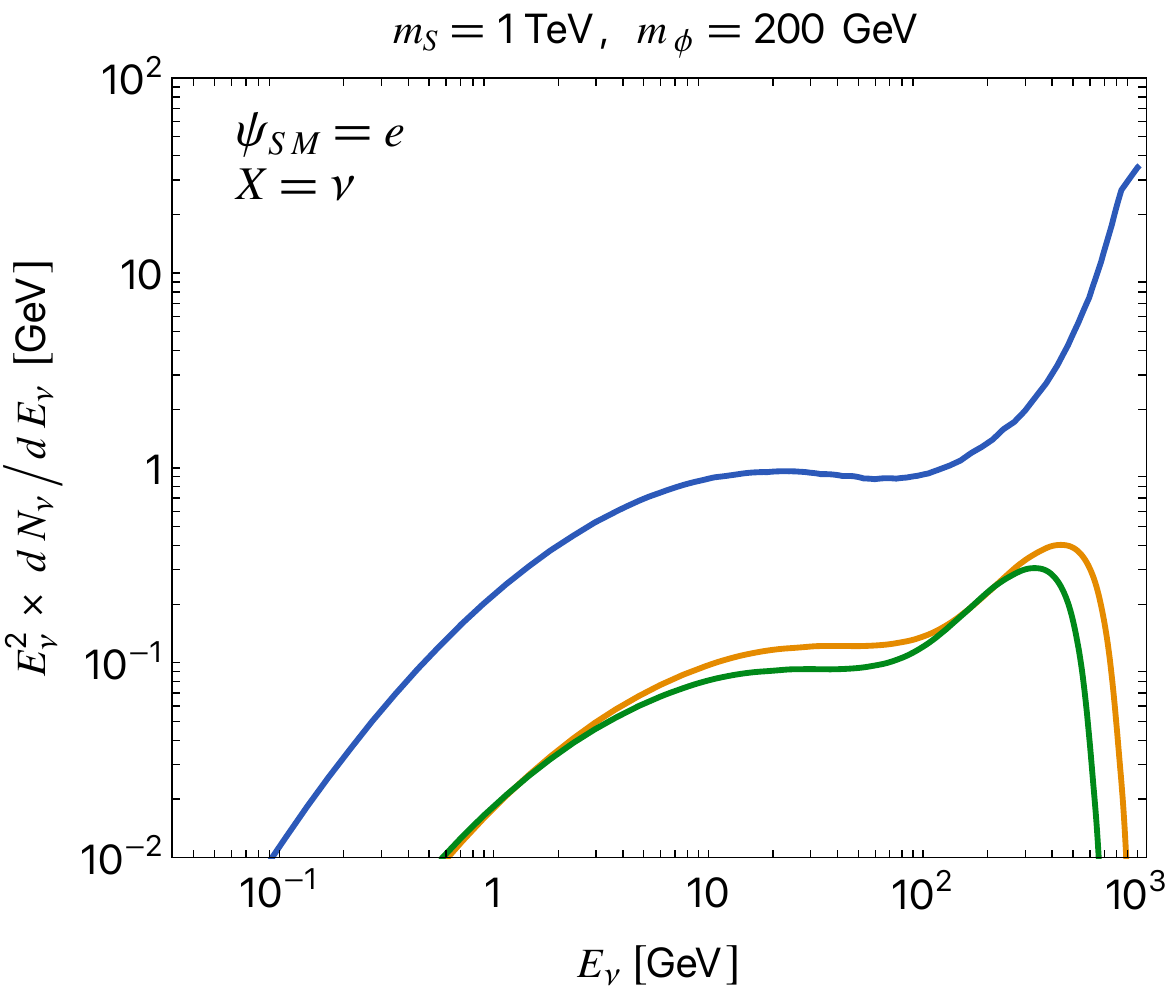}}\quad
    \subfloat[\label{fig:cfr_tau_nu}]{\includegraphics[width=.48\linewidth]{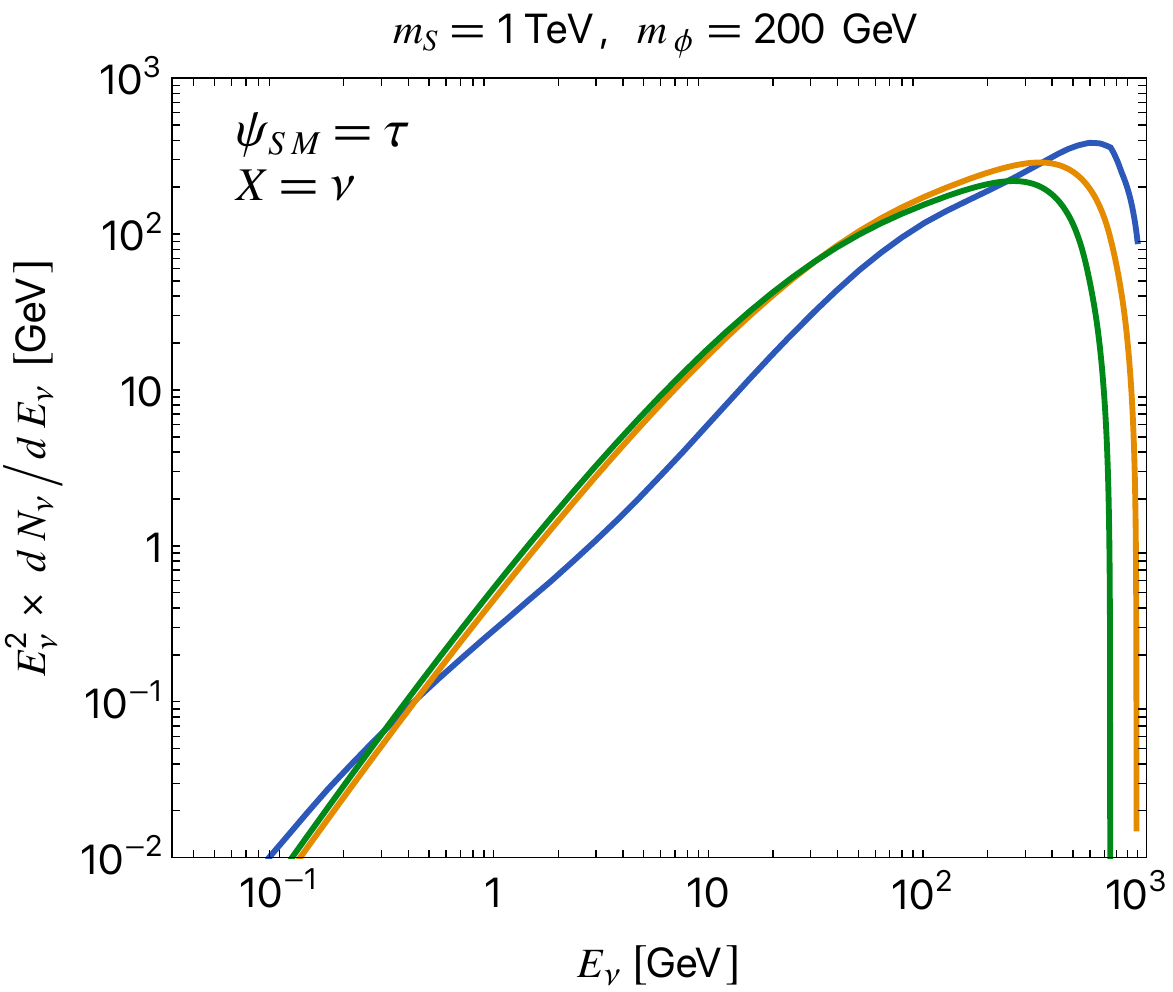}}
    \caption{ \textit{Single process injection spectra: neutrinos.}
    Same as \Fig{fig:cfr_gammas} but for the total neutrino production per single DM collision process. 
    }
    \label{fig:cfr_neutrinos}
\end{figure}

\begin{figure}[t]
    \centering
    \includegraphics[width=0.9\linewidth]{Figures/legenda_ratios.pdf}\\
    \medskip
    \subfloat[\label{fig:r_gamma_nu}]{\includegraphics[width=.48\linewidth]{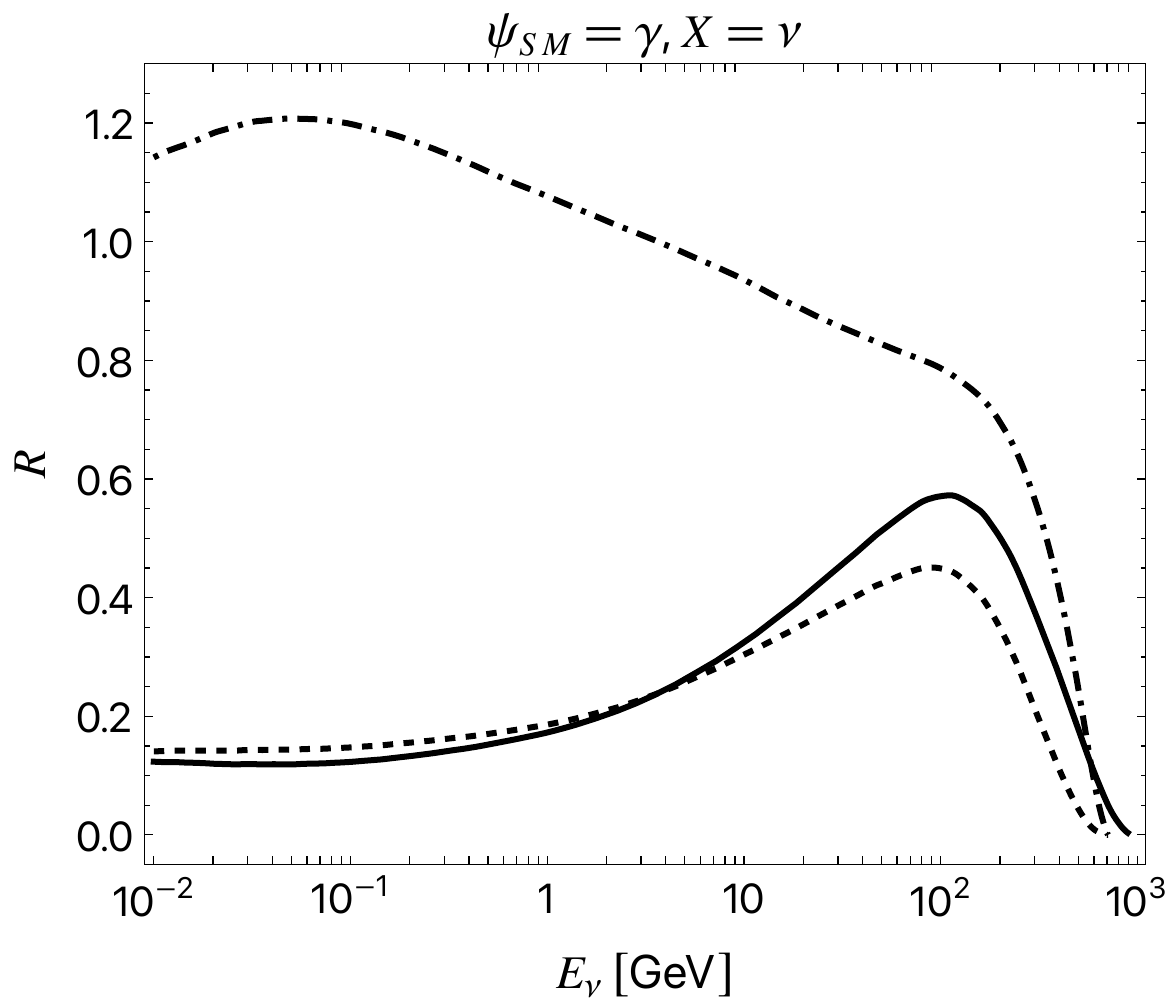}}\quad
    \subfloat[\label{fig:r_b_nu}]{\includegraphics[width=.48\linewidth]{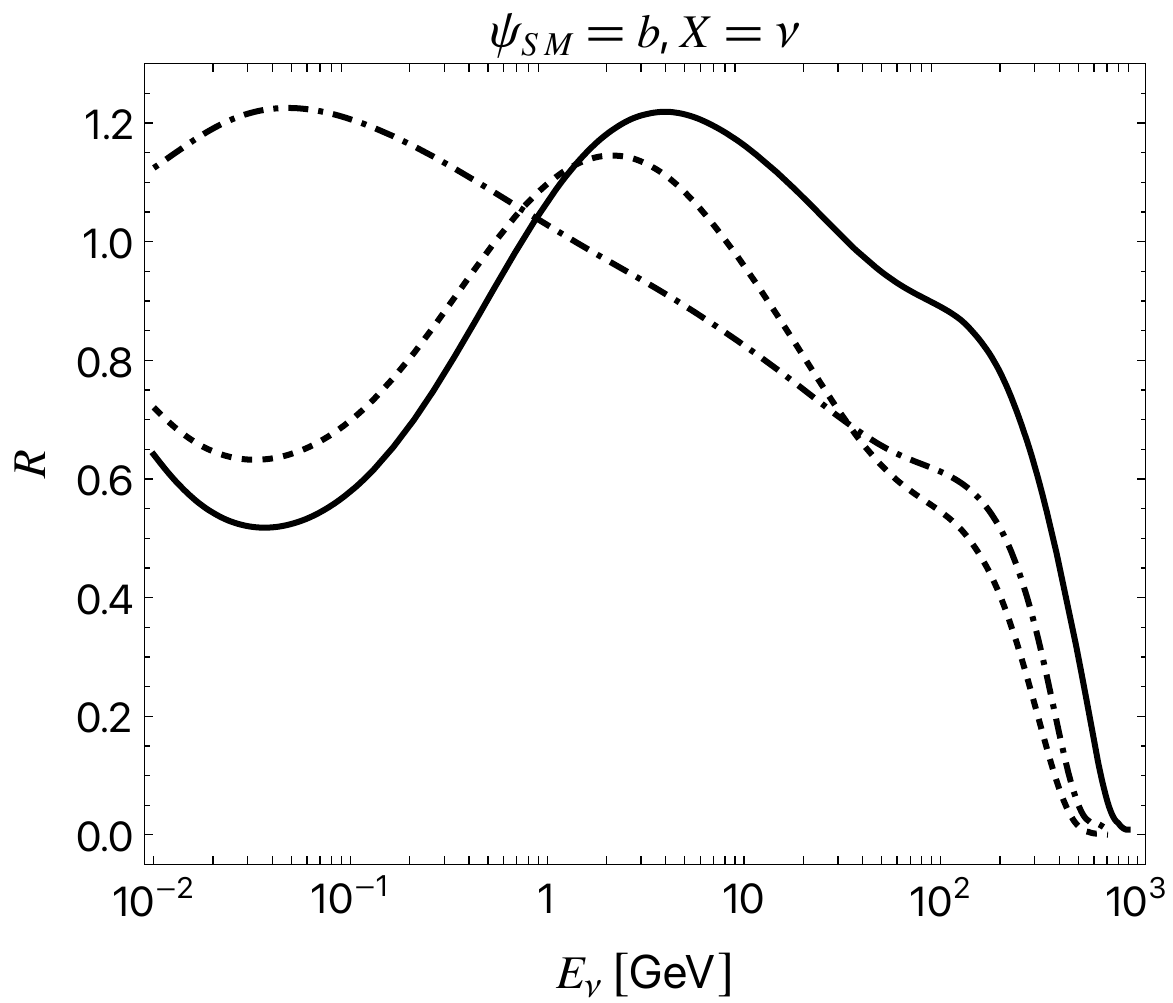}}\\
    \subfloat[\label{fig:r_e_nu}]{\includegraphics[width=.48\linewidth]{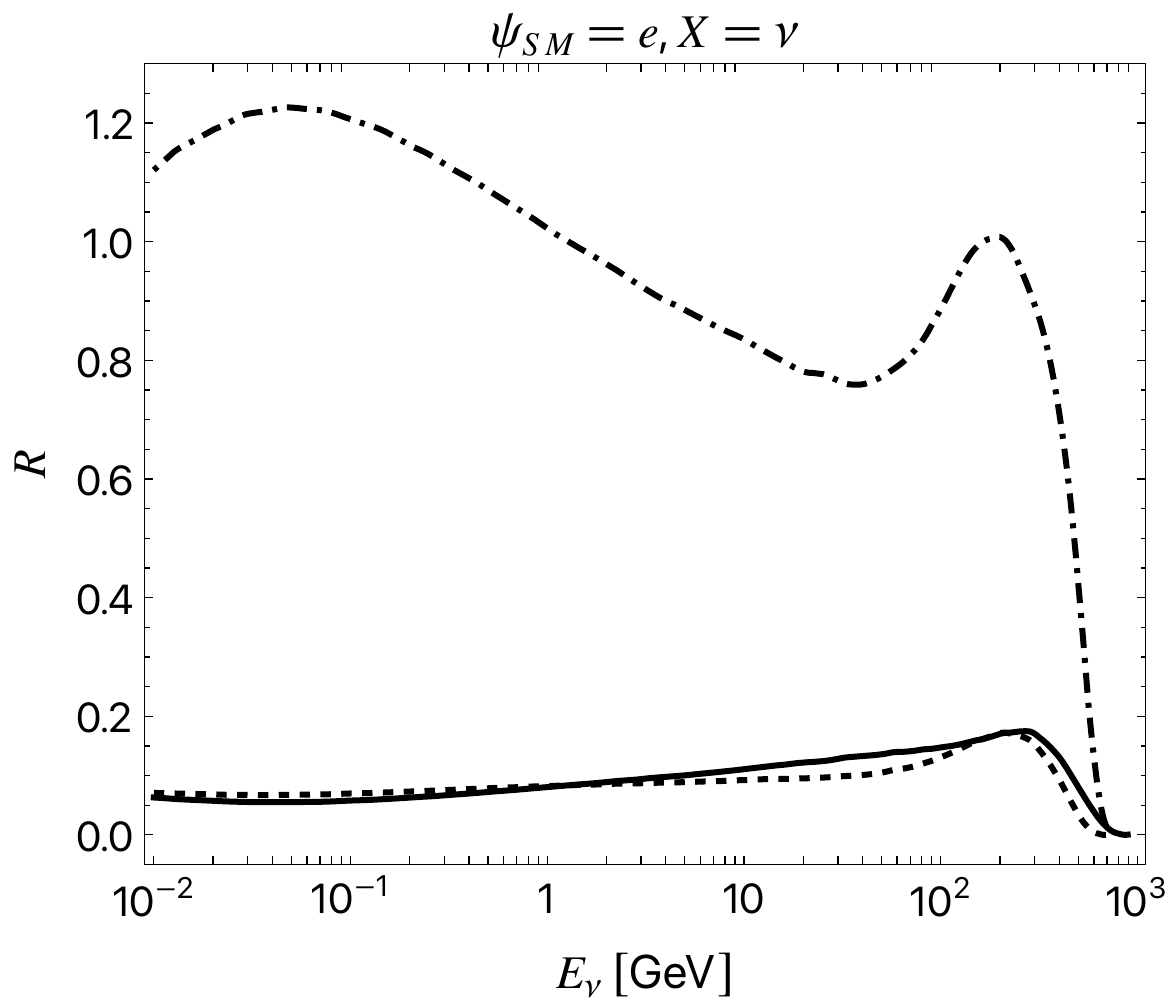}}\quad
    \subfloat[\label{fig:r_tau_nu}]{\includegraphics[width=.48\linewidth]{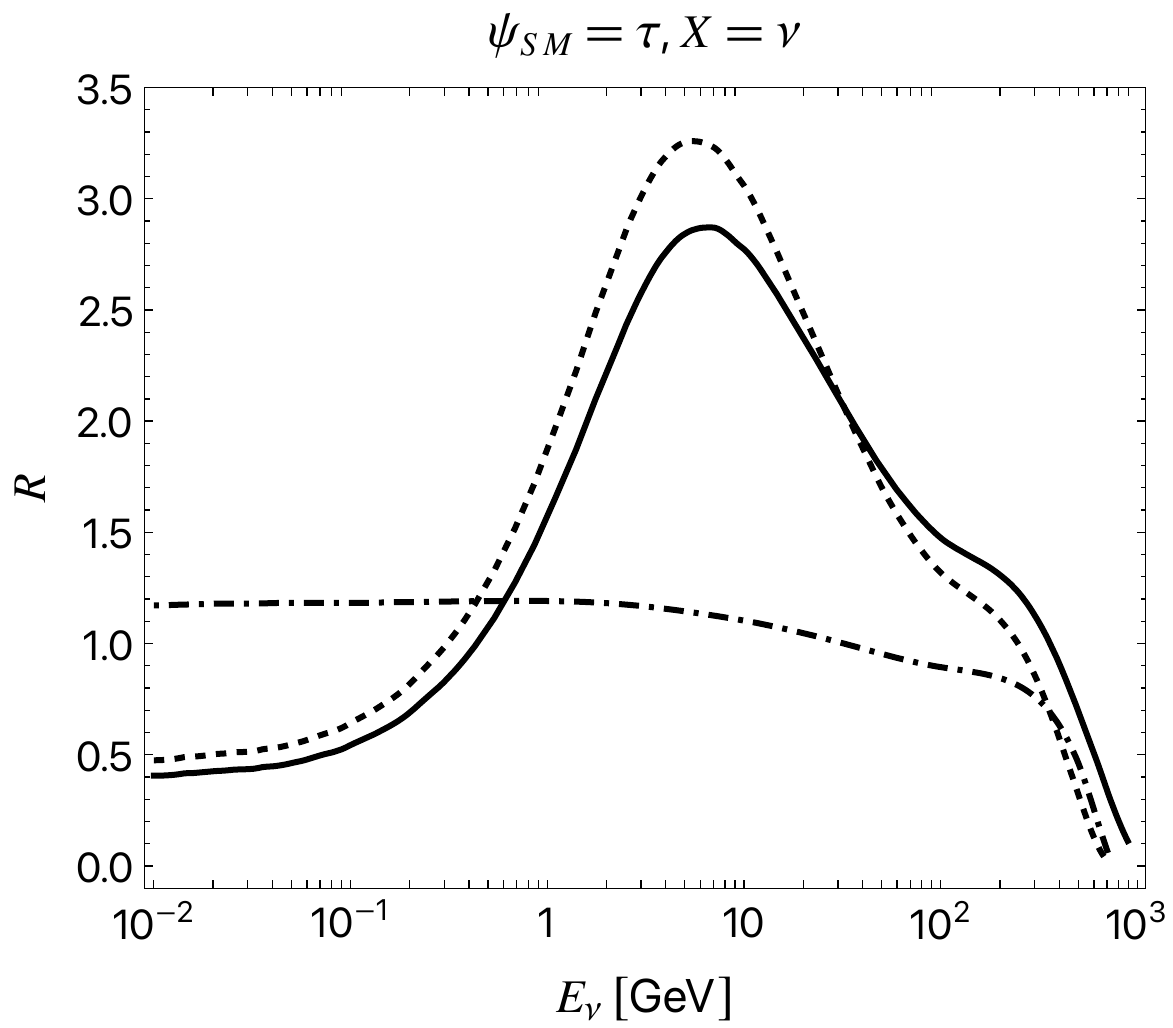}}
\caption{\textit{Ratio of neutrino spectra from different single processes.}
Same as \Fig{fig:ratios_gammas} but for the differential injection spectra of neutrinos per single DM collision process.
}
\label{fig:ratios_neutrinos}
\end{figure}

Following a similar discussion to \Sec{sec:gammas}, we start presenting the neutrino injection spectra resulting from scenarios where DM proceeds $100\%$ via a single collision process (\Fig{fig:cfr_neutrinos}). 
The effects of having one-step cascade reactions are remarkable, with the associated curves evidently suppressed with respect to the direct annihilation signal, with the only exception of the $b$-channel, where observed effects are milder. 
In particular, the spectral modifications are significant when the primary states of the process are either photons or electrons, shown in \Figs{fig:cfr_gamma_nu}{fig:cfr_e_nu} respectively. We find that the shape of the spectra are quite different, and that the peak of the spectrum can be suppressed in the case of one-step (semi-)annihilation by up to about one order of magnitude. 
To better appreciate these features, we turn once again to the ratio of spectral functions in \Fig{fig:ratios_neutrinos}. For $\psi_{\rm SM}=(\gamma,\,e)$, we observe  that the ratio between spectra produced by one-step and direct annihilation processes is suppressed throughout the whole energy range by factors down to $R\sim0.1$. On the other hand, for the other two cases, namely $\psi_{\rm SM}=(b,\,\tau)$, the spectral differences are highly not trivial. In the former scenario, all ratios present an oscillatory behavior with respect to energy, and in different energy ranges. 
This means that, when possibly combining different channels, each one would be dominantly shaping  the signal in different energy ranges. 
Meanwhile, in the latter case, for a leptophilic dark sector coupling mainly to $\tau$s, one-step (semi-)annihilation spectra are only suppressed at the very high and low energy tails, but their ratios $R$ to the direct annihilation show a quite distinctive peak $R\sim3$, meaning a relative enhancement  centered nearly at the same energy, around 10~GeV. 

\begin{figure}[t]
    \centering
    \includegraphics[width=0.9\linewidth]{Figures/legenda_ab.pdf}\\
    \medskip
    \subfloat[\label{fig:mix_gamma_nu}]{\includegraphics[width=.48\linewidth]{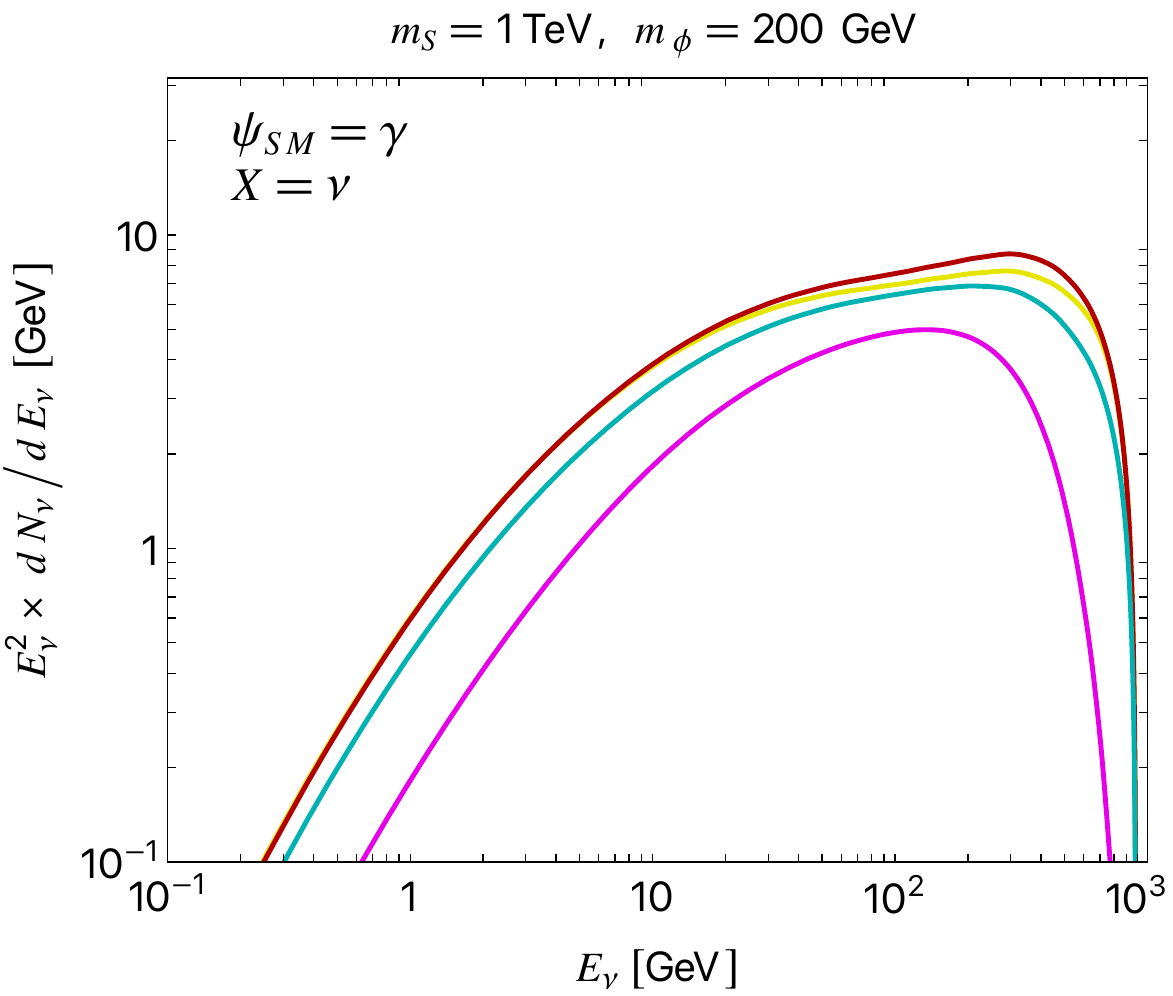}}\quad
    \subfloat[\label{fig:mix_b_nu}]{\includegraphics[width=.48\linewidth]{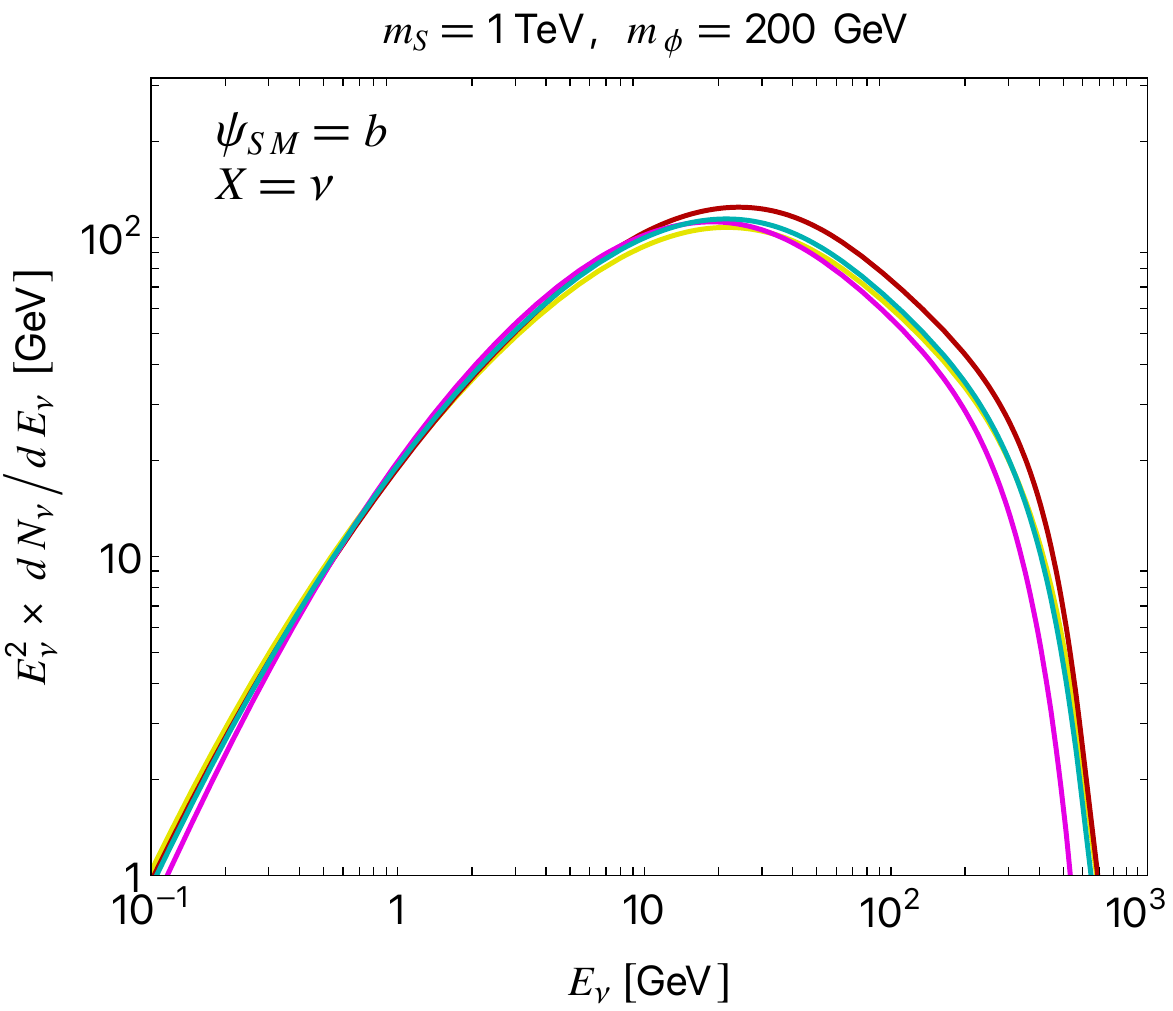}}\\ 
    \subfloat[\label{fig:mix_e_nu}]{\includegraphics[width=.48\linewidth]{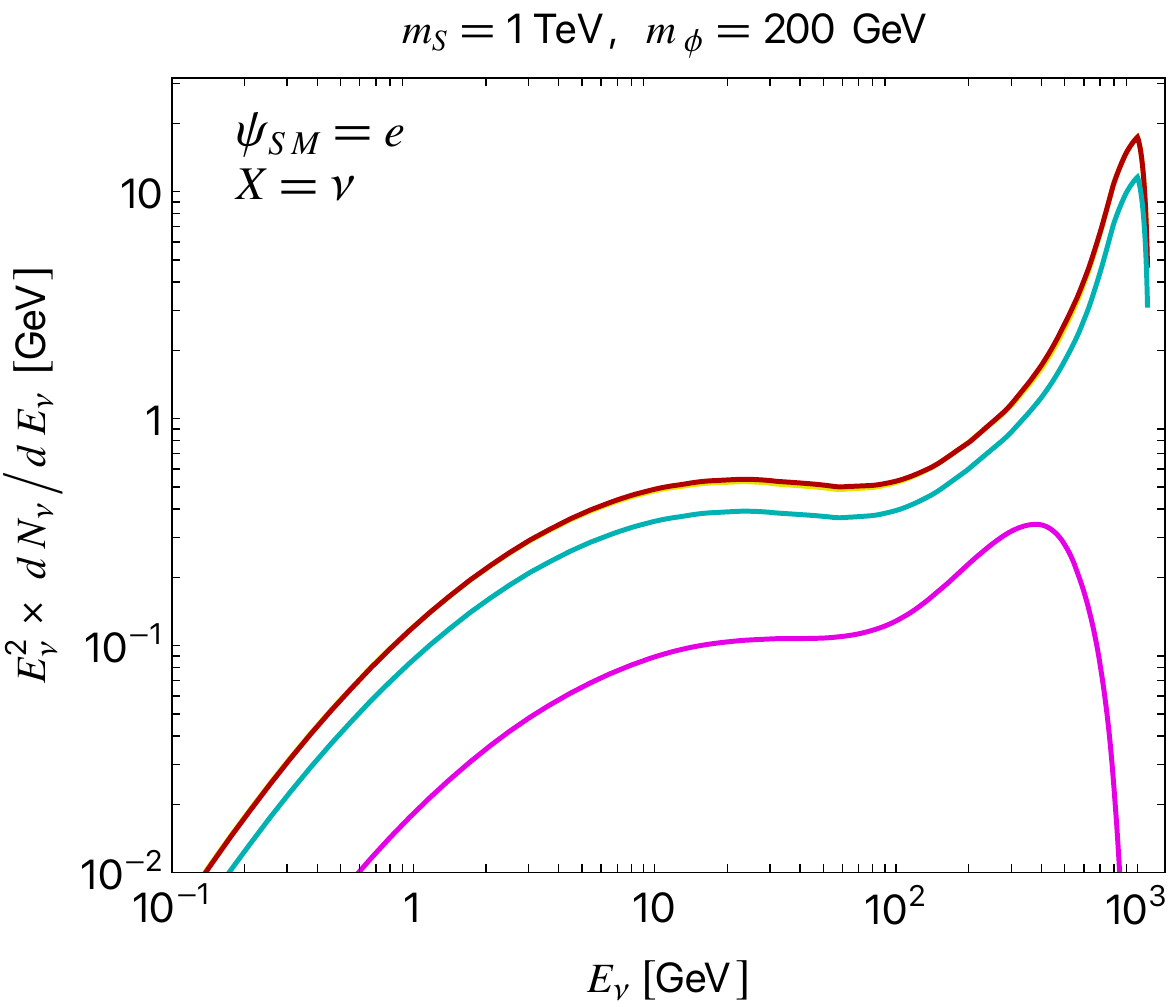}}\quad
    \subfloat[\label{fig:mix_tau_nu}]{\includegraphics[width=.48\linewidth]{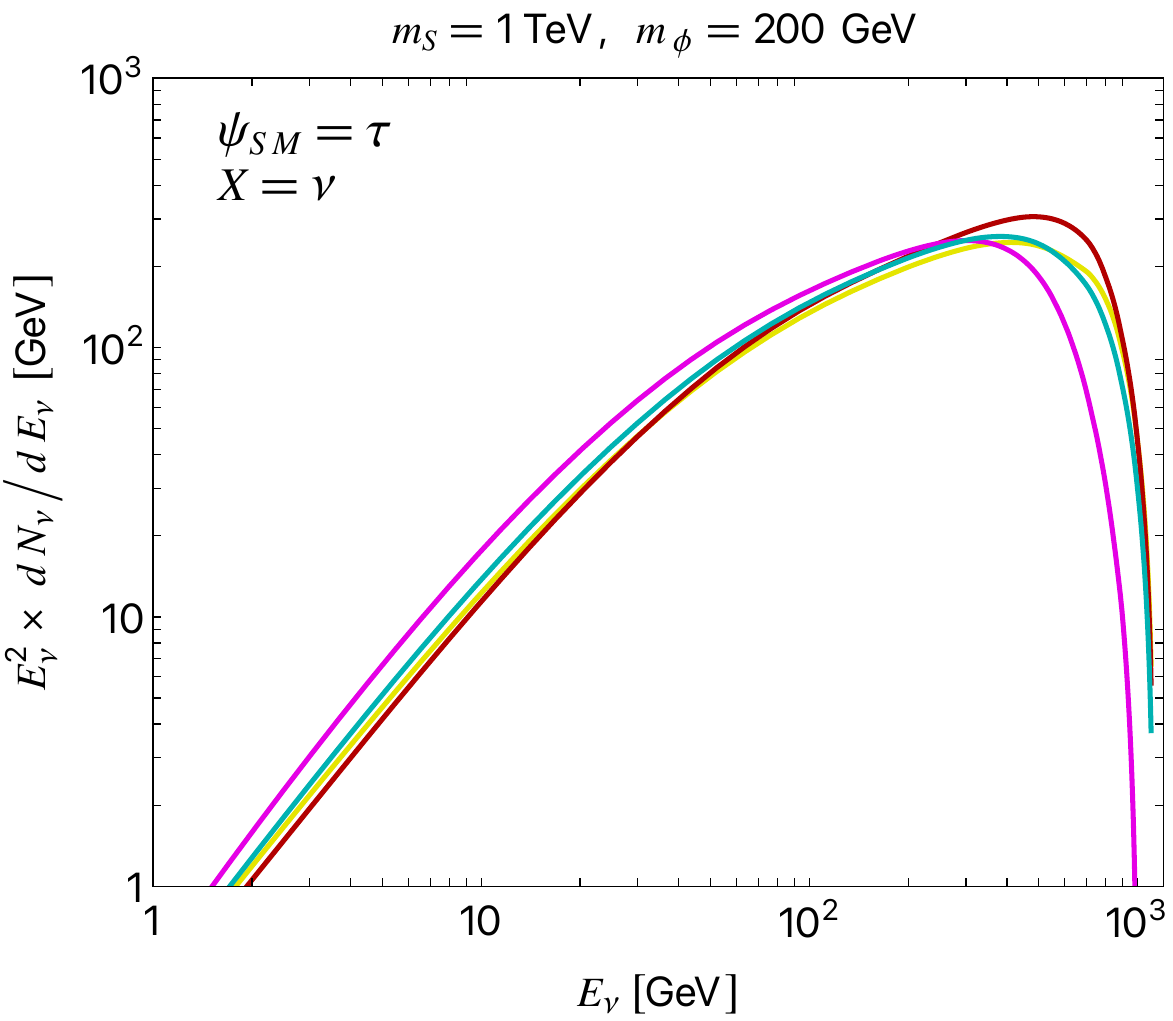}}
\caption{\textit{Injection spectra from a superposition of processes: neutrinos}. Same as \Fig{fig:mixing_gammas} but for the neutrinos per single DM collision process.
}
\label{fig:mixing_neutrinos}
\end{figure}

Finally, we move to the discussion of neutrino spectra produced when a superposition of different processes is considered. The results are shown in \Fig{fig:mixing_neutrinos}. This time, we observe that all combinations give very similar trends, the only exception being the superposition of spectra from one-step cascade processes alone. 
The most interesting case is the injection spectrum resulting from primary $e$ emission, which indeed resembles all the features of neutrinos spectra yielded by one-step cascade reactions. As already shown in \Fig{fig:cfr_neutrinos}, the final energy distributions of neutrinos in such cases is characterized by a severe suppression and remarkable spectral distortions with respect to scenarios where only direct DM annihilation is accounted for in the production mechanisms. 
A framework of leptophilic dark sectors coupled to the first generation and whose ID phenomenology is uniquely dictated by one-step cascade processes thus provides a peculiar case of study, worth of further investigation. 
As for the case of $\gamma$ rays, a simple recasting of existing upper limits computed considering standard annihilation neutrino spectra \cite{Arguelles:2019ouk} is not straightforward, given the non-trivial spectral deviation produced. Nevertheless, given the sizable spectral modifications predicted, particularly for photon and electron SM primary states, current and forthcoming neutrino observations may provide sensitivity to DM interactions beyond standard annihilation.

\section{Charged cosmic rays injection spectra}
\label{sec:chargedCR}


We dedicate this section to the discussion of the resulting injection spectra of charged CRs. Following a similar architecture of the previous sections, we complete our thorough overview of the characterization of differential CR spectra, extending the results to include charged species such as positrons, antiprotons, and anti-nuclei. 

\subsection{Positrons}
\label{sec:positrons}
Charged light antiparticles, such as positrons, coming from DM have been searched for in a number of experiments in the past decades, like PAMELA \cite{Mikhailov:2019umy}, H.E.S.S \cite{HESS:2024etj}, and AMS-02 \cite{AMS:2019rhg}.  
The raise of the positron fraction firstly detected by PAMELA \cite{PAMELA:2008gwm,Mikhailov:2020emn}, and ATIC \cite{Chang:2008aa} collaborations in the GeV-TeV range is now confirmed with unprecedented accuracy by AMS-02. Numerous multi-messenger analysis support an astrophysical origin of this signal, in particular in terms of nearby pulsars (see, e.g., \cite{Evoli:2020szd,Manconi:2020ipm,Orusa:2024ewq}). Despite this, there have been attempts to search for DM-sourced mechanisms to produce such excess positrons, without violating independent constraints. The focus has especially been put on one-step annihilations of dark particles into light sub-GeV leptophilic states \cite{Bergstrom:2008ag,Cholis:2008vb,Cholis:2008wq}. 
We thus find of great interest to investigate what are the distinctive features of CR positrons for extended DM models which include semi-annihilations. 

\begin{figure}[t]
    \centering
    \includegraphics[width=0.5\linewidth]{Figures/legenda_spectra.pdf}\\ \medskip
    \subfloat[\label{fig:cfr_gamma_e}]{\includegraphics[width=.48\linewidth]{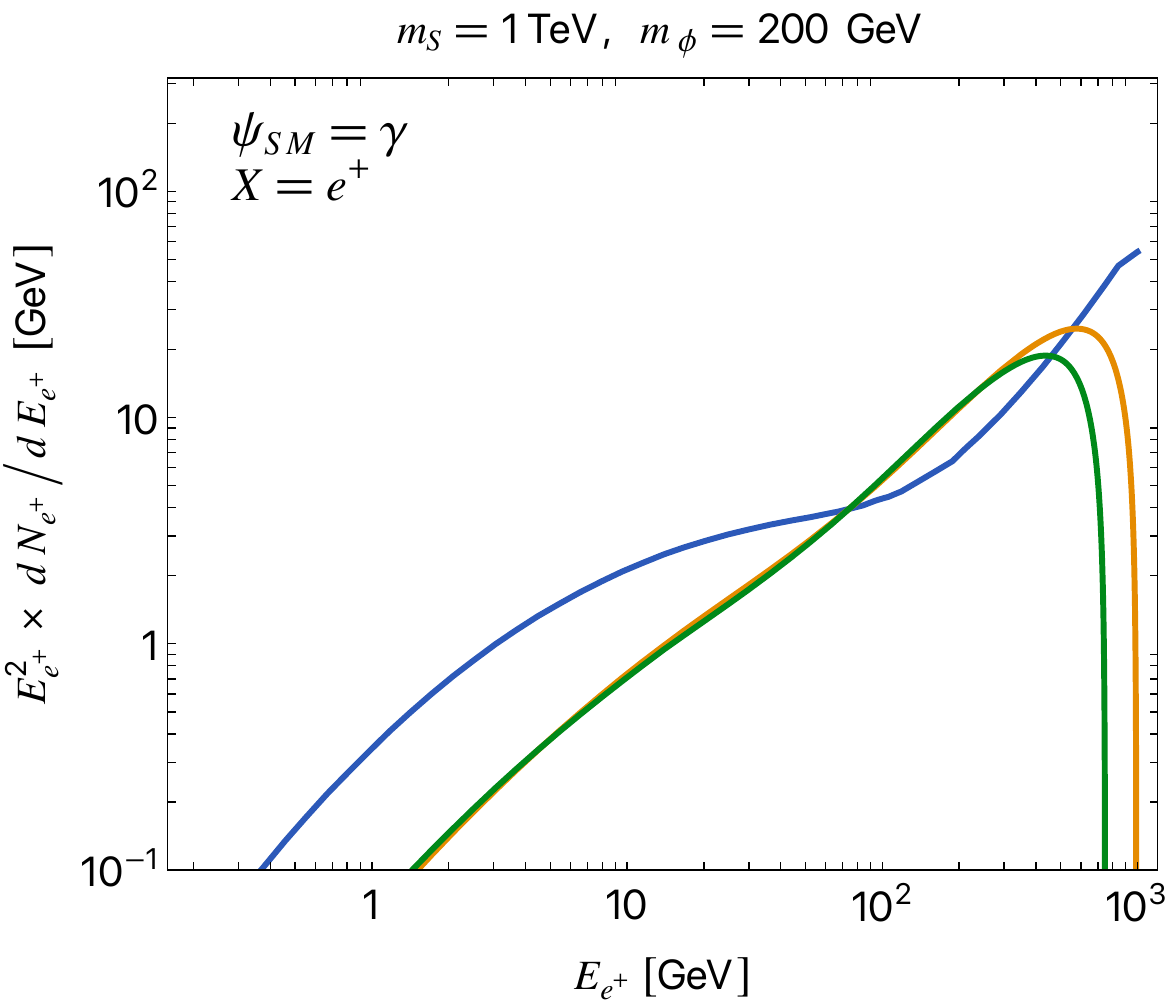}}\quad
    \subfloat[\label{fig:cfr_b_e}]{\includegraphics[width=.48\linewidth]{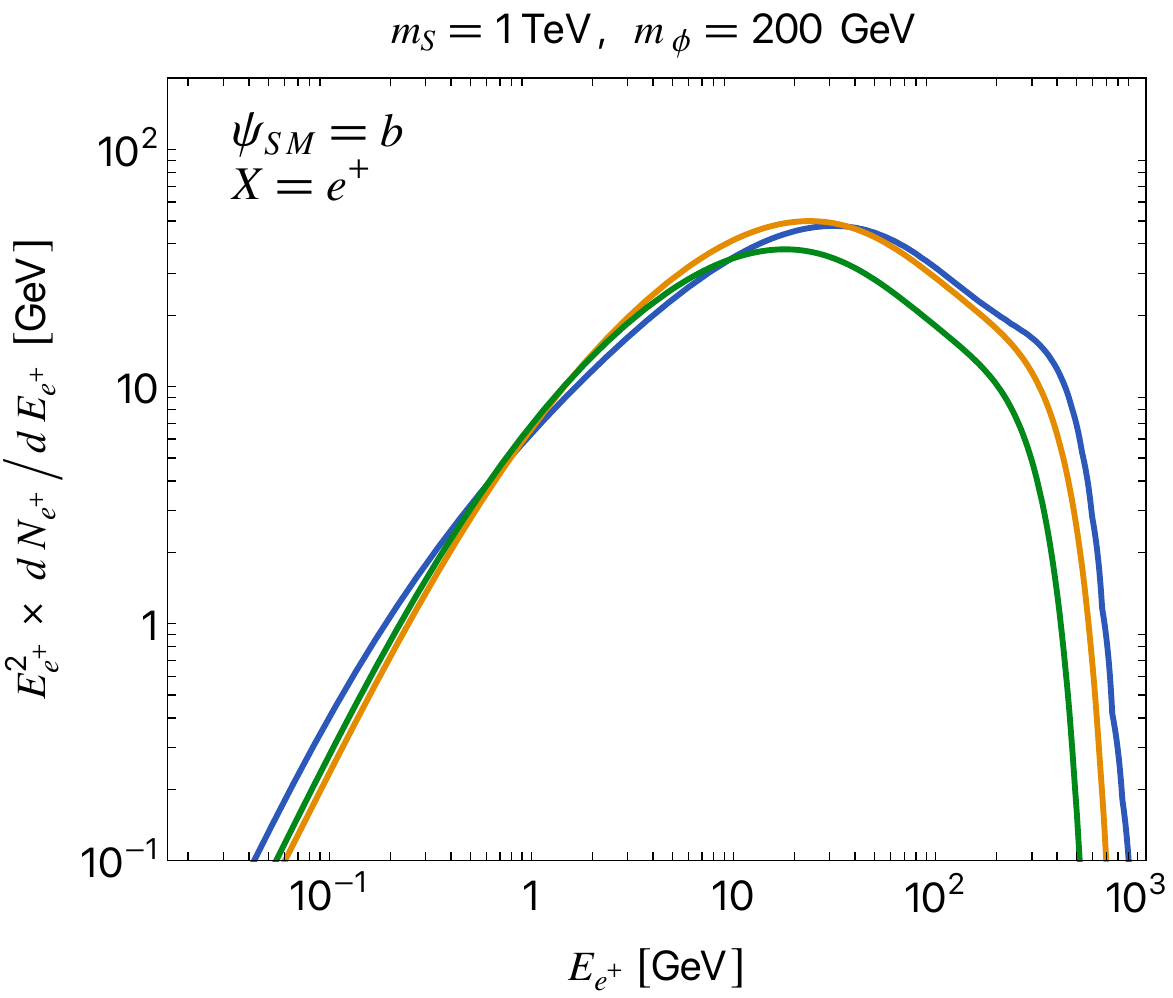}}\\
    \subfloat[\label{fig:cfr_e_e}]{\includegraphics[width=.48\linewidth]{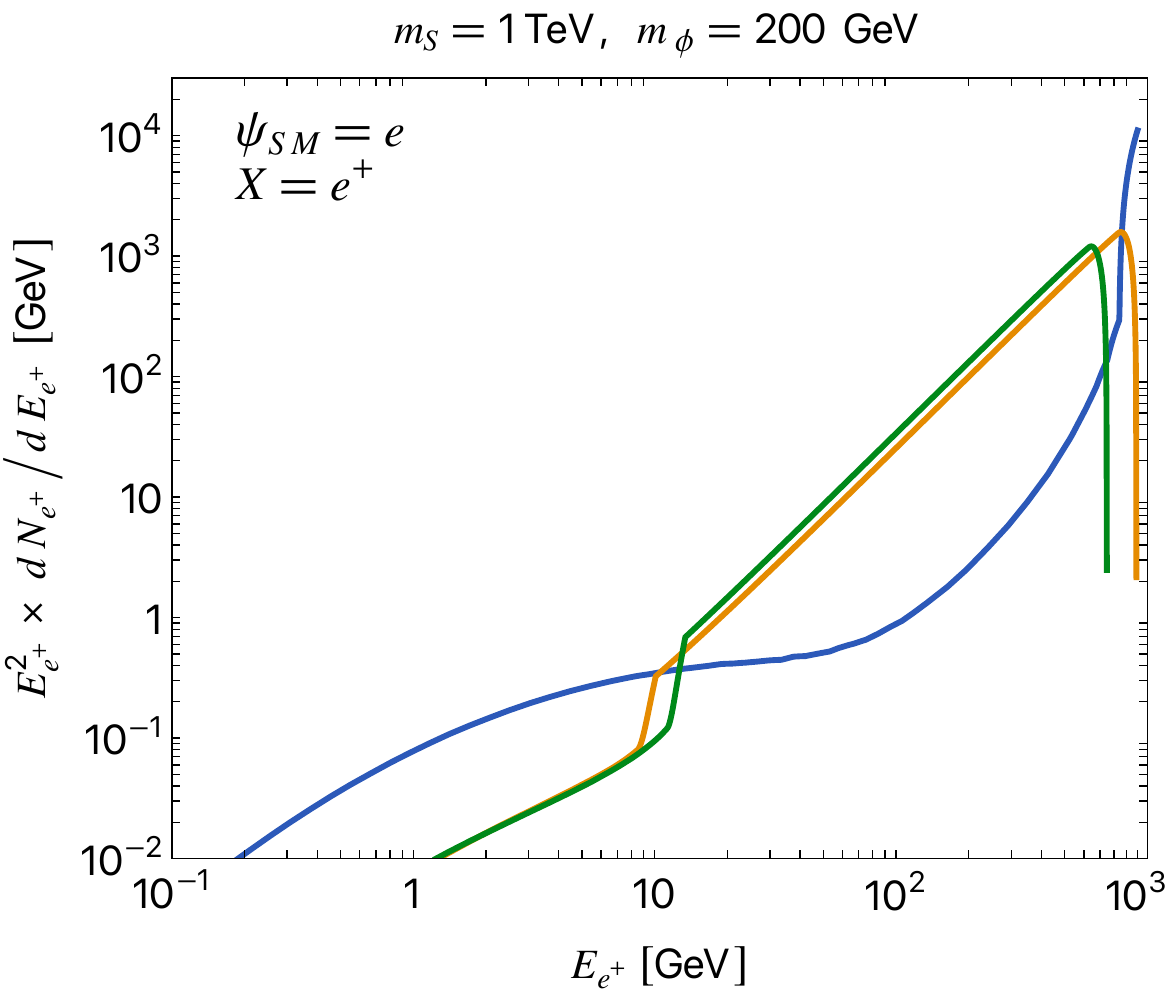}}\quad
    \subfloat[\label{fig:cfr_tau_e}]{\includegraphics[width=.48\linewidth]{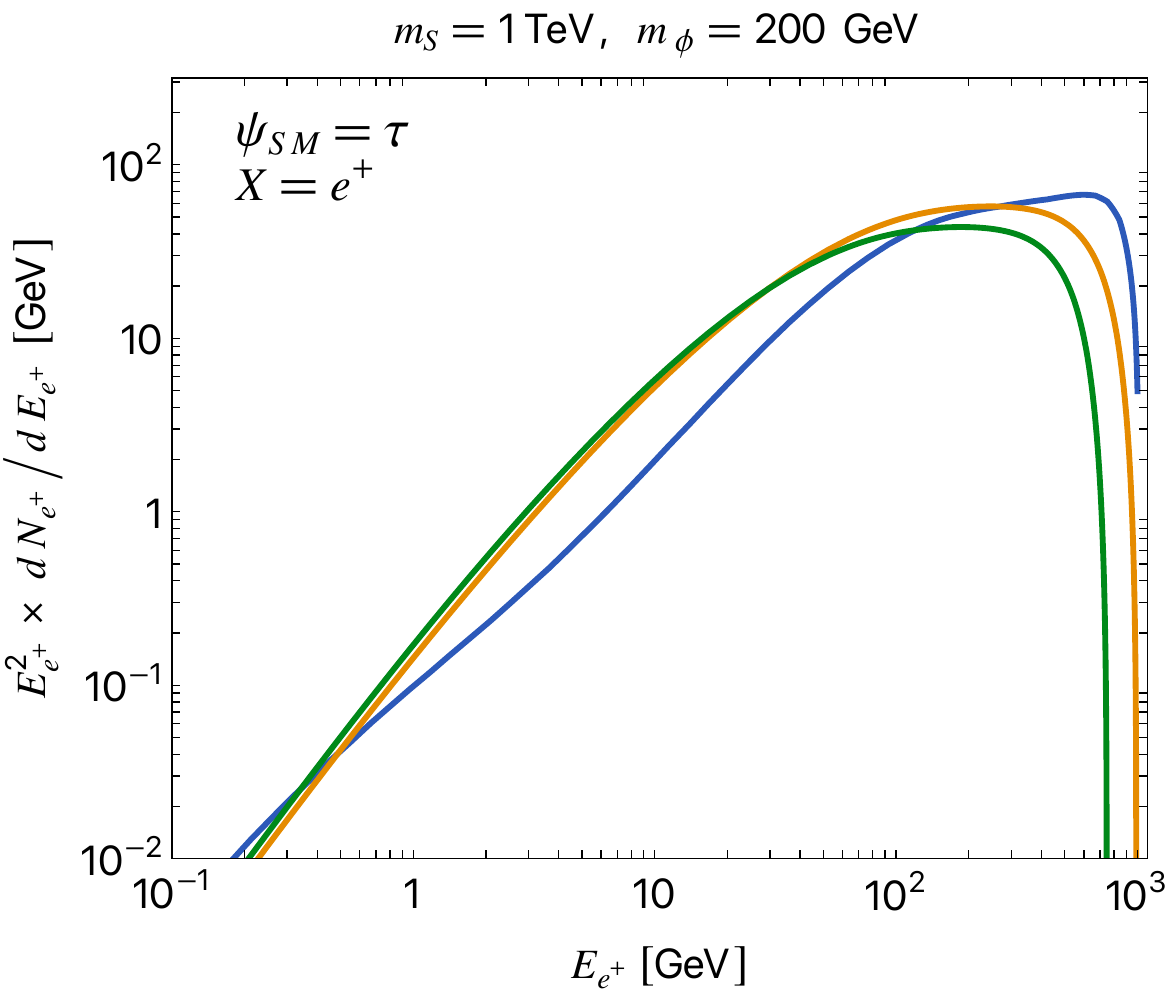}}
    \caption{\textit{Single process injection spectra: positrons.} Same as \Fig{fig:cfr_gammas} but for the spectra of positrons per single DM collision.
    }
    \label{fig:cfr_positrons}
\end{figure}

We start  comparing in \Fig{fig:cfr_positrons} the injection spectra of positrons per single DM pair collision in scenarios where the signal is generated by a single type of scattering process. 
At this stage, we can see that the shape and the characteristics of positron  spectra are quite similar to those discussed for $\gamma$ rays. A feature which appears to be common to all the cases under consideration, with $\psi_{\rm SM}=b$ as the only exception, is that one-step cascade spectra (green and orange) can dominate over the direct annihilation signal (blue lines) in a large energy interval. 
For primary photons (\Fig{fig:cfr_gamma_e}) this is restricted to a region around hundred of GeV for the specific benchmark under consideration, with one-step spectra then decreasing almost linearly with energy. 
A similar argument applies to the scenario where primary and final states are $\psi_{\rm SM}=X=e$ as shown in \Fig{fig:cfr_e_e}. Here, the annihilation spectrum (blue line) is very similar to a line-like emission with a much attenuated tail at low energies, while one-step spectra resemble very much a box-shape, which overwhelms the self-annihilation curve by orders of magnitude in a quite broad energy interval from around 10~GeV up to hundreds of GeV, while indeed remaining suppressed outside the box domain. Remarkable is also the behavior of positrons spectra from primary $\tau$ states (\Fig{fig:cfr_tau_e}): one-step spectra are mildly suppressed at the very high energy end of the spectrum, but they turn out to dominate (or at least to be of the same size) over the self-annihilation scenario basically over the whole remaining energy interval. 

\begin{figure}[t]
    \centering
    \includegraphics[width=0.9\linewidth]{Figures/legenda_ratios.pdf}\\
    \medskip
    \subfloat[\label{fig:r_gamma_e}]{\includegraphics[width=.48\linewidth]{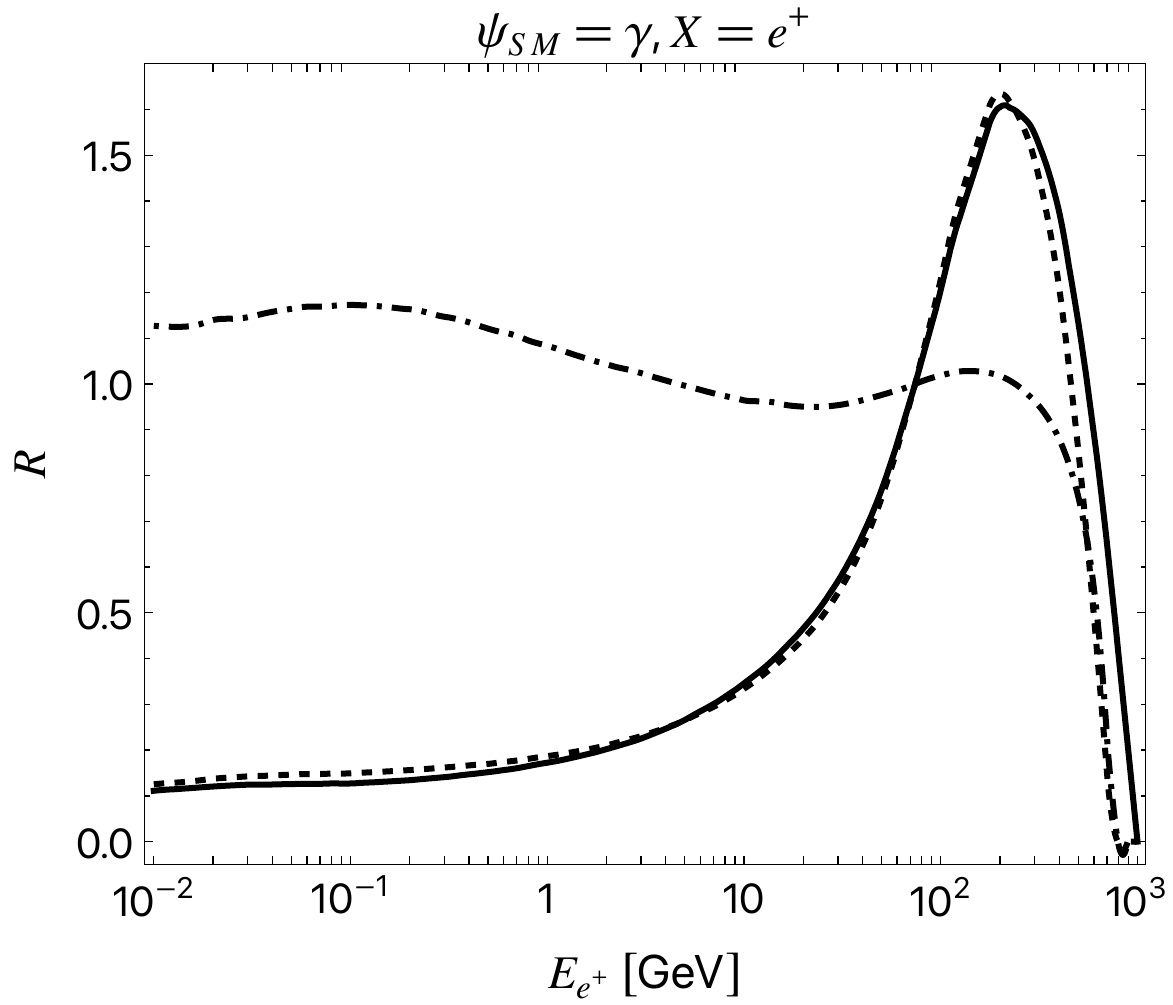}}\quad
    \subfloat[\label{fig:r_b_e}]{\includegraphics[width=.48\linewidth]{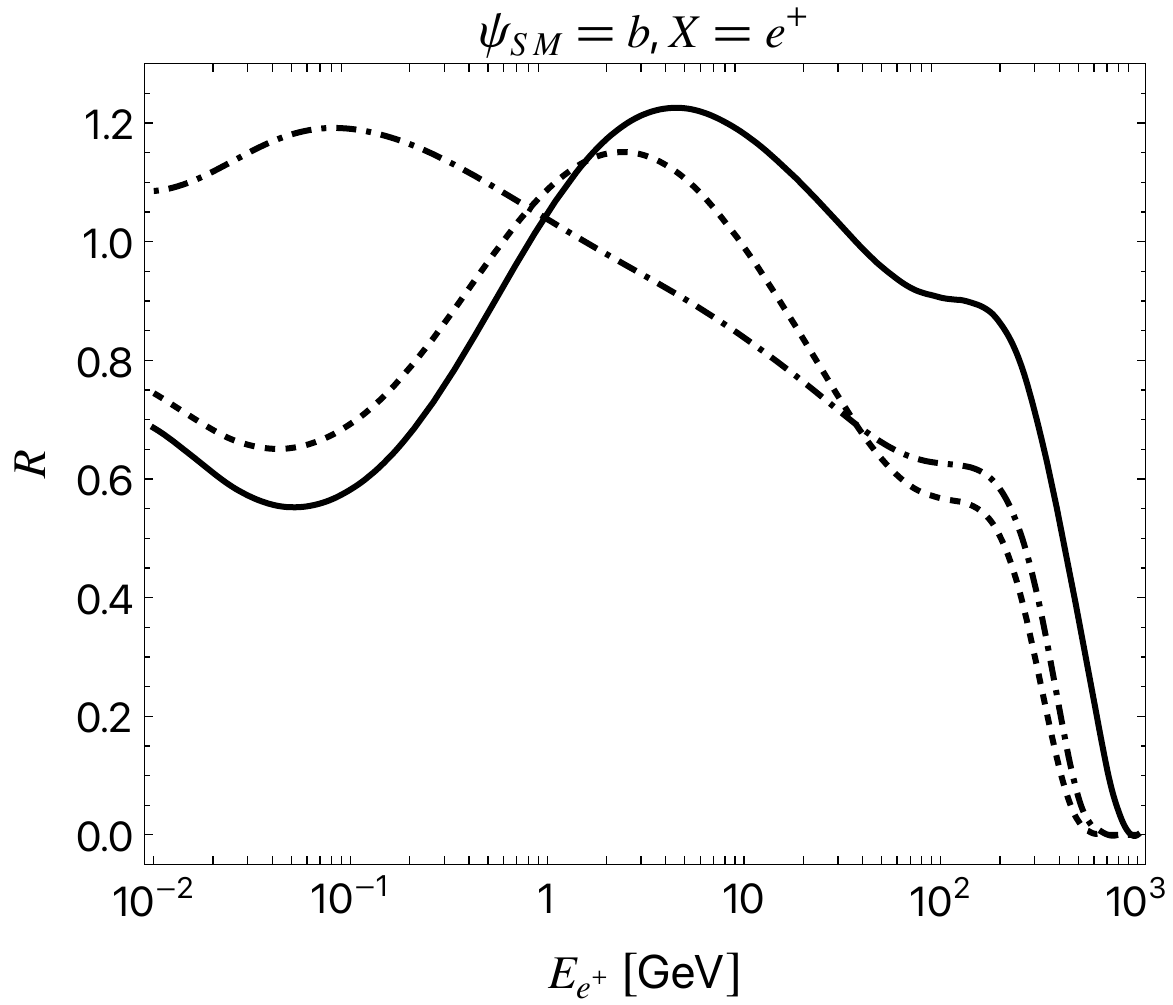}}\\
    \subfloat[\label{fig:r_e_e}]{\includegraphics[width=.48\linewidth]{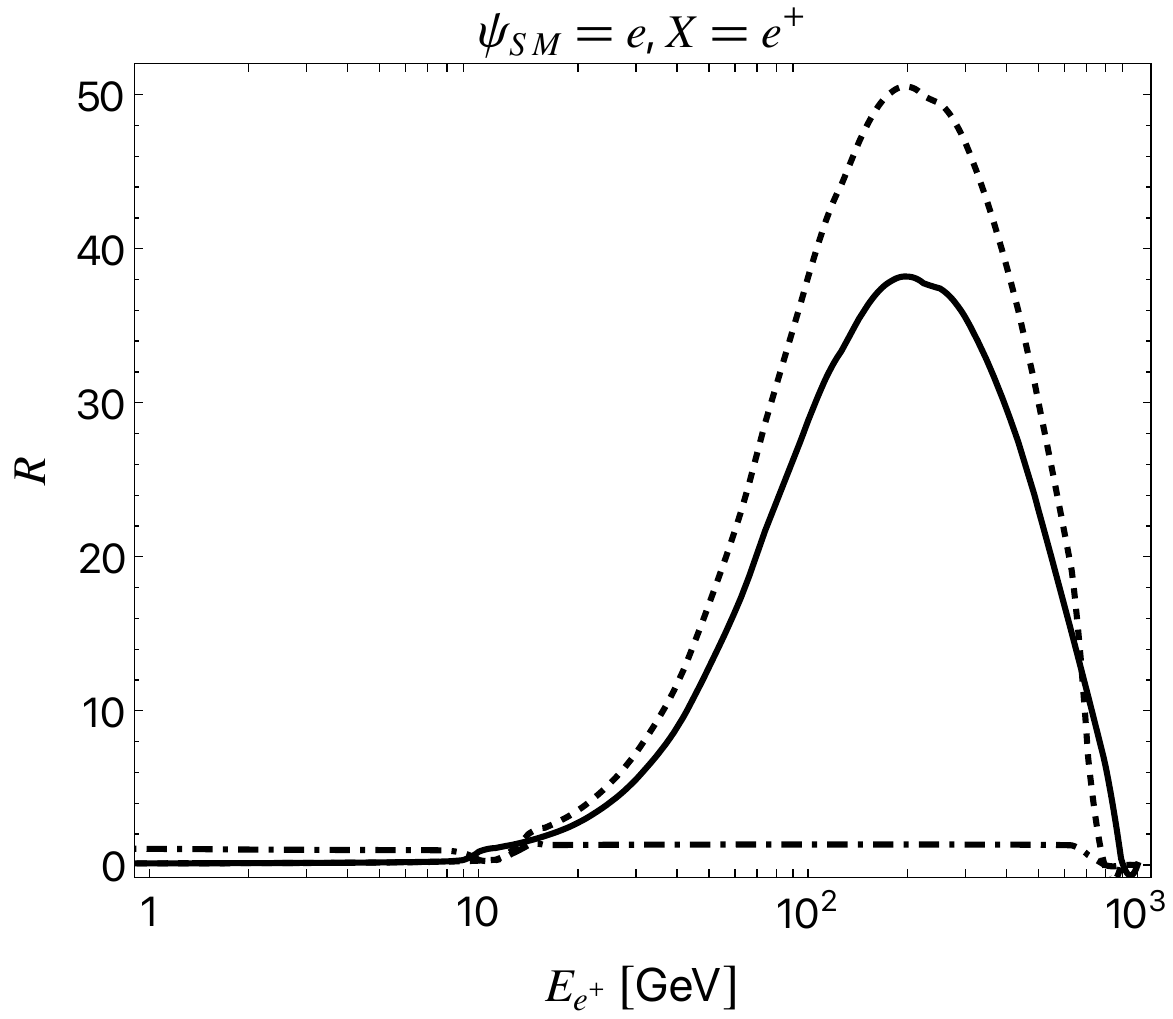}}\quad
    \subfloat[\label{fig:r_tau_e}]{\includegraphics[width=.48\linewidth]{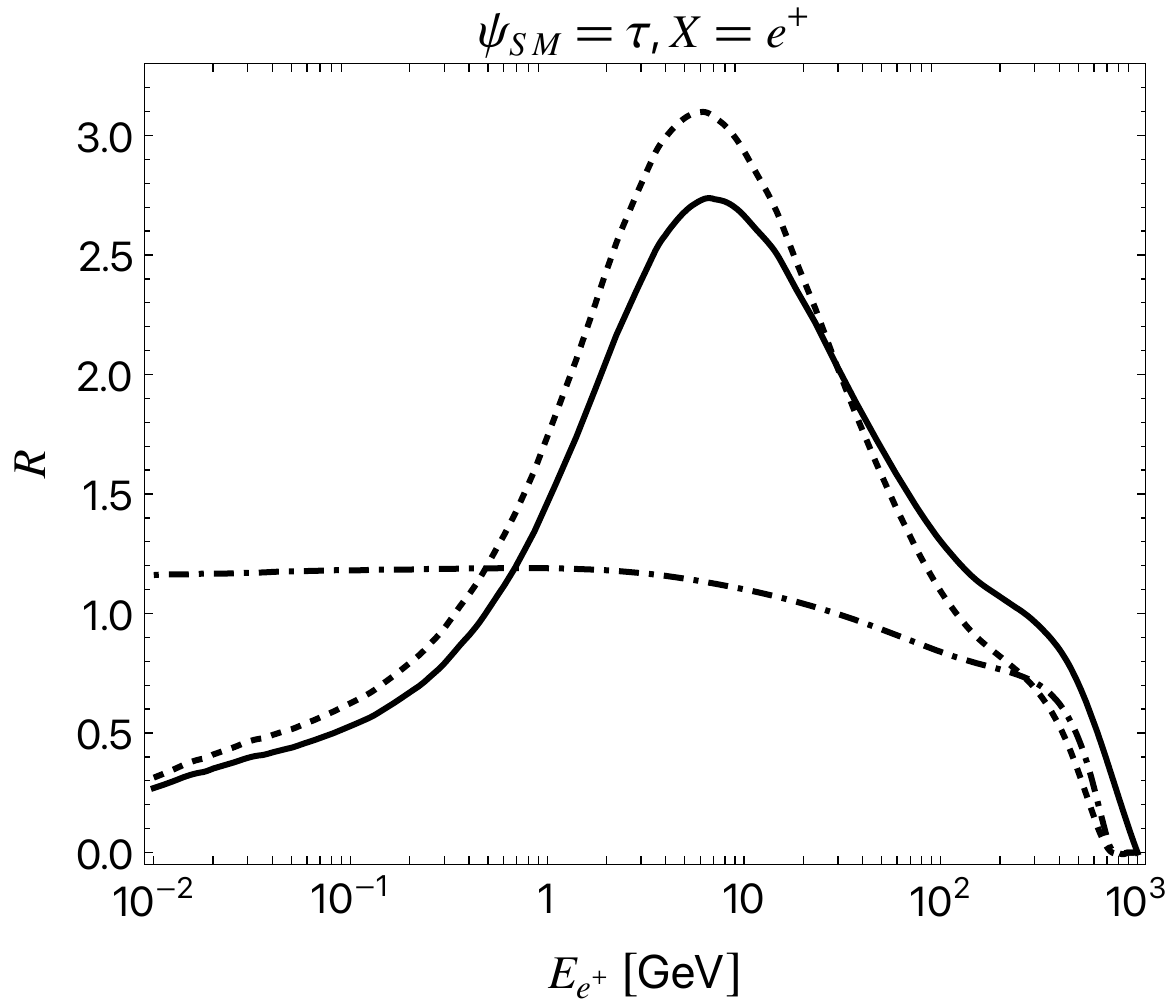}}
\caption{\textit{Ratio of positron spectra from different single processes.} Same as \Fig{fig:ratios_gammas} but for the differential injection spectra of positrons per single DM collision process.
}
\label{fig:ratios_positrons}
\end{figure}

The $\psi_{\rm SM}=b$ case, on the other hand, illustrates that only the semi-annihilation spectra are suppressed at high energies, but eventually all three scenarios end up giving comparable signals both in shape and magnitude. All of these characteristics are further highlighted when looking at the ratios of the positron injection spectra illustrated in \Fig{fig:ratios_positrons}.
Again, the most interesting feature is observed for a leptophilic intermediate state decaying into $\psi_{\rm SM}=(e,\,\tau)$. 
In the former case, one-step (semi-)annihilation spectra result to be up to a factor $(50)40$ larger around hundreds of GeV than the usual self-annihilation scenario while being comparable to each other. In the latter scenario, we notice that one-step cascade generated positron spectra are up to a factor of 3 larger than the annihilation case at around tens of GeV, while being suppressed at the high and low energy tails of the spectrum.

\begin{figure}[t]
    \centering
    \includegraphics[width=0.9\linewidth]{Figures/legenda_ab.pdf}\\\medskip
    \subfloat[\label{fig:mix_gamma_e}]{\includegraphics[width=.48\linewidth]{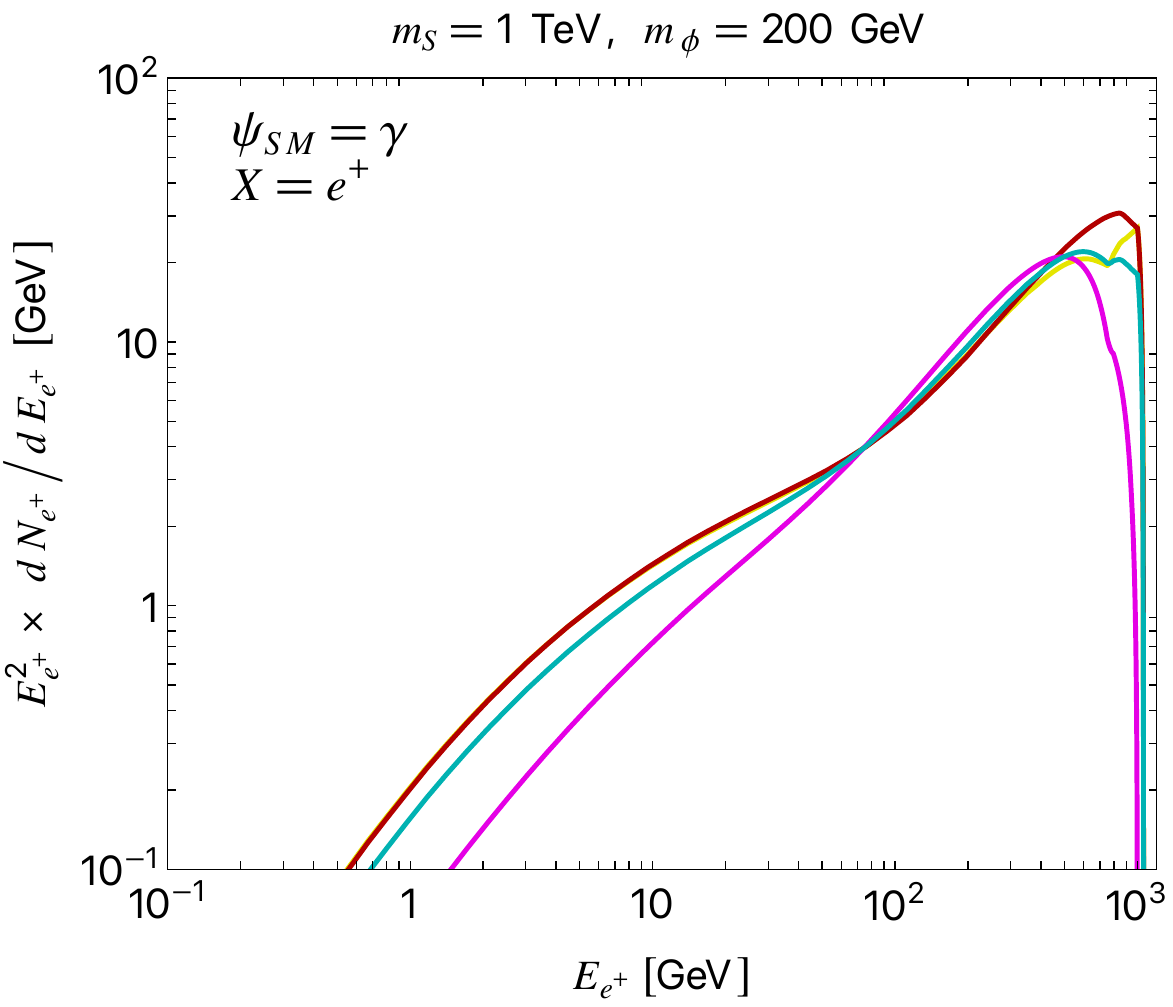}}\quad
    \subfloat[\label{fig:mix_b_e}]{\includegraphics[width=.475\linewidth]{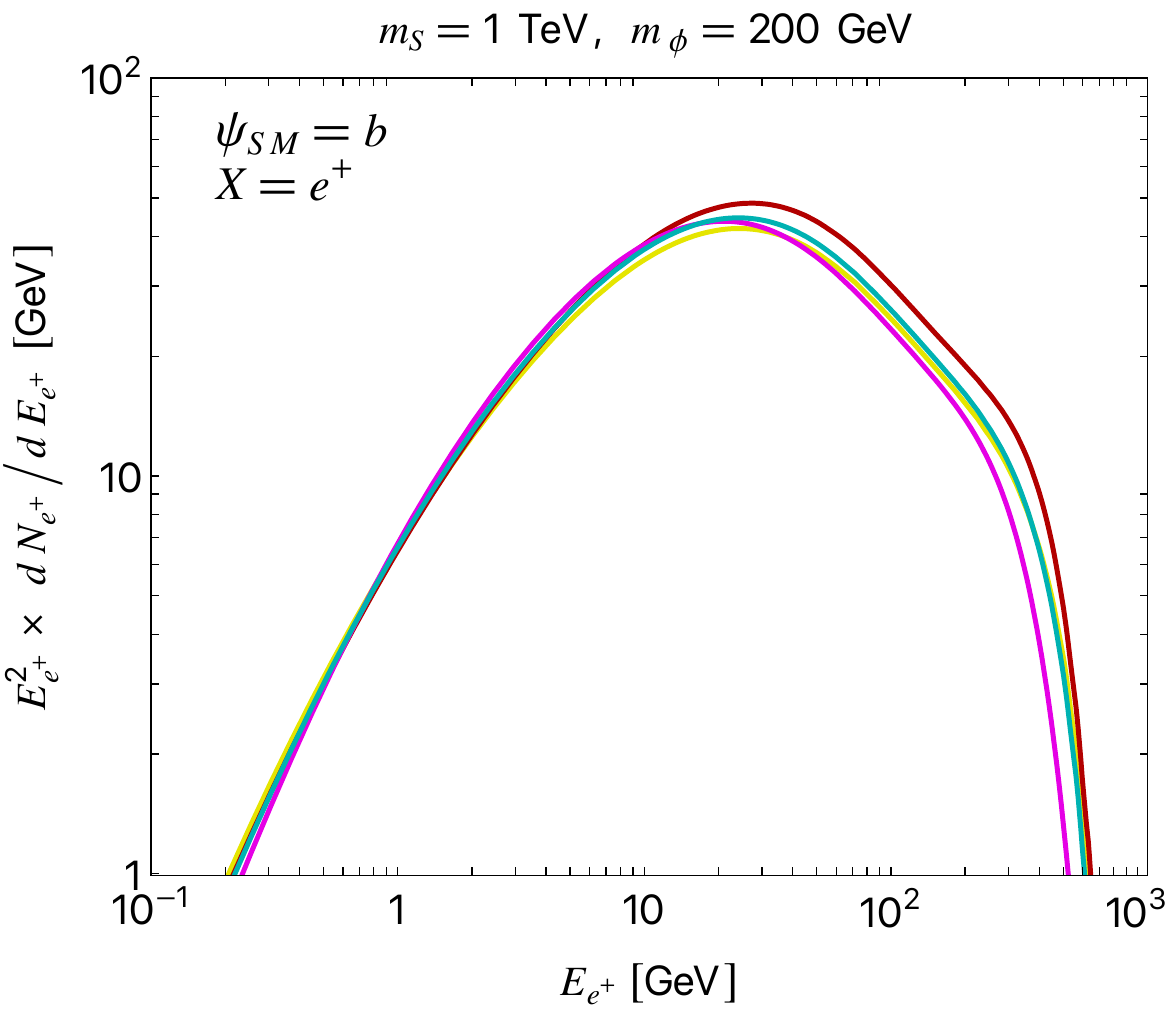}}\\
    \subfloat[\label{fig:mix_e_e}]{\includegraphics[width=.48\linewidth]{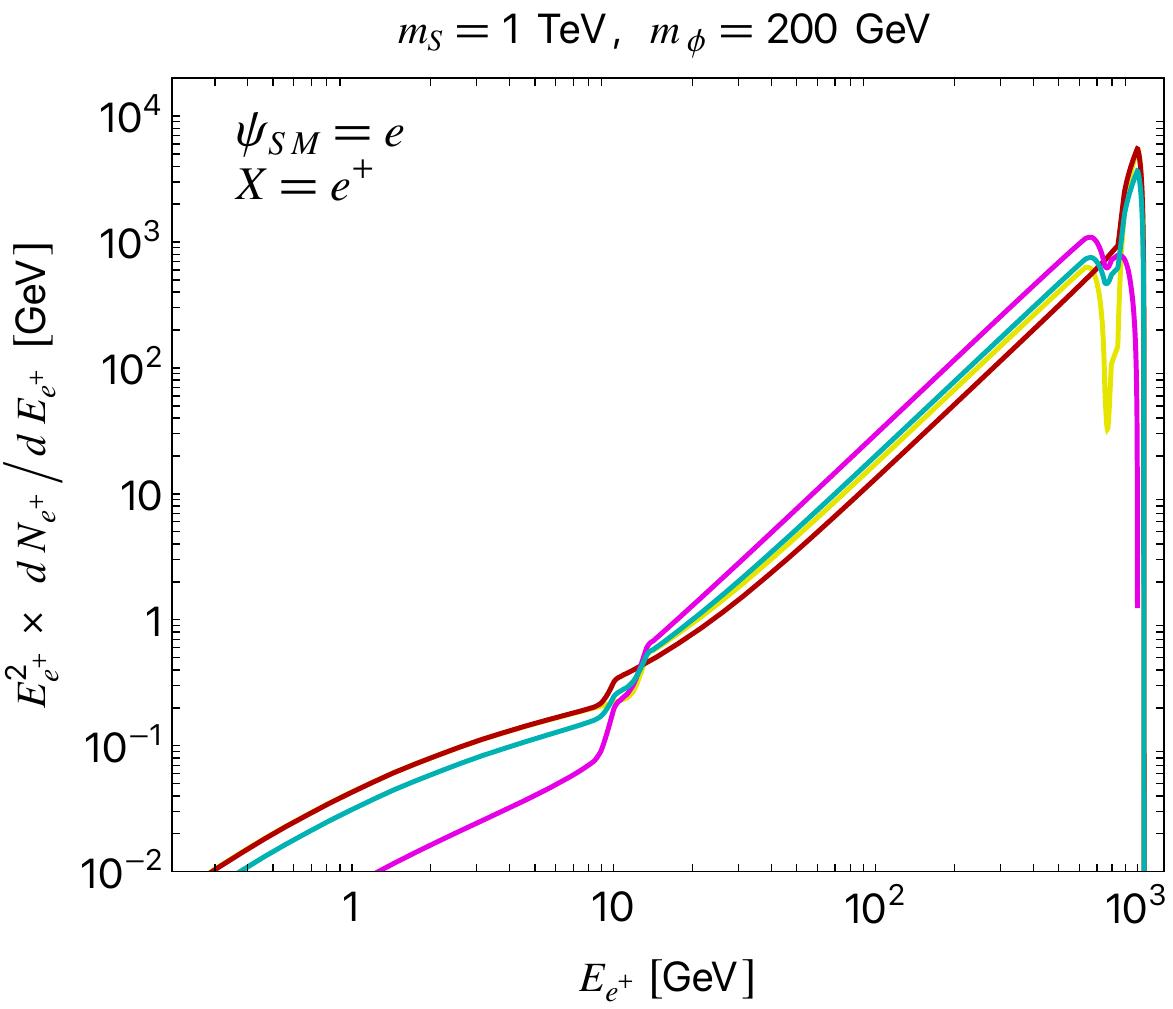}}\quad
    \subfloat[\label{fig:mix_tau_e}]{\includegraphics[width=.48\linewidth]{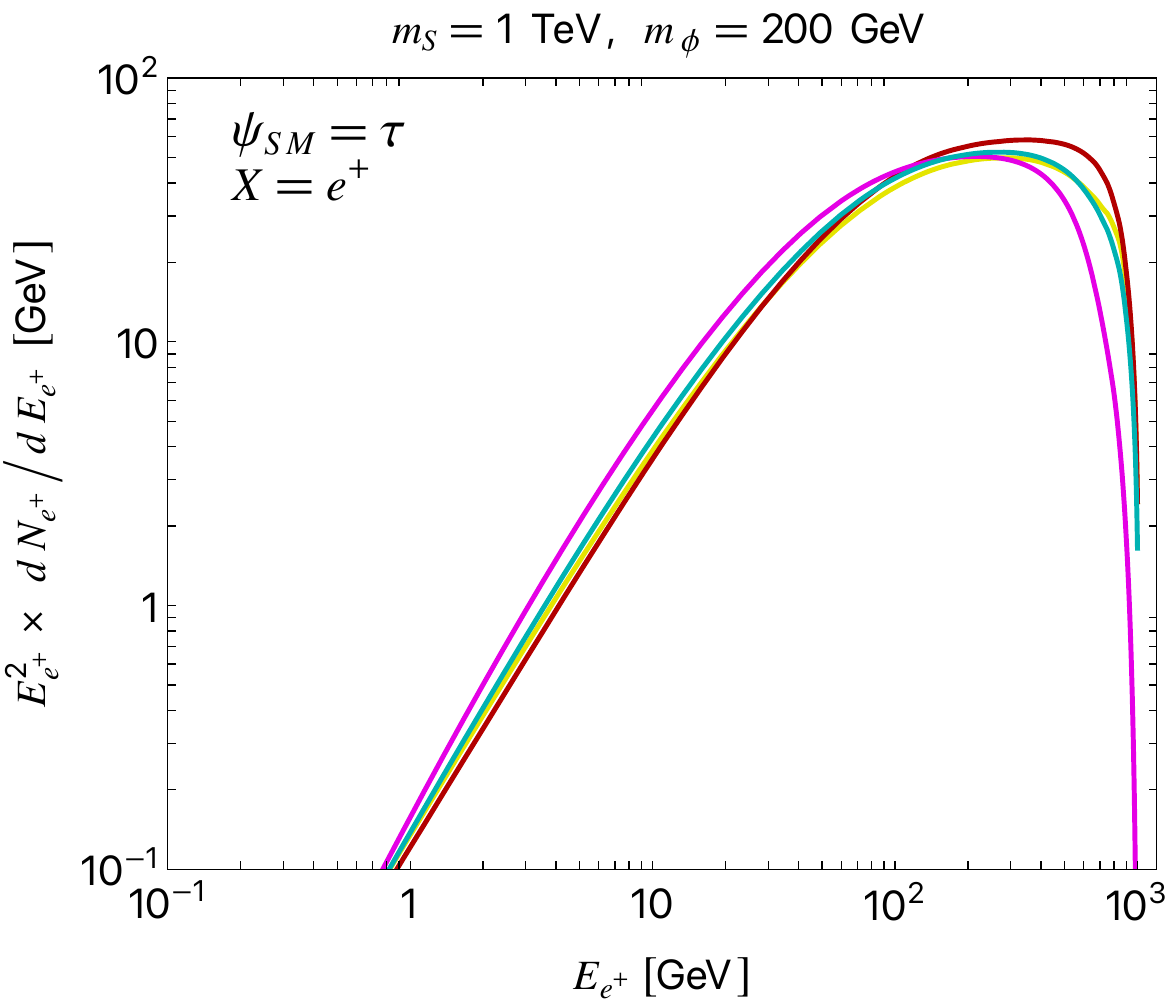}}
\caption{\textit{Injection spectra from a superposition of processes: positrons}. Same as \Fig{fig:mixing_gammas} but for the positrons per single DM collision process.}
\label{fig:mixing_positrons}
\end{figure}

Finally, the scenario of mixed DM phenomenology is illustrated in \Fig{fig:mixing_positrons}. The most interesting aspect we value worth of notice, is that this time the one-step spectral features can be enhanced if these processes have comparable cross sections to the annihilation one. As a consequence, all the mixed scenarios we show, end up revealing no striking features which could make them stand out with respect to the selected benchmark cases of study. The only exception is provided by the $\psi_{\rm SM}=e$ channel, for which the $(\alpha,\,\beta)=(0,\,0.5)$ combination clearly shows the superposition of a box shape and high energy peak inherited from the direct annihilation process. 
Indeed, the full one-step case is still generally the most suppressed one. However, the enhancement previously discussed for positron spectra in the different production channels suggests that the corresponding scattering cross section bounds may differ non-trivially from those currently derived using direct annihilation as the reference spectral template. Quantifying these effects requires a dedicated re-analysis of AMS-02 data after the propagation of positrons in the Galaxy is taken into account, along the lines of the study of box-shaped spectra performed in \cite{Zu:2017dzm}. 

\subsection{Antiprotons and antinuclei}
\label{sec:antiprotons}

Antiprotons are the target of several experimental searches of DM, through e.g.~the AMS-02~\cite{2021PhR...894....1A} and, earlier, PAMELA~\cite{PAMELA:2010kea} data.  
Antiprotons in the Galaxy are expected to be produced by astrophysical processes such as the spallation of primary CRs. A contribution from DM-sourced processes to the observed AMS-02 antiproton flux has been extensively investigated in the past years, finding that potential DM-related excesses are not robust against uncertainties in the CR propagation models, in the spallation cross section  and against the inclusion of the full covariance of experimental data \cite{Boudaud:2019efq,Heisig:2020nse,Calore:2022stf,Balan:2023lwg}. 
In addition, high energy antiprotons from DM are expected to be generally accompanied by the production of high energetic $\gamma$ rays \cite{Arcadi:2017vis,Cholis:2019ejx,Hooper:2019xss}. 

Antinuclei, on the other end, offer one the most striking potential signatures of DM  in galactic environments. For large enough values of the DM mass, an enhanced flux of heavy anti-nuclei, like antideuterons ($\bar D$) or antihelium $\overline{\rm{He}}$, is expected \cite{Donato:1999gy,Serksnyte:2022onw}. 
In particular, for DM masses of tens of GeV, the $\bar D$ DM signal is expected to be peaked at  energies less than about 3 GeV/nucleon, where the astrophysical backgrounds are expected to be subdominant, making this signal a potential smoking gun for DM  events. 
To this end, a number of existing experiments like AMS-02 \cite{AMS:2024idr} and BESS \cite{BESS:2024yma} - which currently sets the stronger upper limits - have been taking data (or will improve) looking for excesses in antinuclei fluxes.  
The relevance of this detection strategy, together with the increasing sensitivity of the experimental apparatus of new and forthcoming facilities (GAPS \cite{Aramaki:2015laa}), have attracted a great attention in recent years.

\subsubsection{Final states mass effects}
So far, we investigated the features of injection spectra of massless or light CRs. Indeed, for dark sector masses in the range of interest, which is given by the typical WIMP mass window, also positrons can be considered to be effectively massless to a good approximation. However, massive and charged CRs like antiprotons provide a strong, complementary test ground for DM ID. 
In this case, it is the kinetic energy of the antiproton which is the relevant quantity, and the injection spectra must be computed as function of this variable. 
In this work, we take into account the full dependence of injection spectra on both intermediate and final states masses, in order to make the discussion as general as possible. This is to allow for any possible separation of scales between the different particles masses involved (still within the kinematic region), without the need to specify our results to be valid only for the large hierarchy regime. 
A full discussion of final states mass effect, and a comparison with previous literature \cite{Elor:2015bho}  is carried in details in \App{app:kinematics}. 
There, we study both qualitatively and quantitatively the differences arising between antiproton injection spectra derived within the large hierarchy approximation, namely when the mediator and final state messenger masses are neglected, and the results we obtained in full generality applying \Eq{eq:dNdE}. Our results are intended to be, eventually, a generalised extension of those presented in the past literature, since they are computed with no initial assumption on the hierarchy between the masses of the particles participating to the cascade reaction. 

\begin{figure}[tb]
    \centering
    \includegraphics[width=0.5\linewidth]{Figures/legenda_spectra.pdf}\\ \medskip
    \subfloat[\label{fig:cfr_gamma_p}]{\includegraphics[width=.48\linewidth]{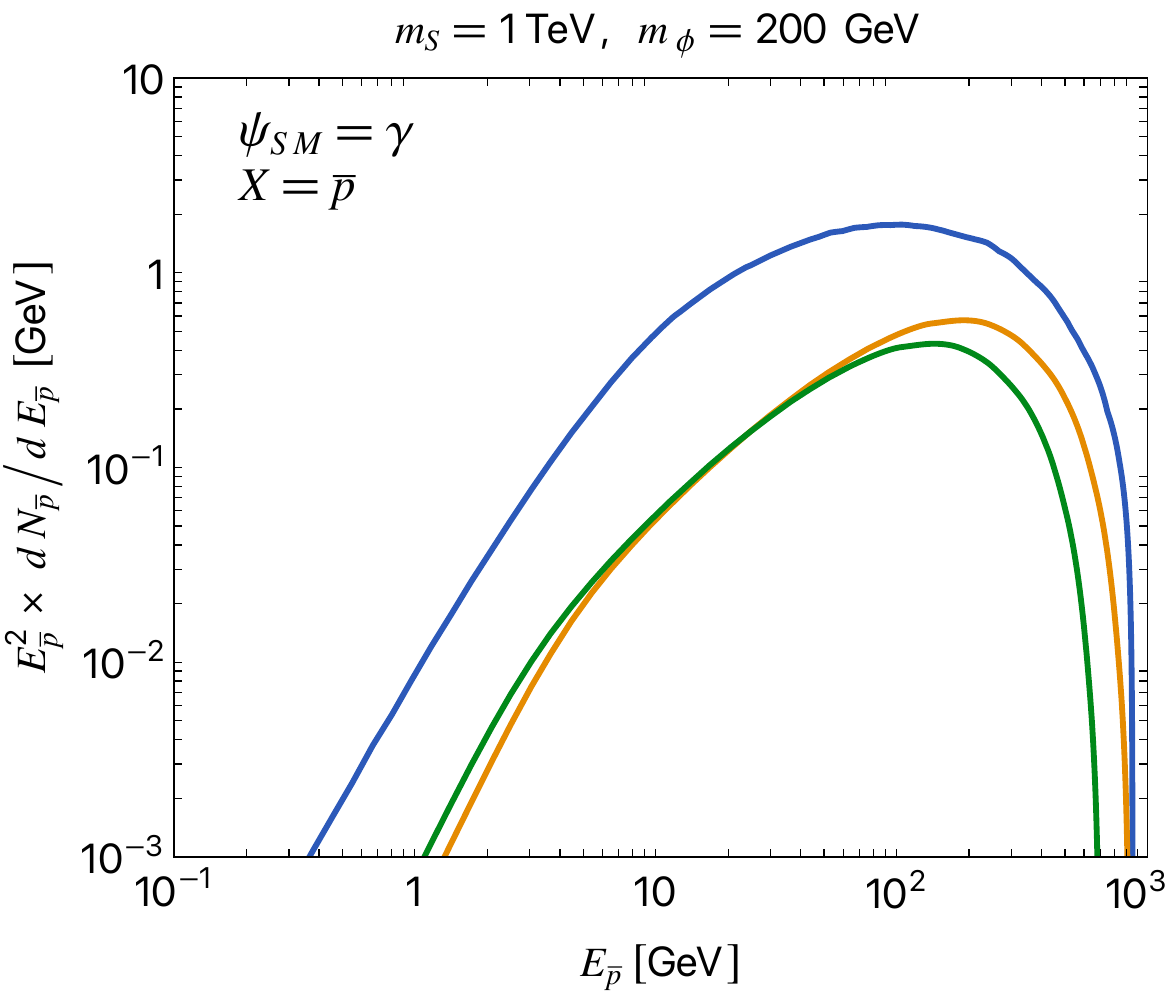}}\quad
    \subfloat[\label{fig:cfr_b_p}]{\includegraphics[width=.48\linewidth]{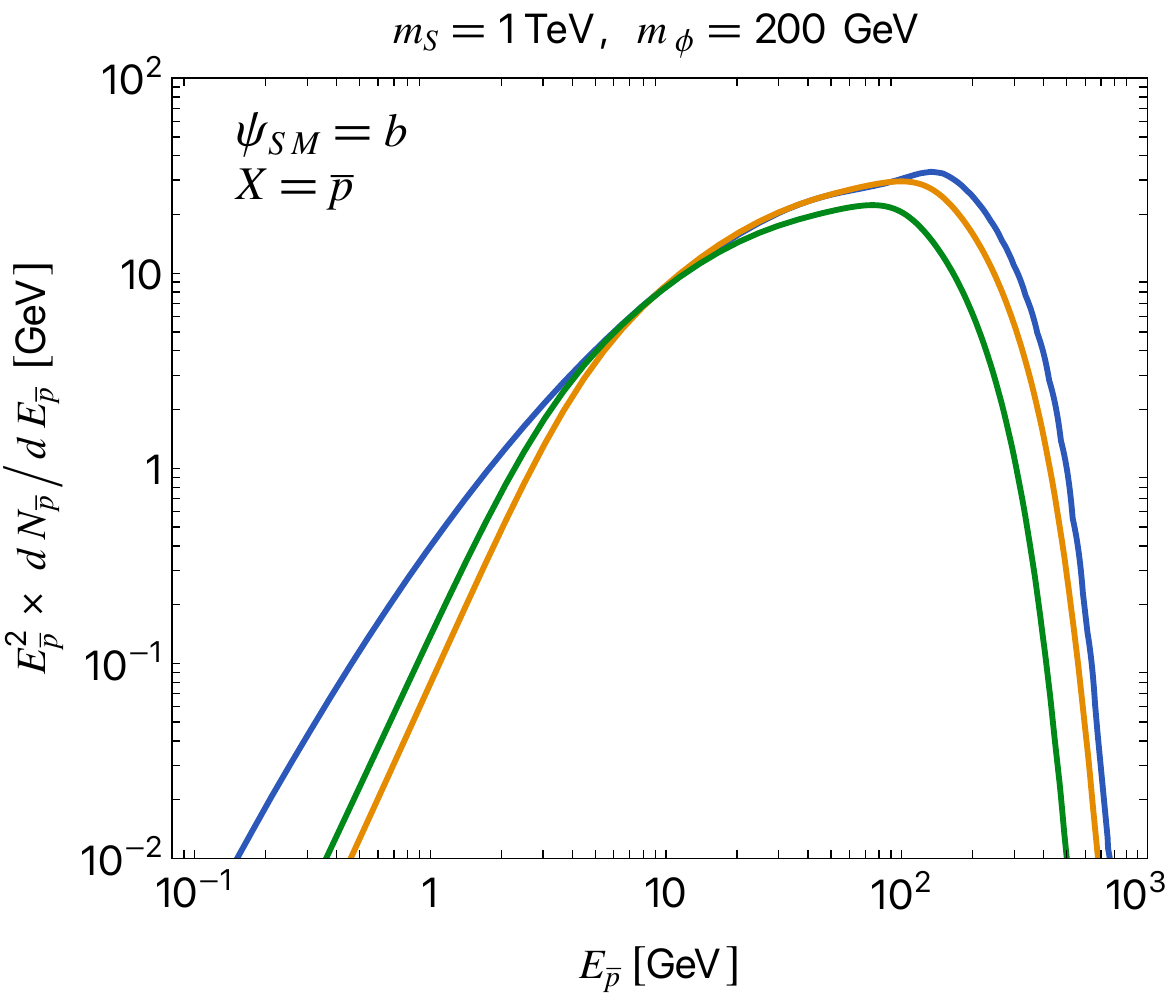}}\\
    \subfloat[\label{fig:cfr_e_p}]{\includegraphics[width=.48\linewidth]{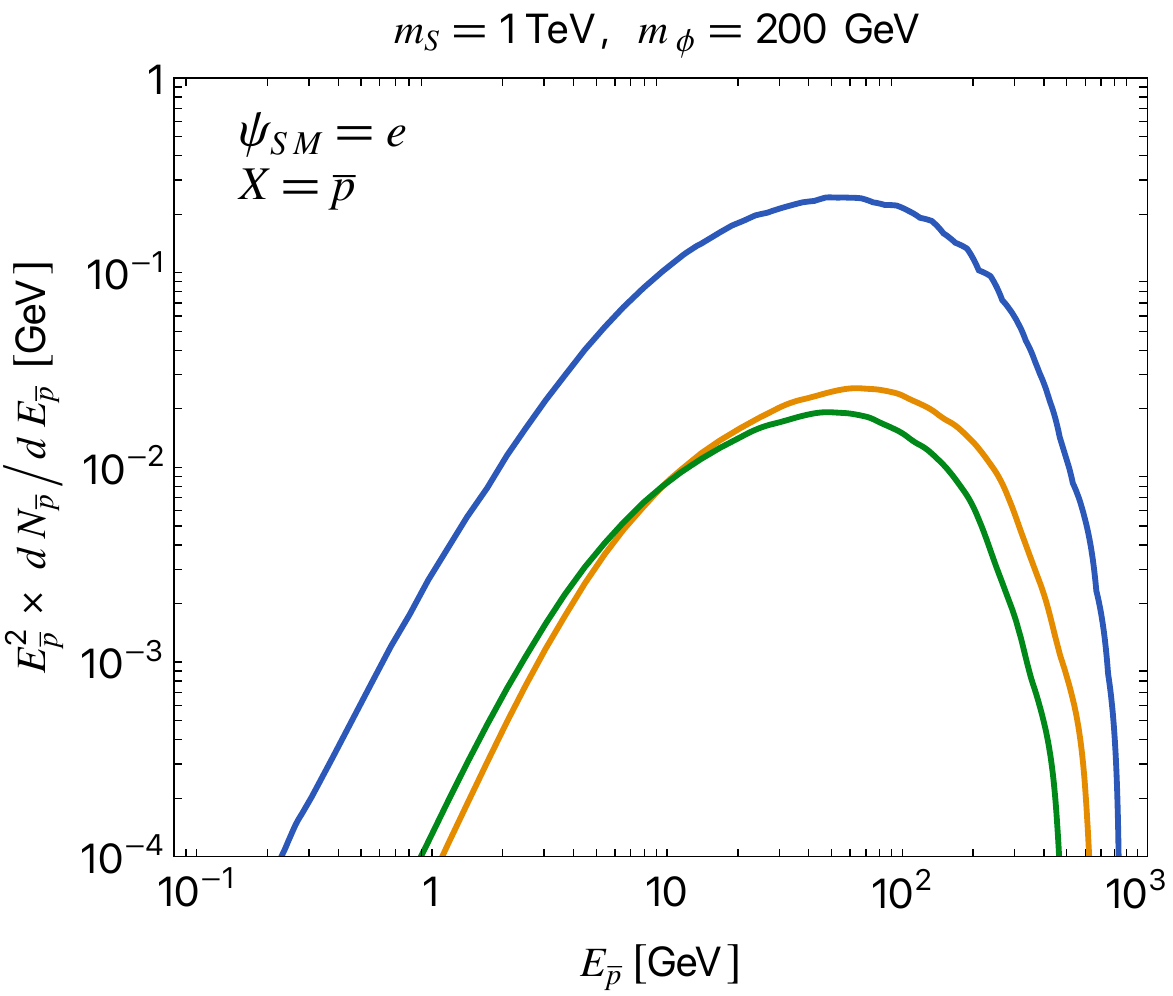}}\quad
    \subfloat[\label{fig:cfr_tau_p}]{\includegraphics[width=.48\linewidth]{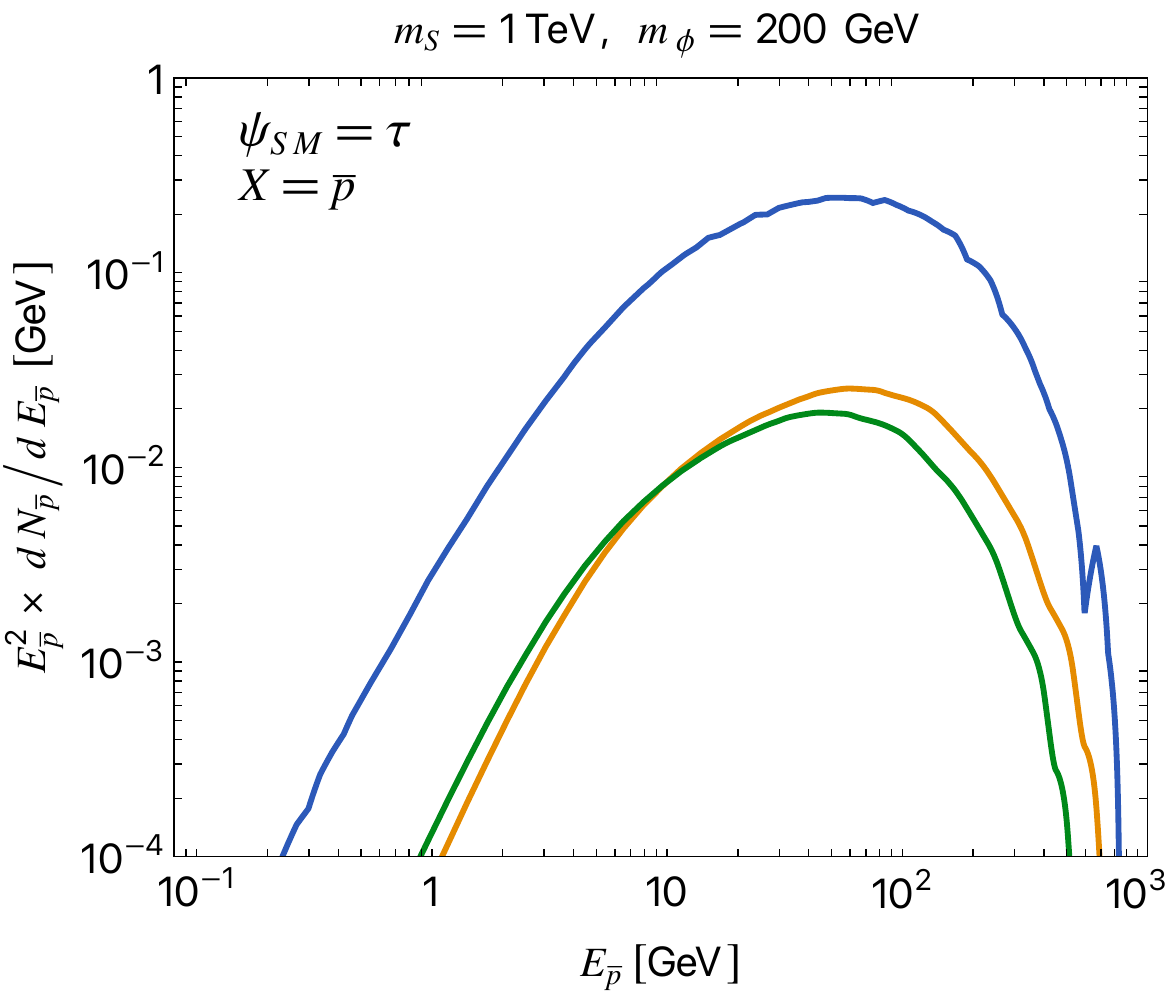}}
    \caption{\textit{Single process injection spectra: antiprotons.} Same as \Fig{fig:cfr_gammas} but for the spectra of antiprotons per single DM collision.
    }
    \label{fig:cfr_antiprotons}
\end{figure}

\subsubsection{Results}
In \Fig{fig:cfr_antiprotons} we show  the comparison between antiproton injection spectra in scenarios where DM phenomenology is sourced 100\% by a single collision process. The results show sizable effects in all the primary channels selected. All injection spectra generated by one-step cascade (semi-)annihilation processes are evidently suppressed with respect to the usual direct self-annihilation signal up to one order of magnitude at their peak in the case of $\psi_{\rm SM}=(e,\,\tau)$. This suppression affects the spectra over the whole energy range, contrary to what observed in all the previously discussed cases. 
This is further evident in \Fig{fig:ratios_antiprotons}, where we see that one-step (semi-)annihilation to annihilation rates ratio is always smaller than one and of the order of $O(10^{-2})$ over a broad energy interval.
Concentrating on the $\psi_{\rm SM}=b$, which results in a higher energy output in antiprotons with respect to the other benchmarks investigated~\cite{Cirelli:2010xx}, the ratio $R$ among the one-step (semi)-annihilation channels and direct annihilation (solid and dotted lines in \Fig{fig:r_b_p}) ranges from $R\sim0.1$ to $O(1)$ across the energy range investigated. Given that AMS-02 antiproton data constrain standard annihilation scenarios up to the TeV scale \cite{Calore:2022stf,Balan:2023lwg}, we expect dedicated propagation modeling of CR antiprotons and likelihood analyses of AMS-02 data  to have sensitivity to the modified injection spectra discussed here.

\begin{figure}[t]
    \centering
    \includegraphics[width=0.5\linewidth]{Figures/legenda_spectra.pdf}
    \\
    \medskip
    \subfloat[\label{fig:cfr_eps_onestep_p}]{\includegraphics[width=.48\linewidth]{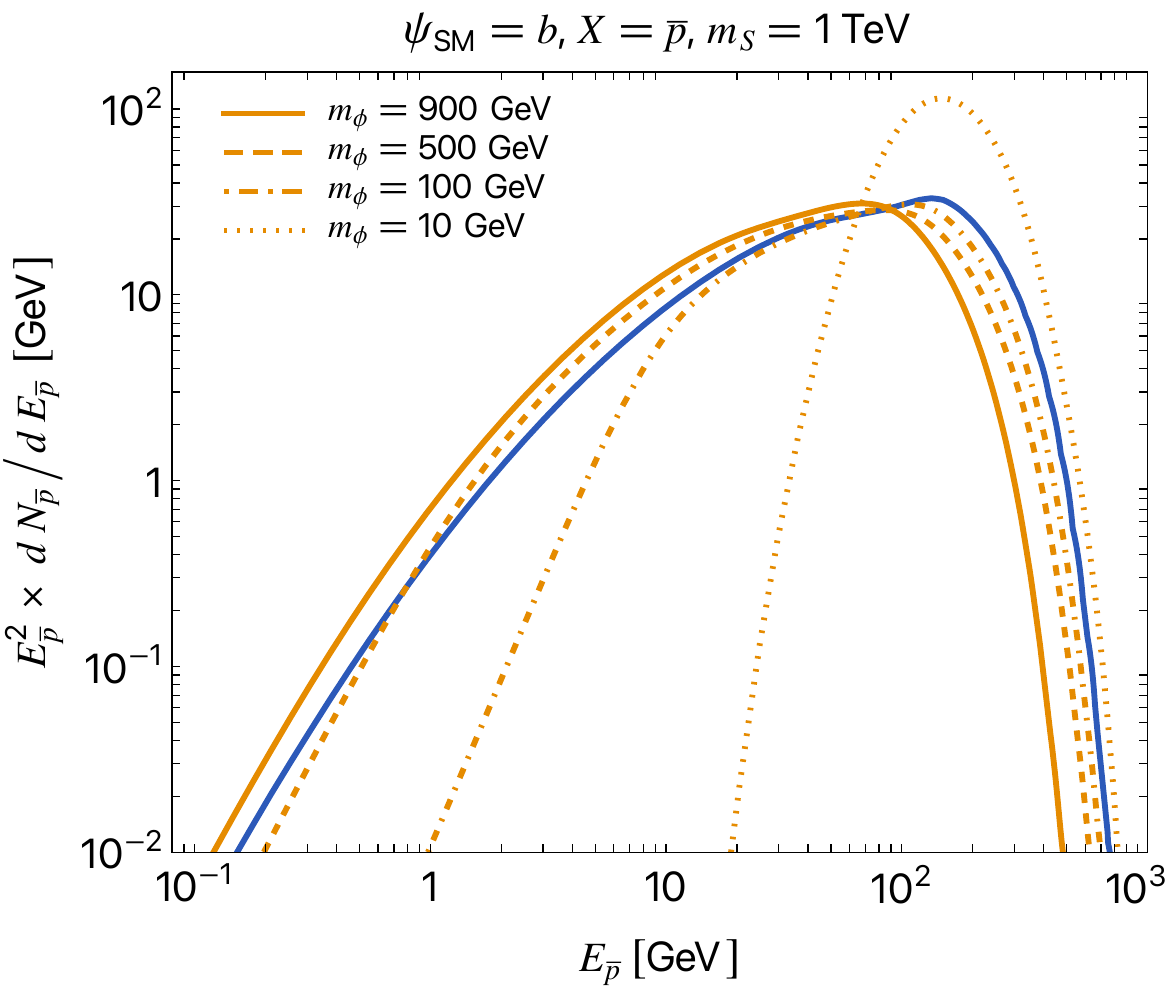}}\quad
    \subfloat[\label{fig:cfr_eps_semi_p}]{\includegraphics[width=.48\linewidth]{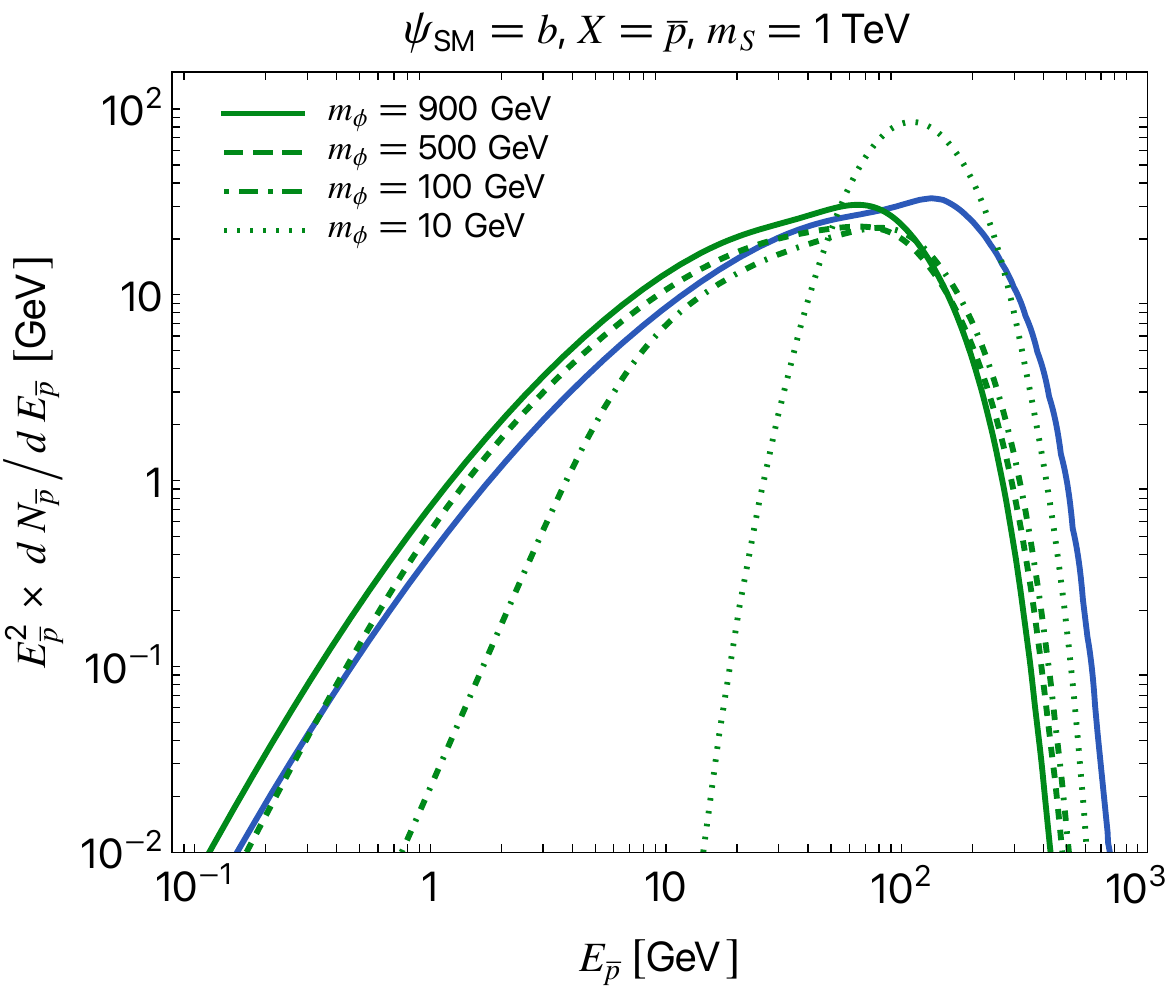}}\\
    \caption{\textit{Single process injection spectra: antiprotons for different mediator-to-DM mass ratios.} Same as \Fig{fig:cfr_massratio_gamma} but for the differential injection spectra of antiprotons.}
    \label{fig:cfr_massratio_p}
\end{figure}

\begin{figure}[tb]
    \centering
    \includegraphics[width=0.9\linewidth]{Figures/legenda_ratios.pdf}\\
    \medskip
    \subfloat[\label{fig:r_gamma_p}]{\includegraphics[width=.48\linewidth]{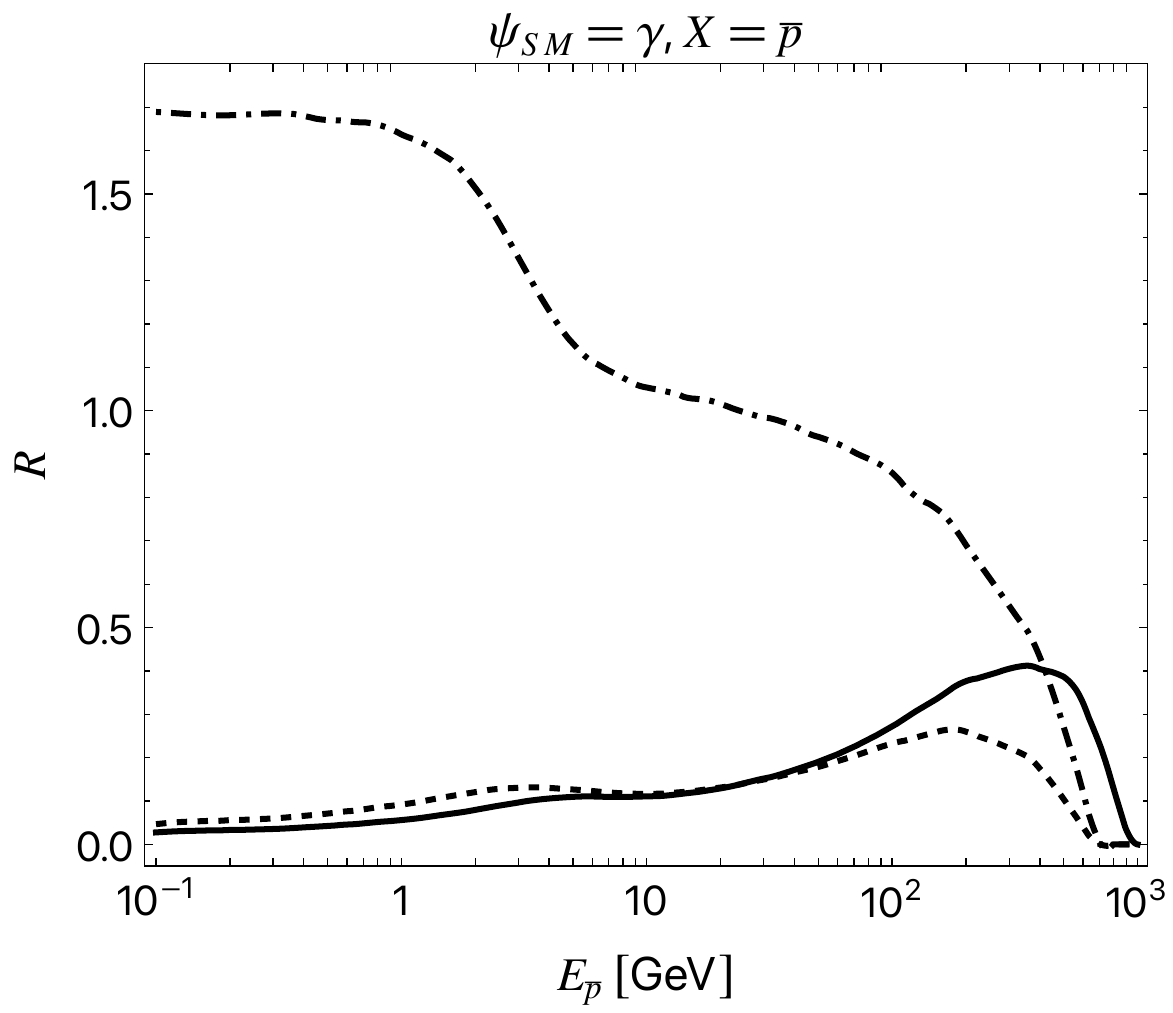}}\quad
    \subfloat[\label{fig:r_b_p}]{\includegraphics[width=.48\linewidth]{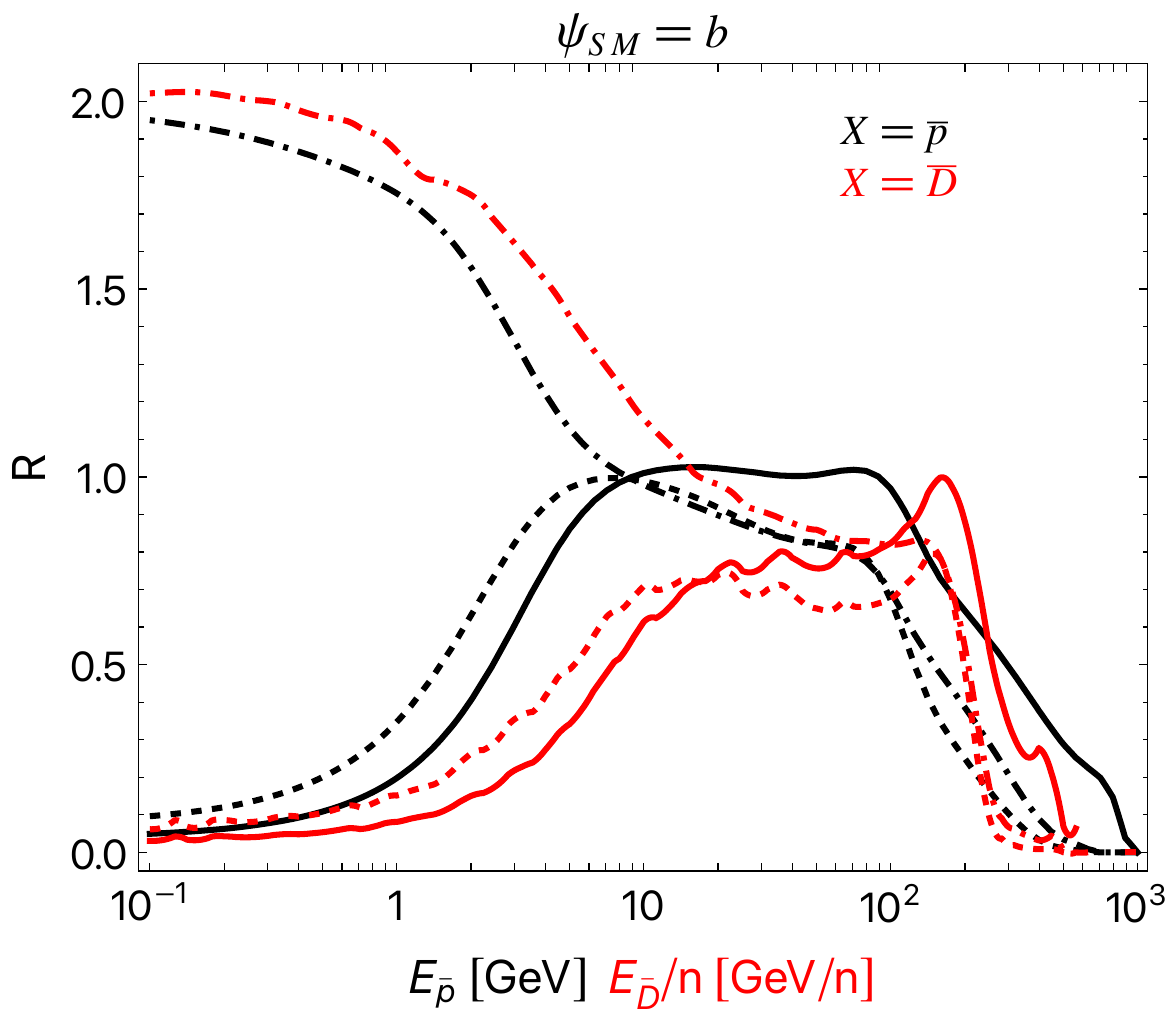}}\\
    \subfloat[\label{fig:r_e_p}]{\includegraphics[width=.48\linewidth]{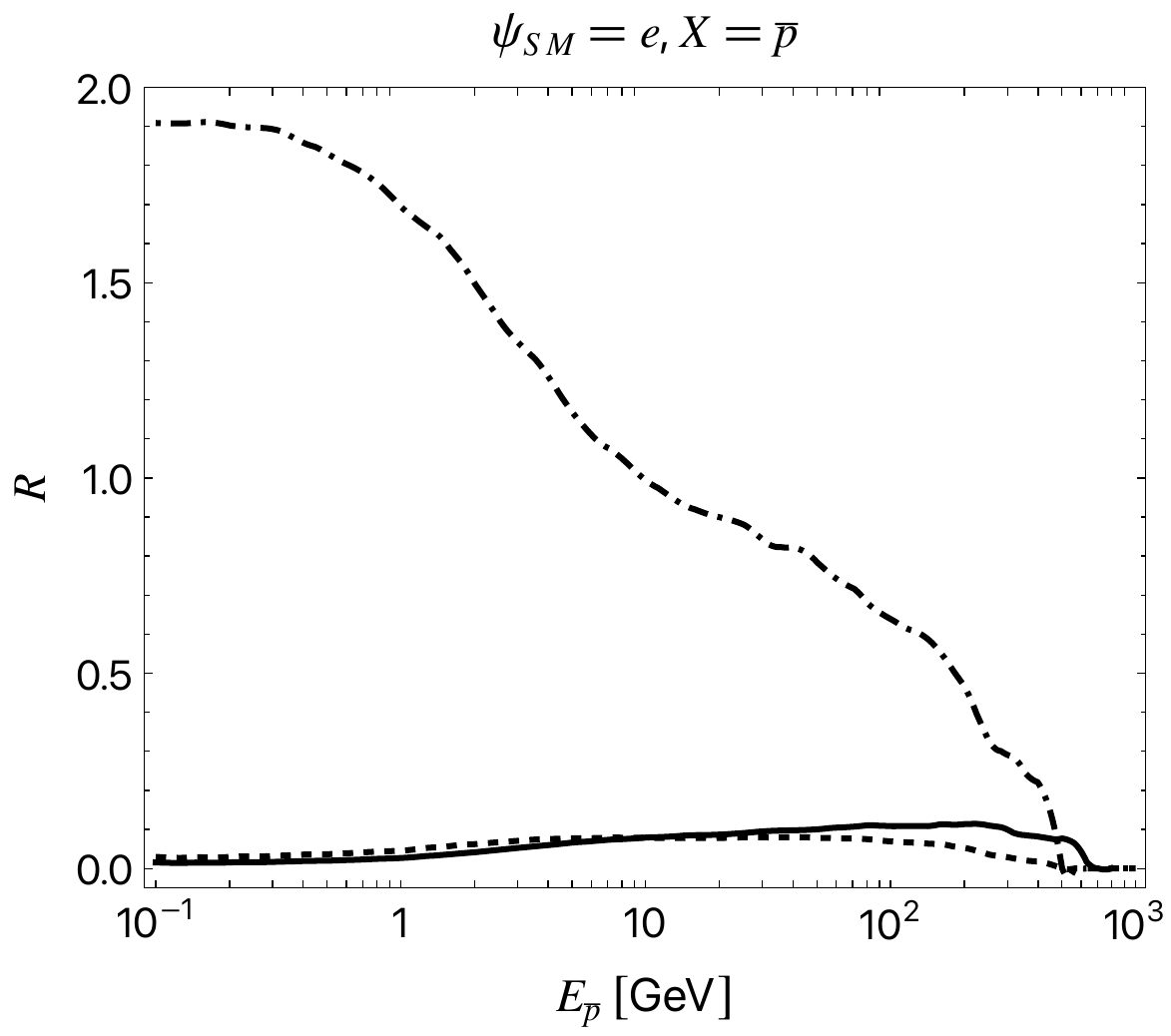}}\quad
    \subfloat[\label{fig:r_tau_p}]{\includegraphics[width=.48\linewidth]{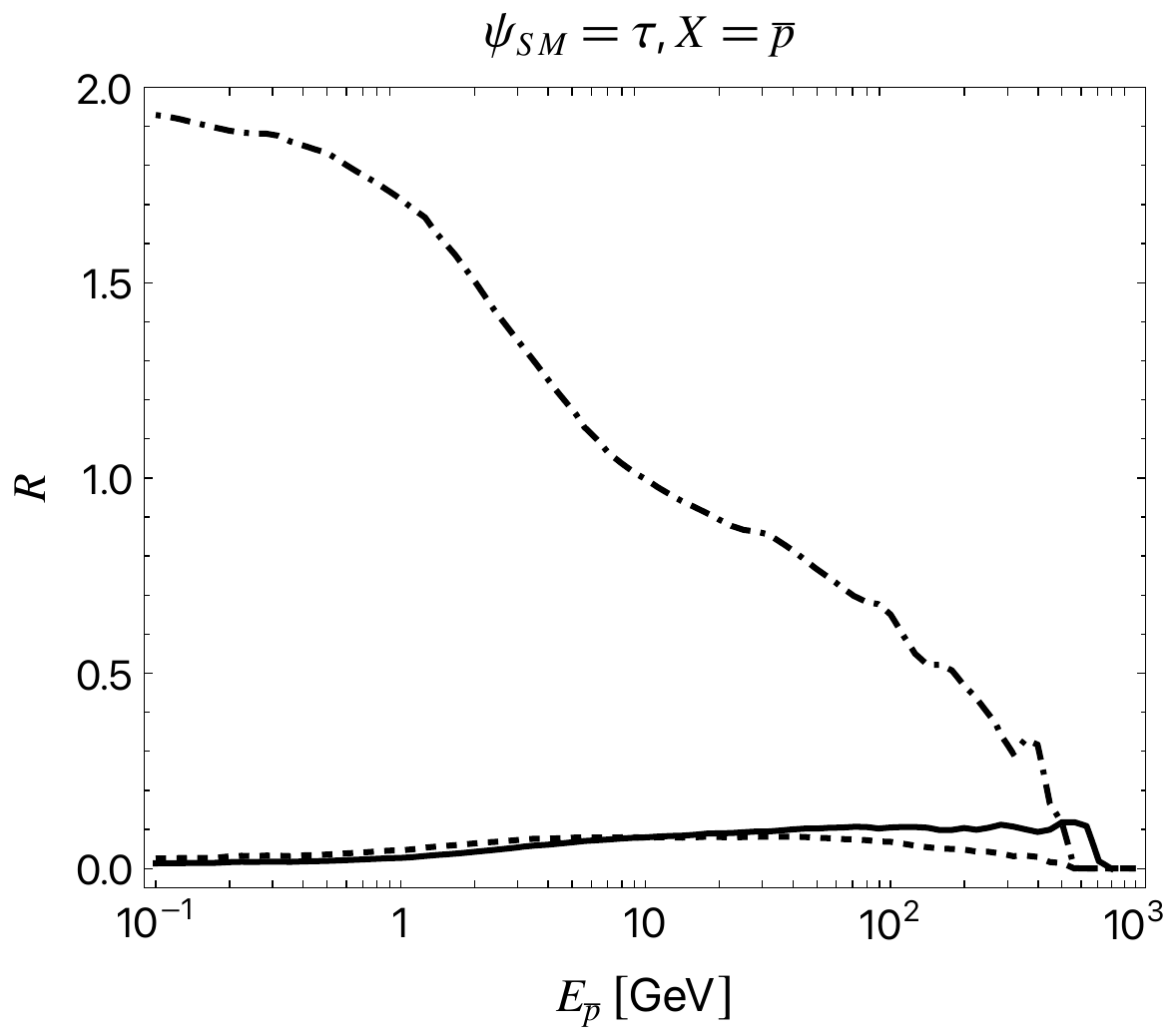}}
\caption{\textit{Ratio of antiproton spectra from different single processes.} Same as \Fig{fig:ratios_gammas} for the differential injection spectra of antiprotons (and antideuterons, panel b) per single DM collision process.}
\label{fig:ratios_antiprotons}
\end{figure}

In a similar spirit to the discussion in \Sec{sec:gammas}, we pick the benchmark $(\psi_{\rm SM}=b,\,X=\bar p)$ as a suitable case to investigate the effect of varying the mediator-to-DM mass ratio. The results are illustrated in 
\Fig{fig:cfr_massratio_p}, where we compare different antiproton injection spectra computed for different mediator masses for fixed DM mass $(\mdm=1\,\TeV)$ and primary channel $(\psi_{\rm SM}=b)$. Heavier mediators imply a slight shift of the injection spectrum to lower energies, compatible with the smaller Lorentz boost imparted to the metastable intermediate state. In this channel we cannot appreciate any significant variation in the normalization of the spectra for different masses as this is mainly ascribed to the DM mass scale. The only modification worth of notice, even though trivial, is that the mediator indeed cannot be too light as the decay $\phi\to\bar bb$ must be kinematically allowed. Closer to the threshold, the primary $b$ quarks generated by the mediator decay are almost at rest and the final CR energy is expected to be dominated by the boost to $\phi$ (up to hadronization effects). For this reason, the curves computed for $m_\phi=10\,\GeV$ show a significantly different shape with respect to all other cases.

\begin{figure}[t]
    \centering
    \includegraphics[width=0.9\linewidth]{Figures/legenda_ab.pdf}\\\medskip
    \subfloat[\label{fig:mix_gamma_p}]{\includegraphics[width=.48\linewidth]{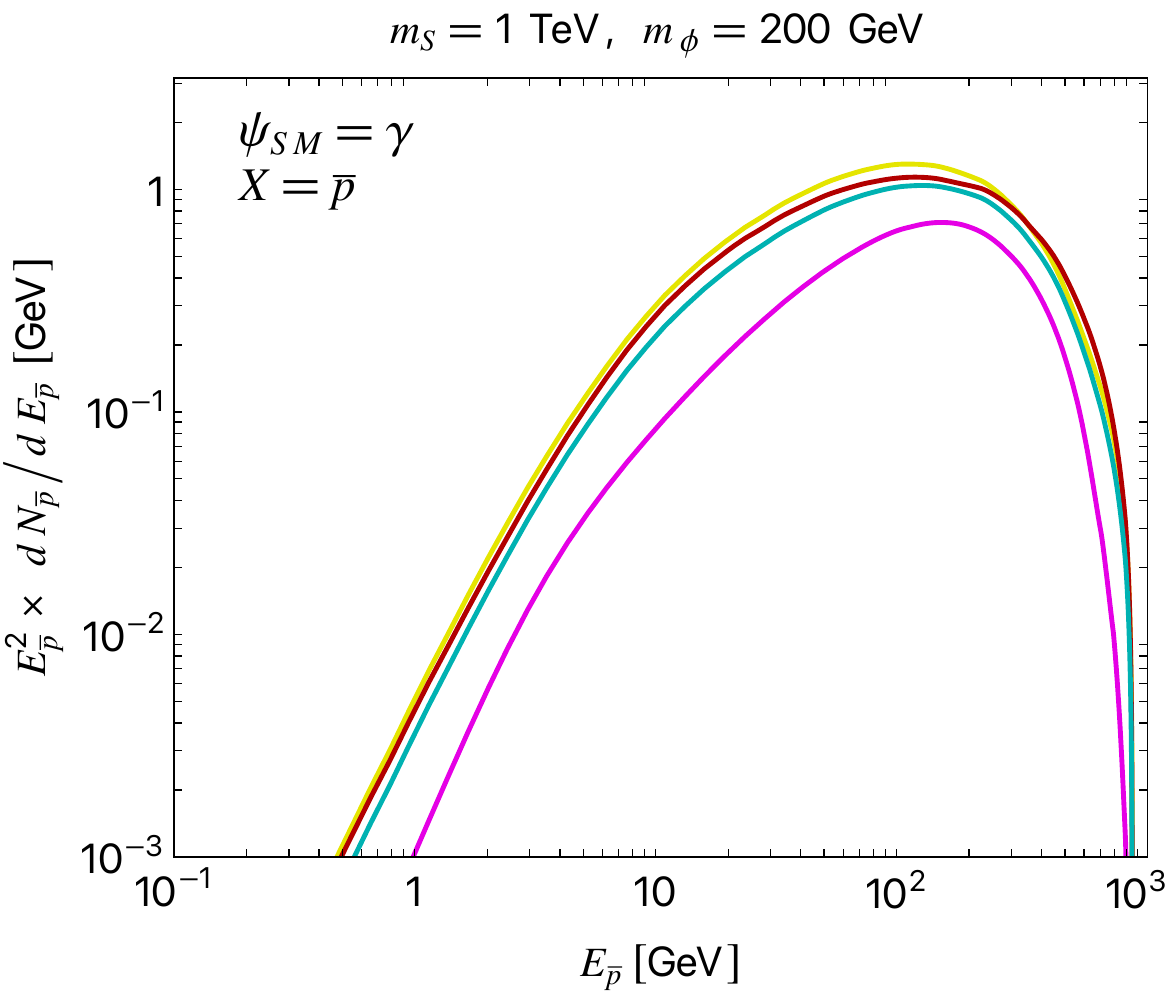}}\quad
    \subfloat[\label{fig:mix_b_p}]{\includegraphics[width=.465\linewidth]{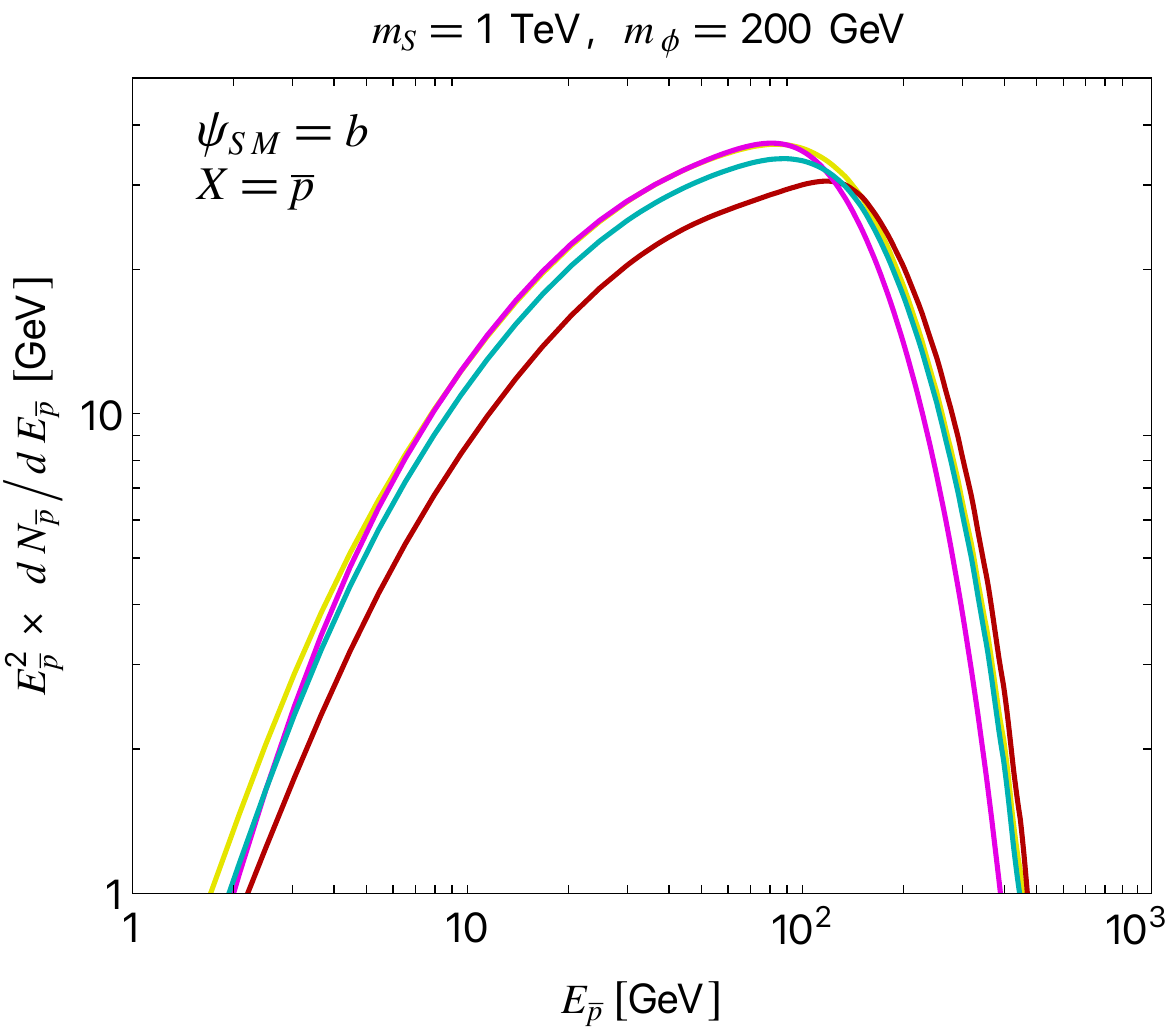}}\\
    \subfloat[\label{fig:mix_e_p}]{\includegraphics[width=.48\linewidth]{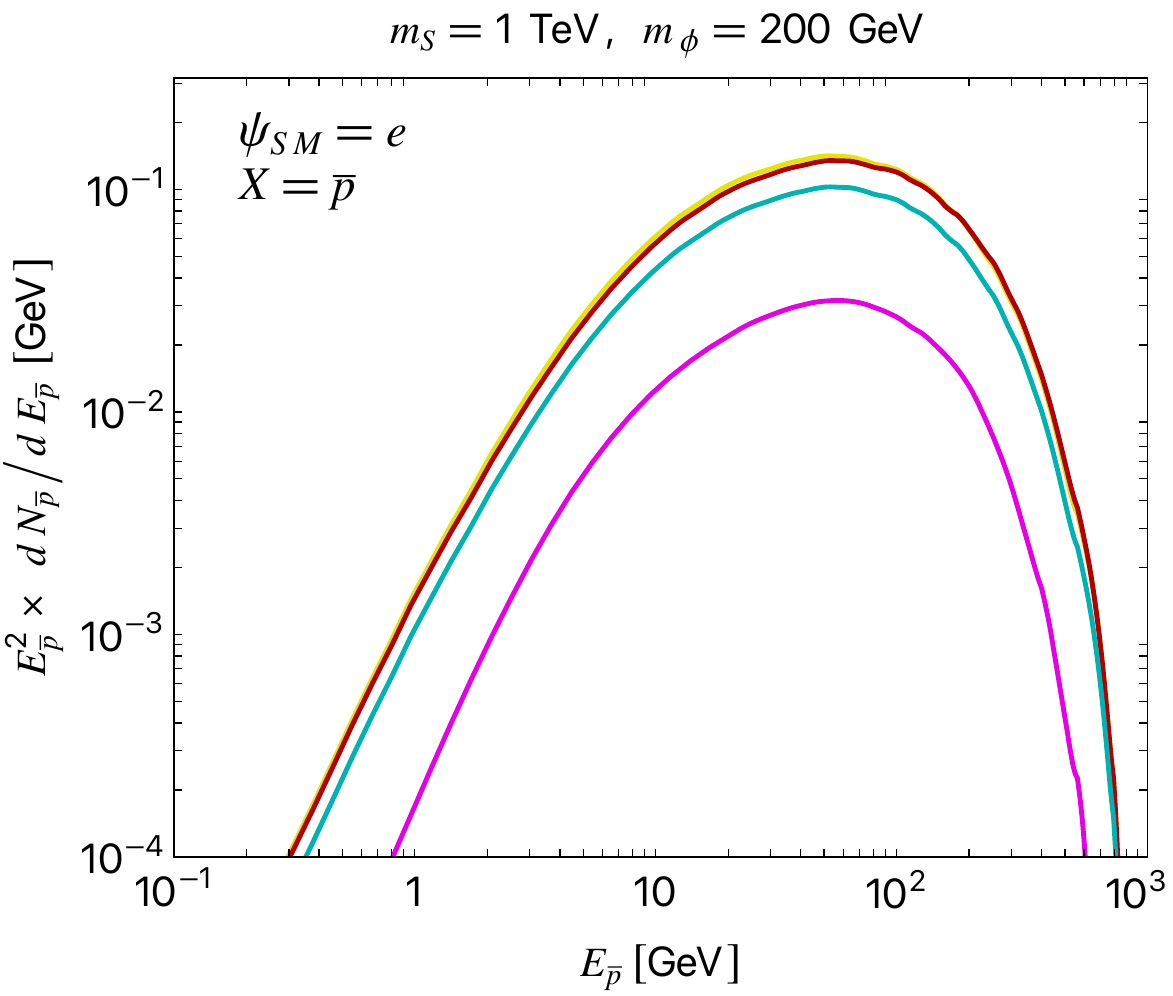}}\quad
    \subfloat[\label{fig:mix_tau_p}]{\includegraphics[width=.48\linewidth]{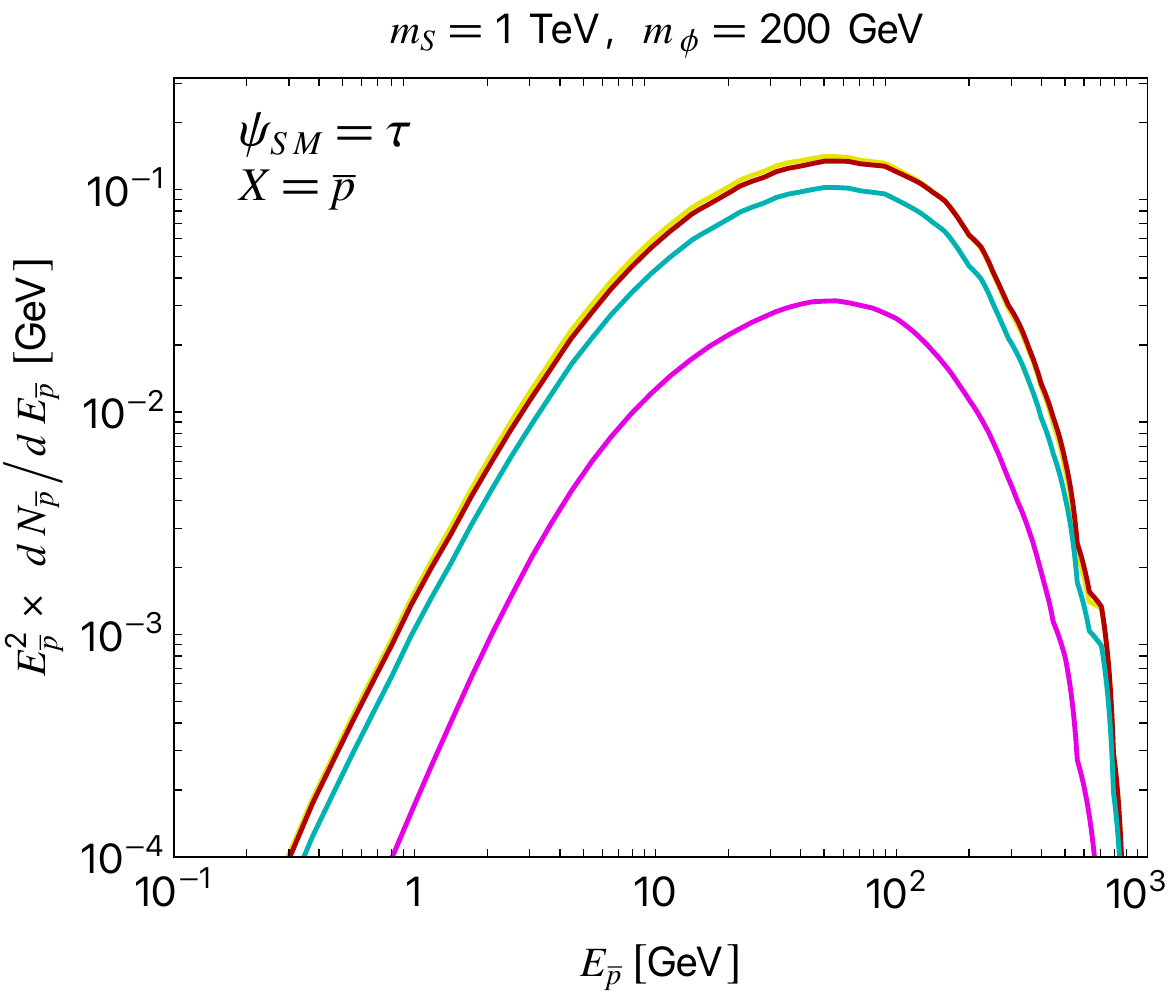}}
\caption{\textit{Injection spectra from a superposition of processes: antiprotons}. Same as \Fig{fig:mixing_gammas} but for the antiprotons per single DM collision process.}
\label{fig:mixing_antiprotons}
\end{figure}

The results in the case of coexistence of two or more collision processes that produce antiprotons are presented in \Fig{fig:mixing_antiprotons}. In this framework, we observe that, as remarked before, if a relevant contribution of DM direct annihilation is present, this is going to shape the major features of the injection spectrum. However, it is still clear that increasing fractional contributions from one-step processes leads to milder and softer antiprotons spectra, with the full one-step (semi-)annihilation case providing the most suppressed signal.

Our framework is easily extended to the case of antinuclei, such as antideuterons. 
Computing the spectra of antinuclei from DM collisions is in principle highly non-trivial, and suffers from large uncertainties on the modeling of the coalescence mechanism, with recent progresses leaning in favor of the Argonne-Wigner scheme \cite{DiMauro:2024kml,DiMauro:2025vxp}.
For the purpose of our work, to compute the zero-step injection spectrum from DM annihilation (or decay) we have relied on the associated dataset of {\tt{CosmiXs}} \cite{DiMauro:2024kml} obtained implementing this coalescence model. 

Concentrating on the antideuteron case, we have computed the differential injection spectra per single DM collision, evaluated the ratio $R$ of the differential spectra and investigated the scenario in which the signal is generated by the superposition of the contribution coming from the different processes under consideration, as done for the other CR messengers. The zero-step injection spectrum of antideuterons is limited to a small number of SM primaries. In particular, only two of the selected benchmark channels, namely $(\psi_{\rm SM}=\gamma,\,b)$ are available. In addition, the production of heavy antinuclei via EW showering is severely suppressed with respect to hadronization evolution. We therefore limited our study to $(\psi_{\rm SM}=b)$. Nevertheless, we observe the same qualitative and quantitative trends as shown for antiprotons. This is clearly summarized in \Fig{fig:r_b_p}, where we overlay the curves for ratios of $\bar D$ injection spectra yielded by different DM collision processes with those for antiprotons. Up to an overall normalization factor and a small shift to higher final CR energies, the antideuterons and antiprotons cases display the same trend in all possible ratios. In short, if a significant contribution from DM direct annihilation exists, it will determine the main characteristics of the injection spectrum. This suggests that current and forthcoming antinuclei searches are expected to be primarily sensitive to standard DM annihilation scenarios for DM masses of tens of GeV (see e.g.~\cite{Serksnyte:2022onw,Heisig:2024jkk}), while modified injection spectra arising from one-step processes may represent subdominant contributions. Assessing the sensitivity to such signatures, including with future facilities such as AMS-100 \cite{Schael:2019lvx}, requires dedicated projections and is left for future work. 

\section{Gamma rays from dwarf spheroidal galaxies}
\label{sec:gamma_flux}
After illustrating the general properties of the injection spectra for different messengers, in this section we provide a specific example of the observable fluxes of $\gamma$ rays for the analyzed scenarios. We concentrate on the $\gamma-$ray flux from DM interactions in dSphs. Since we are interested in $\gamma$ fluxes generated from collision of DM particles with WIMP-scale masses, hence ranging in the GeV-TeV window, these predictions are relevant for CTAO prospects \cite{CTAO:2025gdd} and \textit{Fermi}-LAT data \cite{McDaniel:2023bju}, as their optimal performance are in the $\TeV$ and $\GeV$ windows, respectively.

A fundamental astrophysical input needed for the computation of these fluxes is the integrated $J$ factor. This quantity crucially depends on the modeling of the source, reflected by the choice of both the DM energy density profile distribution, as well as by the value of the angular integration, which is associated to the ``effective'' size of the observed region of interest in the sky.
$J$ factors have been computed, estimated and measured for a number of dwarf galaxies, in particular for Milky Way dwarf spheroidal satellite galaxies, see e.g.~the compilation in \cite{McDaniel:2023bju}. 
While a large number of sources have been characterized, not all candidates are found to be equivalent promising targets in the $\gamma-$ray sky. 
Indeed, in case of observations with specific experiments, it is possible to select few optimal targets. 
For illustrative purpose, in what follows, results are shown selecting a specific, representative source, while we leave the computation of the cumulative signal from a sample of different dSphs, and the comparison to data and/or future observations, to future work. 

In what follows, we motivate the choice of the DraI dSph as a benchmark candidate which suits both CTAO and \textit{Fermi}-LAT observation windows. 
First, the $J$ factor for DraI is ${\rm{Log}}_{10}(J_{\rm Dra})=18.83\pm0.12$, which is among the largest estimated, and suffers from relative small uncertainties due to the modeling of the scaling function used~\cite{Pace:2018tin}.
This value is computed for a Einasto DM density profile, and corresponds to an integration angle $\theta<0.5^\circ$ in the sky, matching the one used for the \textit{Fermi}-LAT combined analysis of Ref.~\cite{McDaniel:2023bju}.  
Following the analysis of Ref.~\cite{CTAO:2025gdd}, it appears that the $J$ factor is  less sensitive to the underlying DM energy density profile with respect to other sources. Specifically (see their figure~(3)), using a cusped (Einasto) or cored (Burkert) $\rho_S(r)$ function yields an integrated $J$ factor compatible within $1\sigma$.
In addition, also the choice of the integration angle is found to not significantly affect the $J$ factor value for this specific dSph, as it shows only a mild dependence on $\theta$ (see figure~(4) of the same Ref.).
Furthermore, its angular extension is mostly concentrated in a small region in the sky, implying that a larger fraction of the signal up to 34.5\% (70.3\%) can be captured within a small observation aperture of $0.5^\circ\, (1^\circ)$ (see figure~(7) and Tab.~(4) of the same Ref.). 
According to the CTAO collaboration analysis in \cite{CTAO:2025gdd}, DraI  is  among the eight optimal targets for $\sim300\,h$ observation time (ranking 7th(4th) for $\theta<0.5^\circ(0.1^\circ)$) after passing a selection based on three criteria: distance, spectroscopic data accuracy and highest $J$ factor. Eventually, together with UMi, this particular dSph was used in Ref.~\cite{CTAO:2025gdd} to derive the strongest forecast for the upper limits on the DM annihilation cross section for different annihilation channels $(b,\tau)$.
Finally, DraI is included in the legacy analysis of \textit{Fermi}-LAT in \cite{McDaniel:2023bju}, and is not among the seven dSph which show a signal excess with respect to the background at a local significance level $\gtrsim2\sigma$ for DM annihilations into either $b$ or $\tau$ channels. 

The total flux of $\gamma$ rays is computed through the general expressions in \Eq{eq:dPhigammagen}, using the value of the astrophysical $J$ factor mentioned above, and implementing the comprehensive injection spectrum of \Eq{eq:phigammagen}. 
For simplicity, only the prompt $\gamma$ rays are considered. Secondary $\gamma$ rays though e.g.~inverse Compton emission are expected to be subdominant, becoming relevant for heavy DM particles and in case of leptonic annihilation channels~\cite{Song:2024vdc}. 
The differential spectra of prompt $\gamma-$ray fluxes are inspected for two different benchmark scenarios: a dark sector that couples uniquely and directly to photons, and a leptophilic dark sector which couples to $\tau$. In both cases, we fix the DM and mediator masses to two different benchmark values, namely we set $\mdm=1\,\TeV,\,m_\phi=200\,\GeV$, to target high energy photons in the preferred observation window of CTAO, whereas we fix $\mdm=10\,\GeV,\,m_\phi=5\,\GeV$ to target \textit{Fermi}-LAT observation window. All the $\gamma-$ray fluxes presented here are normalized to the effective thermally averaged cross section $\Sigma_{\rm eff}$ which correctly reproduces the observed DM abundance.

\begin{figure}[t]
    \centering
    \includegraphics[width=0.5\linewidth]{Figures/legenda_spectra.pdf}\\ \medskip
    \subfloat[\label{fig:cfr_flux_gamma_CTAO}]{\includegraphics[width=0.45\linewidth]{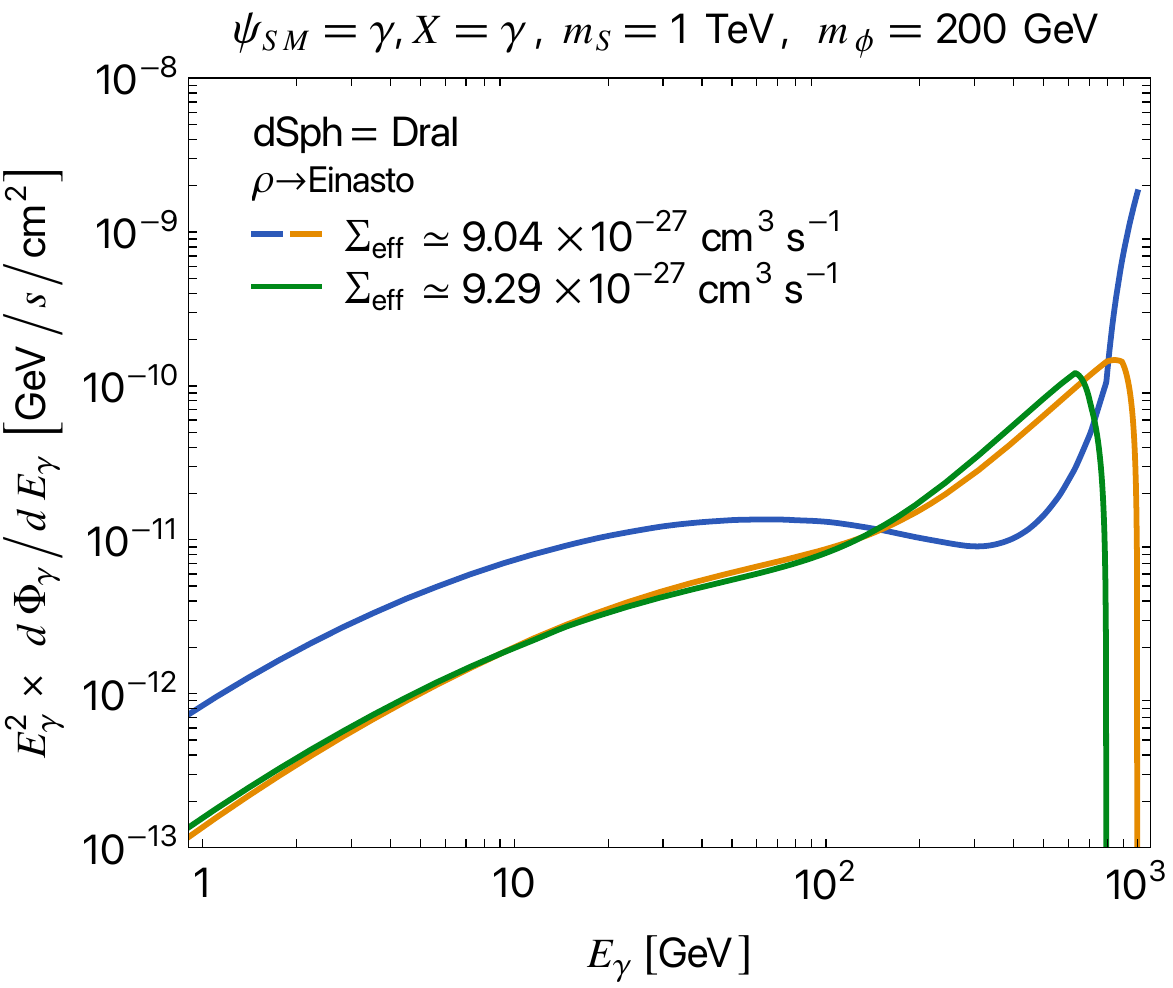}}\quad
    \subfloat[\label{fig:cfr_flux_gamma_FERMI}]{\includegraphics[width=0.45\linewidth]{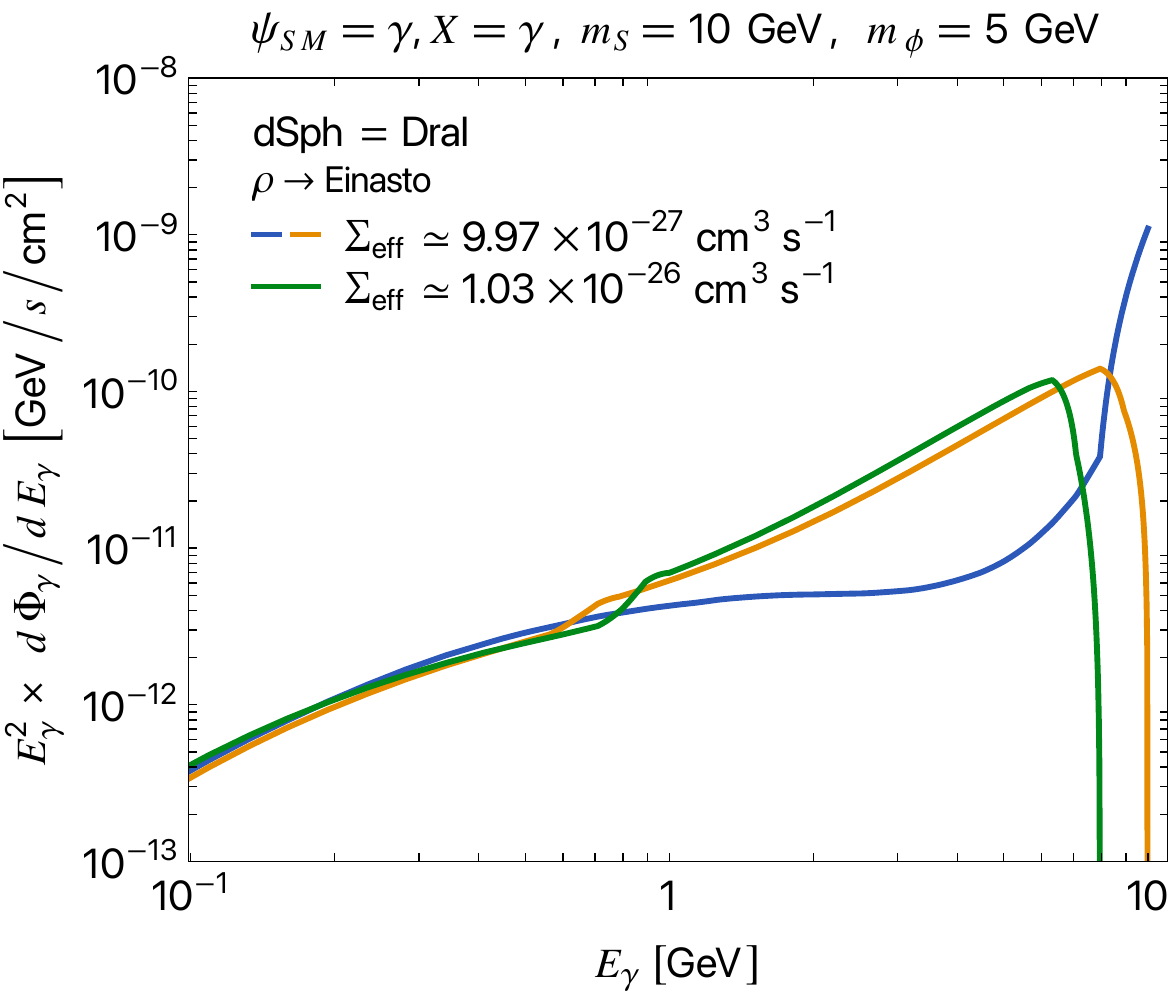}}\\
    \subfloat[\label{fig:cfr_flux_tau_CTAO}]{\includegraphics[width=0.45\linewidth]{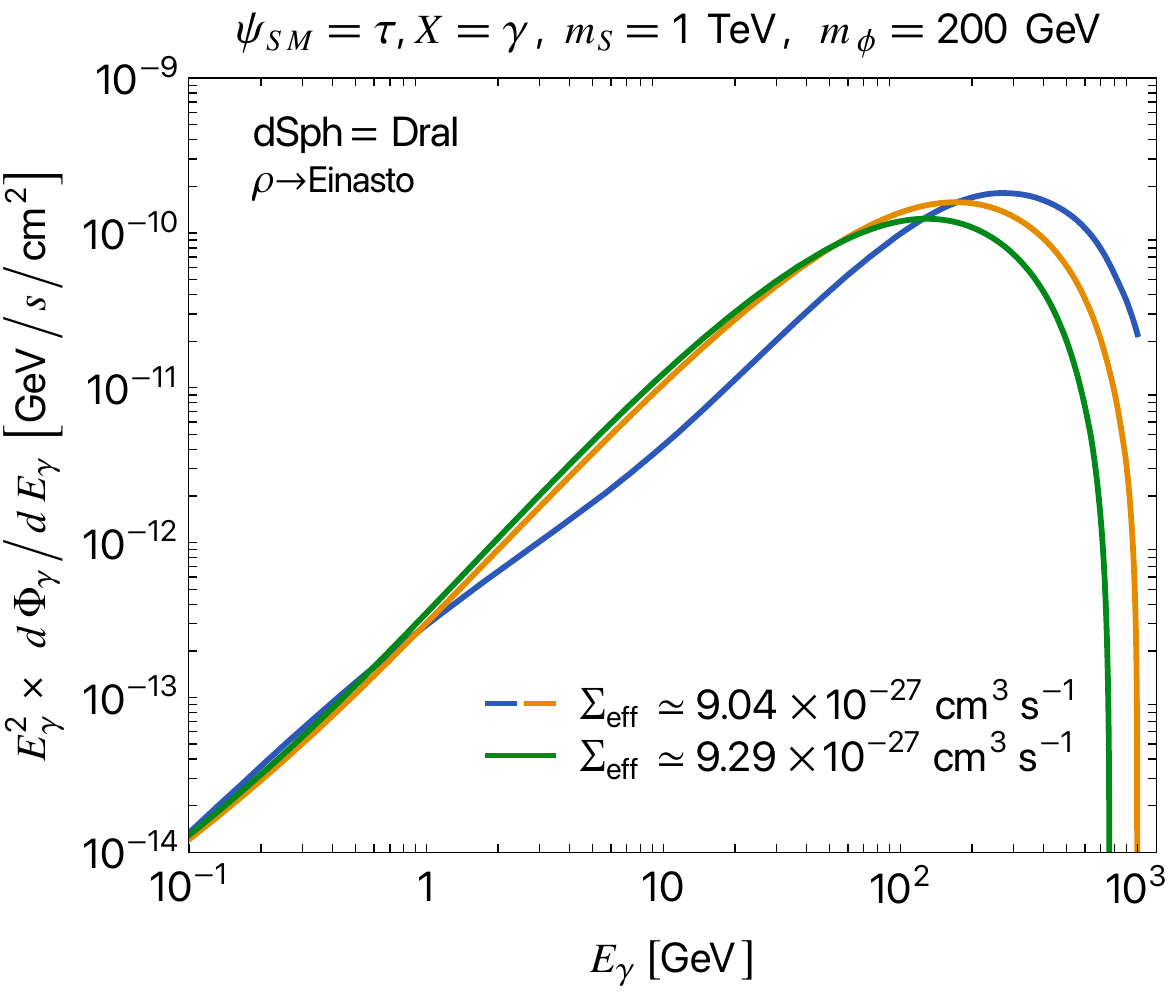}}\quad
    \subfloat[\label{fig:cfr_flux_tau_FERMI}]{\includegraphics[width=0.45\linewidth]{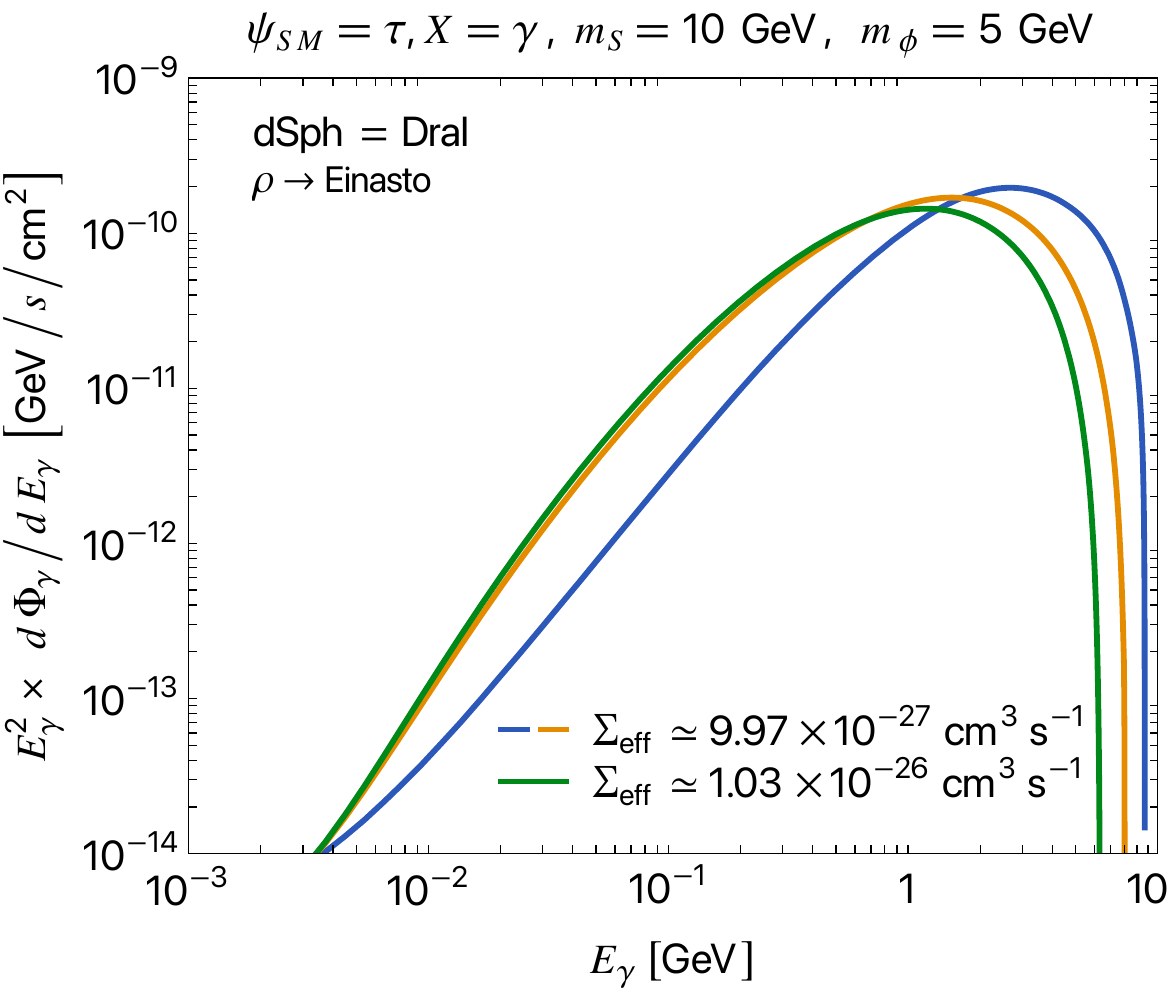}}
    \caption{$\gamma-$ray flux from the dSph DraI generated by DM annihilation (blue lines), one-step annihilation (orange) and one-step semi-annihilation (green). {\it Top} panels show fluxes for photophilic dark sectors, while {\it bottom} panels refer to leptophilic dark sectors coupling to the heavier generation. 
    \textit{Left} panels refer to $\mdm=1\,\TeV,\,m_\phi=200\,\GeV$, while \textit{right} panels to a smaller DM and mediator mass of $\mdm=10\,\GeV,\,m_\phi=5\,\GeV$.
    Each spectrum is normalized for the \textit{effective} thermal cross section $\Sigma_{\rm eff}$ that reproduces the observed DM abundance.}
    \label{fig:cfr_fluxes}
\end{figure}

At first, we only consider the fluxes generated by DM whose phenomenology is completely determined by either one of the three processes under investigation. The results are shown for both photophilic (\Figs{fig:cfr_flux_gamma_CTAO}{fig:cfr_flux_gamma_FERMI}) and leptophilic DM models (\Figs{fig:cfr_flux_tau_CTAO}{fig:cfr_flux_tau_FERMI}). Left (right) panels refer to benchmarks relevant for CTAO (\textit{Fermi}-LAT). As inferred from \Eq{eq:dPhigammagen}, the final CR spectrum of neutral messenger particles is linearly proportional to the product of the $J$ factor times the differential injection spectrum. As a consequence, the spectral features already observed in the $\gamma$-ray injection spectra (shown in \Figs{fig:cfr_gamma_gamma}{fig:cfr_tau_gamma}) are projected onto the physically observable quantity up to a rescaling factor. Indeed, we notice that photophilic dark sectors give rise to box-shaped fluxes for one-step cascade processes, which intensity overcomes that of direct annihilation fluxes throughout the box width, while being suppressed outside. Similarly, we reproduce the behavior observed in the injection spectra of leptophilic DM: direct annihilation fluxes dominate at the high energy end of the spectrum, while they are suppressed towards lower energies. These features are common to both benchmark combinations of DM and mediator masses, i.e. to CTAO and \textit{Fermi}-LAT operational regions. 
More generically, we expect the features observed in \Figs{fig:r_gamma_gamma}{fig:r_tau_gamma} to propagate directly from injection spectra to observable fluxes. 
Detailed simulations and data analysis should be repeated for each specific injection spectrum to assess precise detectability of these peculiar spectral features. Nevertheless, the spectral modifications discussed above, which can induce differences of up to factors of a few between one-step cascade processes and standard direct annihilation, suggest that current and future observations with CTAO and \textit{Fermi}-LAT may provide sensitivity to the collisional origin of DM-induced $\gamma$-ray fluxes. Such effects may also complement the potential capability of future CTAO observations to discriminate between cored and cuspy DM density profiles, whose impact on the resulting CR fluxes is estimated to be of order a few \cite{Hiroshima:2019wvj}. In this respect, the spectral signatures discussed here could provide an additional and non-trivial energy dependence.

\begin{figure}[t]
    \centering
    \includegraphics[width=0.9\linewidth]{Figures/legenda_ab.pdf}\\\medskip
    \subfloat[\label{fig:fluxes_mix_gamma_CTAO}]{\includegraphics[width=0.45\linewidth]{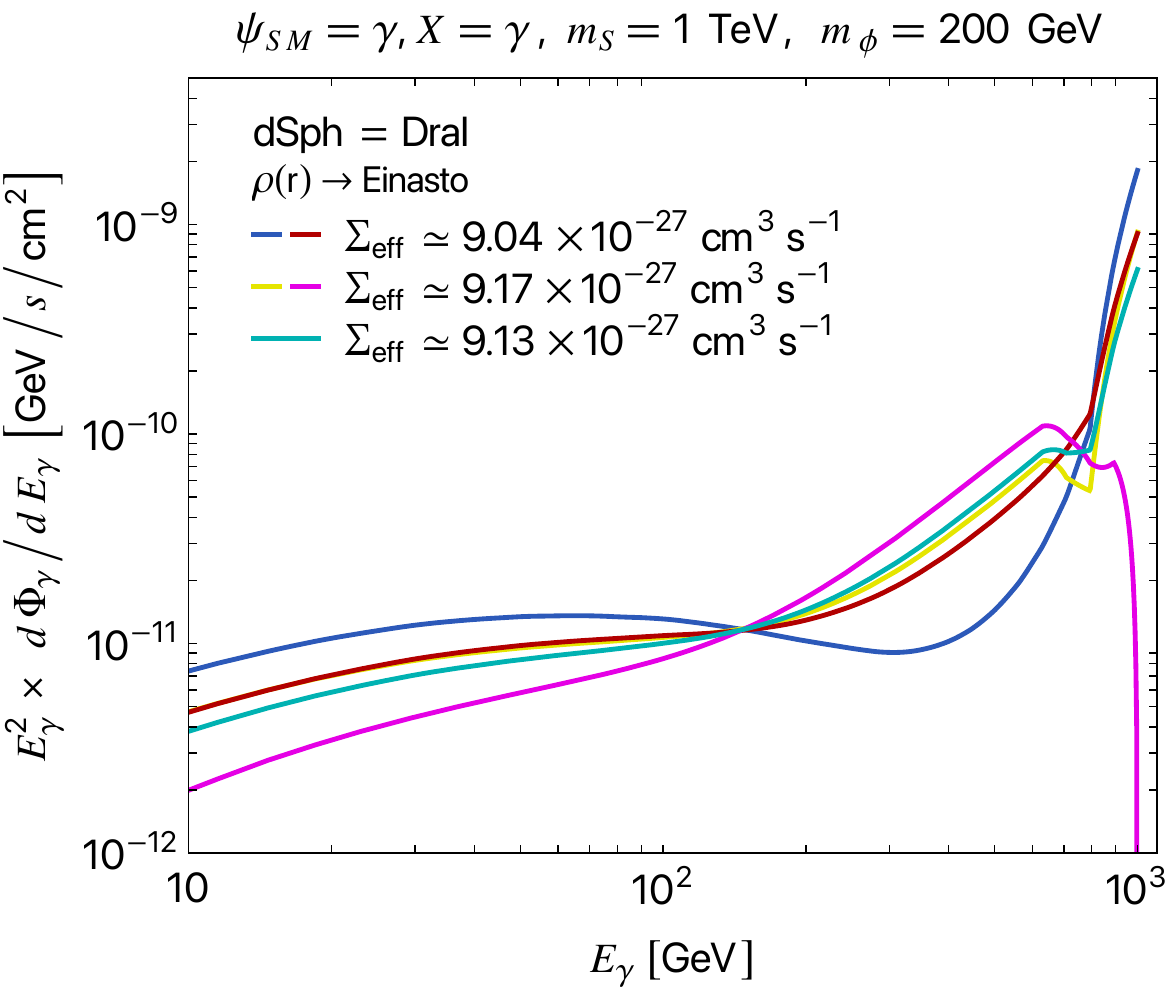}}\quad
    \subfloat[\label{fig:fluxes_mix_gamma_FERMI}]{\includegraphics[width=0.45\linewidth]{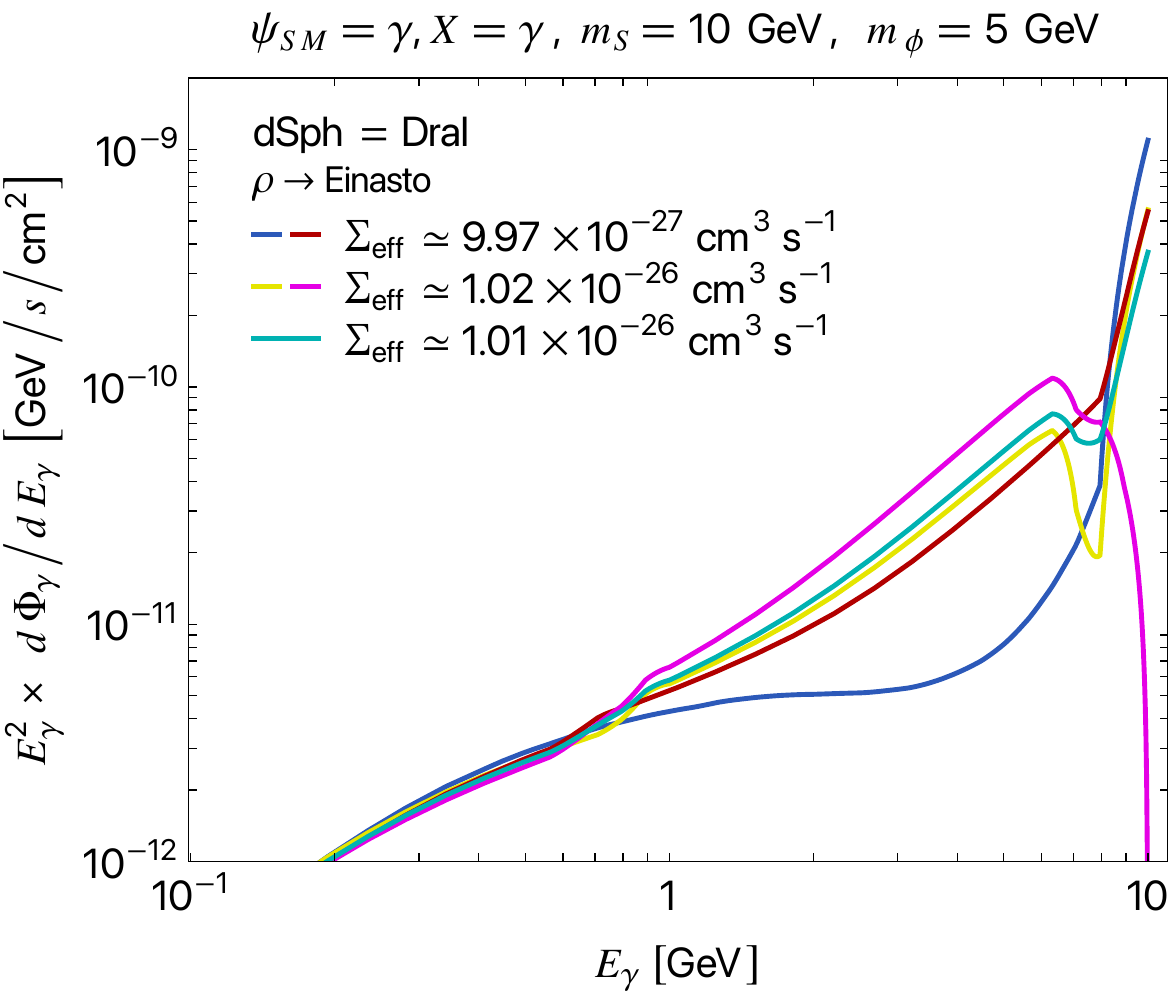}}\\
    \subfloat[\label{fig:fluxes_mix_tau_CTAO}]{\includegraphics[width=0.45\linewidth]{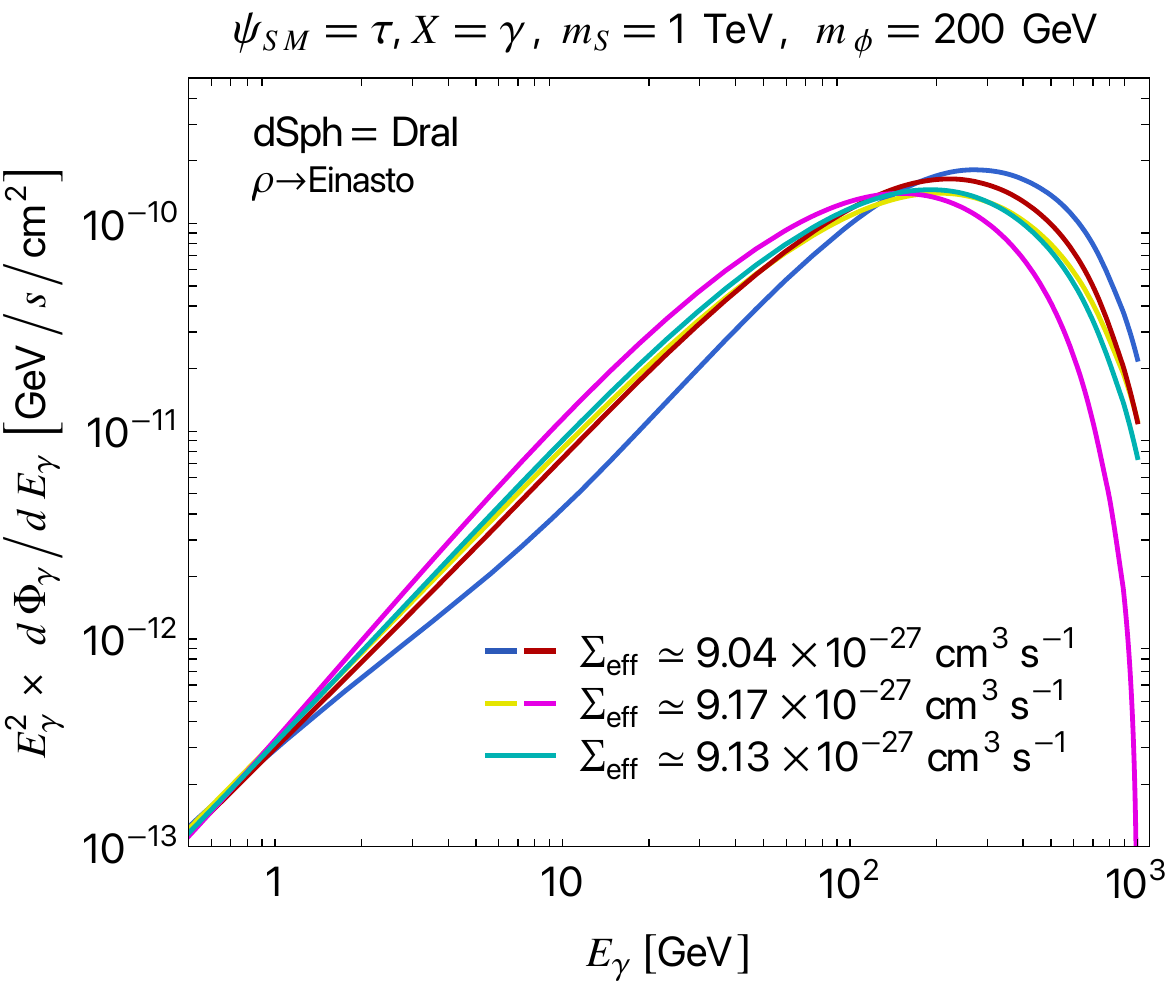}}\quad
    \subfloat[\label{fig:fluxes_mix_tau_FERMI}]{\includegraphics[width=0.45\linewidth]{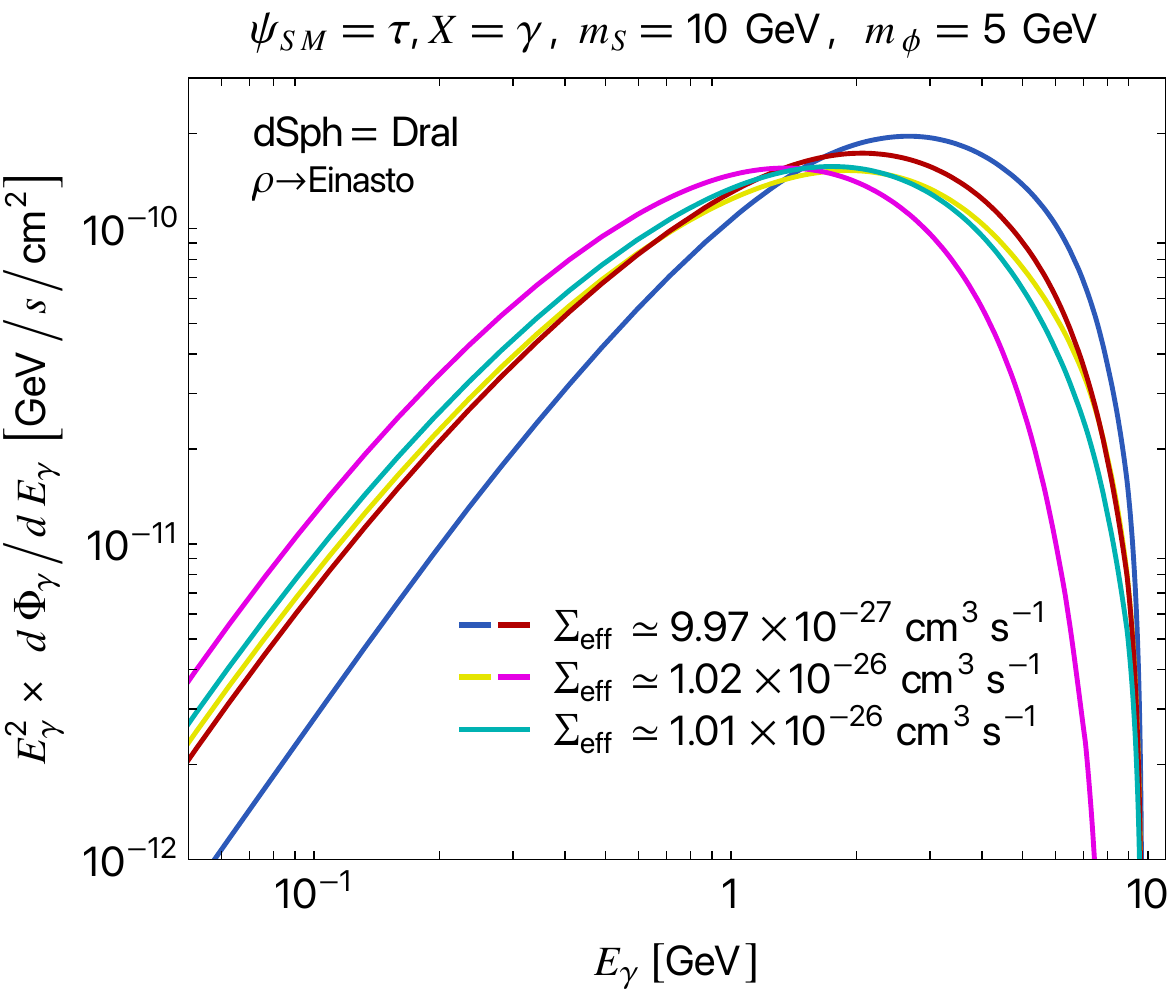}} 
    \caption{Same as \Fig{fig:cfr_fluxes} but for the case in which the signal is shaped by the superposition of the contributions from different collision processes. The standard annihilation is reported in blue for reference, while colored lines correspond to a mix of annihilation, one-step cascade annihilation and semi-annihilation, as defined by the $\alpha,\beta$ parameters (see \Eq{eq:alphabeta}).}
    \label{fig:fluxes_mix}
\end{figure}

We move the discussion to the case of a richer DM phenomenology, which results are shown in \Fig{fig:fluxes_mix}. Here, we allow for the superposition of signals generated by different types of DM scattering processes with different weights. The flux generated by DM direct annihilation is shown for reference as a \textit{blue} line. Also in this case, the $\gamma$-ray flux yielded by each combination of DM binary collisions reflects the spectral features already observed in \Figs{fig:mix_gamma_gamma}{fig:mix_tau_gamma} up to a normalization factor $(\propto J\times\Sigma_{\rm eff})$ for fixed DM mass. Indeed, by varying the relative $\alpha,\beta$ weights for one-step cascade processes, it is possible to shape the spectral functions between the three extremal cases of \Fig{fig:cfr_fluxes}. Even though the size of the effective thermally averaged cross section $\Sigma_{\rm eff}$ only varies within a few \% from one scenario to the other, these features can be clearly distinguished, making evident the differences which arise after combining the injection spectra associated to the three collision types. We thus do not expect a significant improvement in the upper limits on the thermal DM cross section with respect to the standard annihilation scenario. On the other hand, the increased sensitivity and larger statistics of future GeV--TeV $\gamma$-ray observations may eventually allow the distinctive spectral features discussed here to be identified in the presence of a sufficiently significant signal. More generally, the generalized expressions in \Eqs{eq:dPhigammagen}{eq:phigammagen} provide a flexible framework for future reinterpretations of experimental results. By introducing two additional free parameters, $(\alpha,\beta)$, beyond the DM mass $\mdm$ and the overall normalization of the thermal cross section, dedicated analyses could exploit these templates to investigate the nature of DM interactions beyond the standard annihilation paradigm and assess the viability of some of the specific model realizations introduced in the next section.

\section{Models}
\label{sec:models}

The strength of the framework presented so far lies in its fully model-independent nature. Assuming only the presence of multiple $s$-wave annihilation and semi-annihilation processes, we derive the DM relic abundance from thermal freeze-out via \Eq{eq:Omegah2}, in terms of an effective total cross section $\Sigma_{\rm eff}$, while leaving the relative weights of the individual contributions largely unconstrained (see \Fig{fig:relic}). The same set of processes that controls the freeze-out relic density can also be active today in regions of high DM density, giving rise to potentially observable CR fluxes. The differential flux of neutral messengers in \Eq{eq:dPhigammagen} and the source term for charged SM particles in \Eq{eq:chargedX_mixing} are both proportional to the effective total cross section $\Sigma_{\rm eff}$. By contrast, the spectral features are determined by the relative contributions of the individual annihilation and semi-annihilation channels through a generalized injection spectrum. For $\gamma$ rays, this quantity is defined in \Eq{eq:phigammagen}, and the corresponding expression for any other messenger $X$ follows straightforwardly upon replacing $\gamma \to X$.

A natural next step is to identify concrete particle physics constructions that reproduce the effective description outlined above. Constructing explicit Lagrangian realizations serves a twofold purpose. First, it provides a proof of principle that our framework can arise within well-defined UV completions featuring both annihilation and semi-annihilation channels. Second, it allows us to assess how naturally the parameters of the theory can reproduce the relative weights adopted in the previous section to illustrate generalized injection spectra.

In what follows, we present explicit DM theories that realize the effective framework introduced above. Depending on the particle content and interaction structure, these models can generate different combinations of direct and cascade final states, thereby populating distinct regions of the $(\alpha, \beta)$ parameter space defined previously. Rather than treating these scenarios as unrelated cases, it is convenient to interpret them as different limits of a common underlying Lagrangian, discussed in \Sec{sec:modelscalars}, obtained by imposing suitable constraints and exploring specific regions of parameter space. We also comment in \Sec{sec:modelsalternative} on alternative realizations involving different DM candidates and portal interactions. This top-down perspective highlights the robustness of our model-independent approach and provides a coherent interpretation of the results presented throughout this work.

\subsection{An extended scalar sector}
\label{sec:modelscalars}
We consider a minimal extension of the SM with two additional degrees of freedom: a complex scalar $S$, which plays the role of the DM candidate, and a real scalar $\phi$, acting as a metastable mediator. Both $S$ and $\phi$ are singlets under the SM gauge group. In addition to the kinetic terms for the new fields, the relevant part of the Lagrangian is encoded in the renormalizable scalar potential
\be
V(H, S, \phi) = V_{\rm SM}(H) + \Delta V_2(S,\phi,H) + \Delta V_3(S, \phi) \ ,
\label{eq:Vgeneral}
\ee
which we decompose for convenience into three contributions:
\begin{subequations} 
\begin{align} 
\label{eq:V1} 
V_{\rm SM}(H) = & \, - \mu_H^2 |H|^2 + \frac{\lambda_H}{4} |H|^4 \ , \\
\label{eq:V2} 
\Delta V_2(S,\phi,H) = & \, m_S^2 |S|^2 + \frac{1}{2} m_\phi^2 \phi^2 
+ \phi \left(A_S\, |S|^2 + A_H \, |H|^2 + \frac{A_\phi}{3!}\, \phi^2\right) \nonumber \\
& + \frac{\phi^2}{2} \left( \lambda_{\phi H} \, |H|^2 
+ \lambda_{\phi S} \, |S|^2 \right)  + \lambda_{SH}|S|^2|H|^2 + \frac{\lambda_{S}}{4} \,|S|^4 + \frac{\lambda_{\phi}}{4!} \,\phi^4 \ , \\
\label{eq:V3} 
\Delta V_3(S, \phi) = & \, \frac{1}{3!} \left( A_3 + \lambda_3 \phi \right)  (S^3+S^{\dagger 3}) \ .
\end{align}
\label{eq:original_model}
\end{subequations}
The first term corresponds to the SM scalar potential and, as such, depends only on the Higgs doublet field $H$; it therefore requires no further discussion. We now turn to the remaining terms and examine the symmetry structures that can control their presence.

The second contribution, $\Delta V_2(S,\phi,H)$, also involves the scalar singlets $S$ and $\phi$, but does not contain all possible renormalizable operators. We require that the field $S$ appears only through its squared absolute value $|S|^2$ and the quartic self-interaction proportional to $|S|^4$. This condition can be enforced by imposing a global $U(1)_S$ symmetry under which the DM field is charged and transforms as $S \to e^{i\theta} S$. This choice serves a twofold purpose. It ensures the existence of a stable particle, a necessary ingredient for any viable DM candidate, and forbids quadratic terms such as $S^2$ or $S^{\dagger 2}$, which would otherwise induce a mass splitting between the real and imaginary components. While such a splitting would not invalidate our framework\footnote{The lighter mass eigenstate would still constitute a stable DM candidate.}, we choose to avoid this additional complication in our analysis.

The third contribution, $\Delta V_3(S, \phi)$, involves only the new degrees of freedom and is characterized by the presence of cubic DM operators, $S^3$ and $S^{\dagger 3}$, which are responsible for semi-annihilation processes\footnote{One linear combination of $A_3$ and $\lambda_3$ can be made real by redefining the field $S$ through a global $U(1)_S$ rotation, although both parameters can be complex. Since, as discussed below, we focus on a regime where the cross sections are dominated by the cubic terms, we perform a field redefinition such that $A_3$ is real.}. However, such operators are present only if the symmetry responsible for DM stabilization allows them. The global $U(1)_S$ symmetry introduced above clearly forbids these terms, but they can arise if the DM particle is instead stabilized by an Abelian \ZT~symmetry. Notably, the \ZT~symmetry still forbids the unwanted terms $S^2$ and $S^{\dagger 2}$ discussed in the previous paragraph, while remaining compatible with all operators appearing in $\Delta V_2(S,\phi,H)$.

We consider two main scenarios. In the first case, the DM particle is stabilized by a global Abelian $U(1)_S$ symmetry, and the scalar potential therefore contains only the contributions from \Eqs{eq:V1}{eq:V2}. In this setup, semi-annihilation processes are forbidden by the underlying symmetry structure (i.e., $\beta = 0$), and only DM annihilations are present, including both direct channels and one-step cascades producing two metastable mediators. In the second case, the DM particle is stabilized by an Abelian \ZT~symmetry, and the scalar potential includes the additional operators appearing in \Eq{eq:V3}. In this scenario, semi-annihilations are allowed (i.e.~$\beta \neq 0$) and can play an important role in both freeze-out and ID phenomenology.

The purpose of this section is to provide a proof of existence of particle physics frameworks realizing our scenarios, rather than a detailed exploration of the full parameter space. We therefore focus on the region where both the DM and mediator masses are much larger than the electroweak symmetry-breaking (EWSB) scale, with the requirement that $m_S > m_\phi$ in order to kinematically allow one-step cascades. In other words, we consider the region of parameter space characterized by the mass hierarchy $m_S \gtrsim m_\phi \gg m_Z$. This choice renders the discussion as transparent as possible and allows us to derive simple analytical expressions.

We assume that quartic couplings, while necessary to ensure the stability of the  EWSB vacuum and the boundedness of the scalar potential, do not contribute significantly to the interaction rates. The smallness of the $S$-$H$ mixing, i.e.\ $\lambda_{SH} \ll 1$, is required by stringent bounds on Higgs-portal interactions of DM candidates, particularly from direct detection searches \cite{Arcadi:2019lka,Arcadi:2024ukq}. Similarly, we focus on regions of parameter space where the quartic couplings $\lambda_{\phi H}$ and $\lambda_{\phi S}$ are small. Dark sector self-interactions proportional to $\lambda_S$ and $\lambda_\phi$ do not affect our analysis, and the quartic interaction proportional to $\lambda_3$ is assumed to give a negligible contribution to semi-annihilation amplitudes. In this regime, the cubic couplings $\{A_S, A_H, A_3\}$ dominate the transition rates, and the dark sector couples to the visible sector primarily through the mediator $\phi$. Given the mass hierarchy introduced above, the metastable mediator dominantly decays into pairs of Higgs bosons and longitudinal weak gauge bosons, $W^\pm$ and $Z$. Therefore, the processes responsible for the production of primary SM particles are DM annihilations and one-step cascade reactions ending into these states. This ensures that one-step DM (semi-)annihilations can act as an exotic source of CR fluxes via $\phi$ decays. 

We report below the corresponding thermally averaged cross sections, valid in the region of parameter space defined above and in the $s$-wave limit
\begin{subequations}\label{eq:X_sections}
\begin{align}
\braket{\sigma_{SS^\star\to HH^\star}\vmol} \simeq & \, \frac{A_S^2\,A_H^2}{16\pi\,m_S^2\,(4 m_S^2 - m_\phi^2)^2}\ , \\
\braket{\sigma_{SS^\star\to \phi\phi}\vmol} \simeq & \, \frac{A_S^4}{16\pi\, m_S^2\,(2 m_S^2 - m_\phi^2)^2} \sqrt{1 - \frac{m_\phi^2}{m_S^2}}\ ,\\
\braket{\sigma_{SS\to S^\star\phi}\vmol} \simeq & \, \frac{27 \, A_S^2\,A_3^2}{256 \pi\,m_S^2 (3m_S^2-m_\phi^2)^2} \left(1 + \frac{1}{9} \frac{m^2_\phi}{m^2_S} \right)^2 \sqrt{1 - \frac{10}{9} \frac{m^2_\phi}{m^2_S} + \frac{1}{9} \frac{m^4_\phi}{m^4_S}} \ .
\end{align}
\end{subequations}
Notably, all processes predicted by the scalar potential in \Eq{eq:Vgeneral}, with explicit contributions given in \Eq{eq:original_model}, possess a non-vanishing $s$-wave component and therefore fully satisfy the assumptions underlying our analysis.

Given this general setup, we now organize the remainder of this section to show how all phenomenologically relevant scenarios can be realized within this minimal model. By imposing suitable hierarchies among the couplings or modifying the underlying symmetry structure, we identify the four benchmark scenarios illustrated in figures~\ref{fig:mixing_gammas}, \ref{fig:mixing_neutrinos}, \ref{fig:mixing_positrons}, \ref{fig:mixing_antiprotons}, and \ref{fig:fluxes_mix}, corresponding to the cases discussed in the previous sections. Each benchmark is associated with a specific choice of the effective parameters $(\alpha,\,\beta)$.  For each scenario, we derive analytical expressions that can be directly mapped onto the notation introduced in \Sec{sec:rates}. Imposing the relic density constraint $\Sigma_{\rm eff}\sim10^{-26}\,\cms$ fixes the overall scale of the cubic couplings for a given DM mass, while the dimensionless ratios among the couplings, encoded in $(\alpha,\,\beta)$, determine the CR spectral shape. In what follows, we provide explicit expressions valid in the limit $m_\phi \ll m_S$, so that the DM mass becomes the only relevant scale in the problem.

\begin{itemize}

\item \textbf{Global $U(1)_S$.} We begin by considering the phenomenological scenario described by the first two terms of the scalar potential in \Eq{eq:original_model}. As discussed above, this setup, together with the absence of the $\Delta V_3(S,\phi)$ contribution containing cubic DM couplings, can be realized by stabilizing the DM particle via a $U(1)_S$ symmetry. As a consequence, DM semi-annihilations are forbidden, and their contribution to both freeze-out and ID signals vanishes (i.e., $\beta = 0$). Both early- and late-time phenomenology are therefore governed by direct and one-step annihilations. The freeze-out condition, corresponding to fixing the value of $\Sigma_{\rm eff}$, reads
\be
\left. \left( \frac{1.2 \, {\rm TeV}}{m_S} \right)^2  \left( \frac{A_S}{m_S} \right)^2  \left[\left( \frac{A_H}{m_S} \right)^2 + 4 \left( \frac{A_S}{m_S} \right)^2 \right] \right|^{\rm relic}_{U(1)_S} \simeq 1 \ .
\ee
The weights of the two processes determine the shape of the resulting CR injection spectra, with the relative contribution of one-step annihilations given by
\be
\left.\alpha\right|_{U(1)_S} = \frac{4 A_S^2}{4 A_S^2 + A_H^2} \ .
\ee
We focus on the regime in which the two processes contribute comparably to the generation of CR fluxes (i.e., $\alpha = 0.5$). Imposing this condition leads to $A_H \simeq 2 A_S$, indicating that no large hierarchy between the two couplings is required.

\item \textbf{Cascades.} A first variation of the global $U(1)_S$ scenario discussed above arises when the stabilizing symmetry acting on the DM field is replaced by an Abelian \ZT~discrete symmetry. From this point on, all scenarios considered will adopt this symmetry to stabilize the DM field, with the full scalar potential of \Eq{eq:original_model} including interactions involving three DM fields. As a result, semi-annihilation processes are allowed. We focus here on the regime in which direct annihilations are suppressed. Inspecting the cross sections in \Eq{eq:X_sections}, this corresponds to the condition $A_H \ll \{A_S, A_3\}$. In this case, the phenomenology is dominated by DM annihilations into pairs of mediators and by semi-annihilations producing a single light mediator in the final state. The freeze-out condition in this scenario then leads to the relation
\be
\left. \left( \frac{1.2 \, {\rm TeV}}{m_S} \right)^2  \left( \frac{A_S}{m_S} \right)^2  \left[3 \left( \frac{A_3}{m_S} \right)^2 + 4 \left( \frac{A_S}{m_S} \right)^2 \right] \right|^{\rm relic}_{\rm Cascades} \simeq 1 \ .
\ee
Present-day CR fluxes targeted by ID searches are therefore sourced exclusively by one-step cascade processes. By construction, the relative contributions of the two processes satisfy $\alpha + \beta = 1$, so that specifying one fraction is sufficient. We choose to express the semi-annihilation weight, as it directly enters the freeze-out computation
\be
\left.\beta\right|_{\rm Cascades} = \frac{3 A_3^2}{3 A_3^2 + 4 A_S^2} \ .
\ee
We identify the region of parameter space in which the two processes contribute comparably to DM phenomenology (i.e., $\alpha = \beta = 0.5$), which implies $A_3 \simeq 1.2\, A_S$ and indicates that, again, no large hierarchy between the couplings is required. 

\item \textbf{Direct+Semi.} We now consider the last remaining option among those in which two out of the three classes of processes contribute to the CR flux. This corresponds to the case where one-step annihilations are absent, which requires $A_S \ll \{A_H, A_3\}$. The phenomenology of this scenario is therefore governed solely by DM direct annihilations into SM states and by semi-annihilations with a lighter metastable mediator in the final state. In this setup, the freeze-out condition reads
\be
\left. \left( \frac{1.2 \, {\rm TeV}}{m_S} \right)^2  \left( \frac{A_S}{m_S} \right)^2  
\left[\left( \frac{A_H}{m_S} \right)^2 + 3 \left( \frac{A_3}{m_S} \right)^2 \right] \right|^{\rm relic}_{\rm Direct+Semi} \simeq 1 \ .
\ee
We note that pair production of mediators does not appear in the freeze-out condition, as in this scenario we have $\alpha \ll 1$. The only relevant parameter is therefore the relative contribution of semi-annihilations
\be
\left. \beta \right|_{\rm Direct+Semi} = \frac{3 A_3^2}{3 A_3^2 + A_H^2} \ .
\ee
However, the scaling of the cross sections in \Eq{eq:X_sections} highlights a subtlety: all thermally averaged cross sections are proportional to powers of the coupling $A_S$, making it challenging to suppress a single contribution independently. Nevertheless, it is possible to compensate for the smallness of $A_S$ (required to suppress $\phi$ pair production) by appropriately adjusting the other trilinear couplings $A_3$ and $A_H$. In particular, this requires enhancing the portal coupling to the SM Higgs. Imposing the condition $\beta = 0.5$ then leads to the relation $A_H \simeq 1.7 \, A_3$, showing once again that only an $\mathcal{O}(1)$ hierarchy between the couplings is needed to realize this configuration.

\item \textbf{Anarchic.} Finally, we consider the case in which no restriction on DM interactions is imposed. In this setup, the scalar potential in \Eq{eq:original_model} predicts a scenario in which DM direct annihilations into SM states, one-step annihilations, and semi-annihilations with light mediators in the final state are all present. This corresponds to the most general phenomenological picture, which forms the backbone of this work. We refer to this framework as the \textit{anarchic} scenario, since the relative contributions of the three processes are not fixed by the relic-density constraint on the effective cross section $\Sigma_{\rm eff}$, as long as their sum matches the thermal value through the condition
\be
\left. \left( \frac{1.2 \, {\rm TeV}}{m_S} \right)^2  \left( \frac{A_S}{m_S} \right)^2  
\left[\left( \frac{A_H}{m_S} \right)^2 + 3 \left( \frac{A_3}{m_S} \right)^2 + 4 \left( \frac{A_S}{m_S} \right)^2 \right] \right|^{\rm relic}_{\rm Anarchic} \simeq 1 \ .
\ee
Starting from the thermally averaged cross sections in \Eq{eq:X_sections}, we compute the branching ratios of the $SS^\star\to \phi\phi$ and $SS\to S^\star\phi$ processes, which correspond to the coefficients $\alpha$ and $\beta$, respectively (see \Eq{eq:alphabeta}). This leads to a generalization of the previous benchmark scenarios:
\begin{subequations}
    \begin{align}
      \left. \alpha \right|_{\rm Anarchic}  & =\frac{4 A_S^2}{3 A_3^2+A_H^2+4 A_S^2} \ ,\\
      \left. \beta \right|_{\rm Anarchic} & = \frac{3 A_3^2}{3 A_3^2+A_H^2+4 A_S^2}\ .
    \end{align}
\end{subequations}
We can now identify the conditions under which the three processes contribute with comparable weights. Since all cross sections scale with powers of the coupling $A_S$, it is convenient to factor it out and express the remaining couplings in relative terms. Imposing $\alpha = \beta = 1/3$ leads to the relations $\{A_H, A_3\} =\{2,\, 2 / \sqrt{3}\} A_S$. As expected, the three couplings must be of the same order of magnitude, and in practice they must lie within a factor of $\mathcal{O}(1)$ of the $S$-$\phi$ coupling, which sets the overall scale.

\end{itemize}

\subsection{Other microscopic realizations}
\label{sec:modelsalternative}

The explicit construction described in \Eq{eq:original_model} provides a convenient and well-motivated microscopic framework capable of accommodating all the benchmark scenarios discussed in the main body of this paper. The realization of each benchmark is, of course, far from unique. Here, we present additional examples that lead to similar phenomenology, thereby extending the landscape of models encompassed by our fully model-independent approach.

Following the logic of the previous discussion, we begin by presenting alternative realizations without semi-annihilations. In this context, a fermionic DM candidate stabilized by a \ztwo~symmetry and interacting through a massive dark photon portal to the visible sector provides an exemplary model~\cite{Hambye:2008bq}. In this framework, semi-annihilations are forbidden both by symmetry considerations and by angular momentum conservation, due to the spin structure of the dark-sector particles. As a consequence, the model naturally lies in the $\beta = 0$ plane. A non-vanishing renormalizable kinetic mixing between the dark and SM photons ensures that the mediator is unstable and guarantees the existence of one-step cascade annihilation processes that can source additional ID signals. For sufficiently heavy DM and dark photon states with sizable couplings, the relic abundance can be set via the thermal freeze-out mechanism~\cite{Hambye:2019dwd}. The thermally averaged cross sections for DM annihilation into SM particles and into dark photon pairs are both $s$-wave in the non-relativistic limit. The kinetic mixing parameter controls the annihilation into SM states, while dark photon pair production is determined solely by the dark gauge coupling and the DM charge. It is therefore possible to exploit the different scaling of the two thermal averages and identify regions of parameter space in which the model realizes a phenomenological scenario with $\alpha = 0.5$.

We now turn to scenarios in which all CR fluxes are produced via cascade processes. In the previous section, this framework was obtained by assuming a suppressed cubic coupling $A_H$. However, it is also possible to construct models in which standard DM annihilations into SM pairs are naturally absent. In particular, Refs.~\cite{DEramo:2025xef,Guo:2023kqt} present realizations in which such channels are forbidden by symmetry and/or kinematic constraints (e.g.~on-shell conditions). In these scenarios, freeze-out is entirely driven by (semi-)annihilations into mediator states, and the resulting ID phenomenology is correspondingly determined by one-step cascade processes. Ref.~\cite{DEramo:2025xef}, in particular, provides an appealing example in which cascade processes are the sole source of exotic CR fluxes. In this setup, the SM is extended by an axion-like particle (ALP) $\phi$, which mediates interactions with a complex scalar field $S$. As the only field charged under a dark \ZT~symmetry, $S$ is stabilized and constitutes the DM candidate. This structure ensures the presence of cubic DM self-interactions associated with semi-annihilations. Interactions between the mediator and the visible sector establish a common thermal bath, but do not affect the DM freeze-out dynamics; instead, they determine the SM final states $\psi_{\rm SM}$ produced in the cascade decays relevant for ID signals. Notably, DM annihilation into pairs of SM particles via $\phi$ exchange vanishes identically once the external DM legs are taken on-shell. As a consequence, the model lies on the $\alpha + \beta = 1$ slice of parameter space (see \Eq{eq:alphabeta}). The thermally averaged cross sections for both annihilation channels are $s$-wave dominated, and one can identify regions of parameter space where $\alpha \simeq \beta$, ensuring that both processes contribute comparably to the relic abundance and to the resulting CR spectra. Furthermore, this framework allows also to naturally realize the limit $\beta = 1$. Indeed, Ref.~\cite{DEramo:2025xef} neglects ALP pair production and retains only DM semi-annihilations. This choice is motivated by the fact that, when considering effective operators up to dimension 5, the cross section for one-step annihilations is suppressed by an additional power of the ALP decay constant compared to semi-annihilations. However, this process also receives contributions from dimension-6 operators and its suppression is not guaranteed a priori. It was shown that in specific UV completions these additional contributions can lead to an exact cancellation. 

The scenario featuring direct annihilations together with one-step semi-annihilations is particularly challenging to realize in a natural way. Higgs-portal interactions between the dark and visible sectors provide a viable framework for such a realization (with our model belonging to this class of theories). By suitably adjusting the couplings and/or introducing mass hierarchies, it is in principle possible to suppress the contribution from annihilations into pairs of metastable mediator states. This setup can effectively lead to a two-component DM scenario with a stable interaction carrier~\cite{DiazSaez:2021pmg}. In this case, DM conversion processes contribute to the freeze-out dynamics but become irrelevant at later times for ID searches. More elaborate constructions involving scalar multiplets, also relying on Higgs-portal interactions, can realize scenarios featuring both annihilation and semi-annihilation channels~\cite{Beauchesne:2024vbo}. However, if the SM Higgs boson appears in the final state of the semi-annihilation process, these models typically exhibit a strong dependence on Higgs and Higgs-portal physics, which can be regarded as a potential drawback.


\section{Conclusions}
\label{sec:conclusions}

The experimental program aimed at testing the thermal freeze-out paradigm is rather mature, with decades of dedicated efforts pursuing multiple complementary strategies. A key feature of this production mechanism is that the predicted DM relic abundance depends on particle masses and couplings that are, at least in principle, accessible to experiments. In particular, the relic abundance scales inversely with the annihilation cross section, establishing a direct connection with early-universe dynamics. Requiring that a given DM candidate is not overproduced (i.e., $\Omega_S h^2 \lesssim 0.12$ in \Eq{eq:Omegah2}) implies a \textit{minimum interaction strength} for a viable cosmological history. This lower bound places the WIMP paradigm under increasing tension in light of the latest experimental results.

In this work, we have explored scenarios in which DM particles are fully or partially \textit{secluded} from the visible sector. If thermalization in the early universe is primarily driven by DM annihilations into non-SM mediator states, constraints from direct detection and collider searches become largely decoupled from those governing the relic abundance. The situation differs for ID, as mediators are typically unstable and eventually decay into visible states. In this case, the \textit{signal rate} is set by the cross section into mediators, while the mediator couplings to SM particles determine the \textit{signal shape}, a separation that can help alleviate tensions with current experimental bounds. The general setup is captured by the three diagrams illustrated in \Fig{fig:diagrams}. Focusing on the region of parameter space where cascade processes are kinematically allowed, we investigate scenarios where linear combinations of these processes contribute to both early-universe dynamics and present-day ID signals. Our approach, described in detail in \Sec{sec:rates}, is largely model-independent and relies only on the assumption that all three classes of processes possess non-vanishing $s-$wave contributions.

To investigate the ID signals sourced by these processes, we derived the CR differential injection spectra per unit energy for a single DM collision, using the benchmark $\mdm=1\,\TeV,\,m_\phi=200\,\GeV$. The approach is general and extends to lighter or heavier DM as well as to arbitrary mediator-to-DM mass ratios, with potentially observable signatures from X-rays to high-energy CRs. By considering representative choices for the relative weights $\alpha, \beta$ defined in \Eq{eq:alphabeta}, we characterized how the injection spectra for both neutral and charged messengers are shaped. Overall, including one-step cascade processes leads to sizable spectral changes relative to the current and near-future detector sensitivities across all cosmic messengers. The main findings regarding the expected signatures for the CR species investigated, namely $\gamma$ rays, neutrinos, positrons, antiprotons and antinuclei, are summarized below.
\vspace{-0.1cm}
\begin{itemize}
\item \textbf{$\gamma$ rays}. One-step cascade (semi-)annihilations introduce non-trivial spectral modifications for all SM primary states investigated. The most significant ones are found for the $\psi_{\rm SM} = \gamma$ case, with suppression and enhancement of $O(5)$ which are not only overall rescaling factors, but introduce distinctive features throughout the inspected energy range. We have shown how the trends observed for the $\gamma$-ray injection spectra are projected to observable fluxes for an illustrative example, the $\gamma$-ray flux from the DraI dwarf galaxy and two benchmark combinations of DM and mediator masses relevant for GeV and TeV searches with \textit{Fermi}-LAT and CTAO, respectively. \vspace{-0.1cm}
\item \textbf{Neutrinos}. For $\psi_{\rm SM} = (\gamma,\,e)$, the ratio of spectra from one-step to direct annihilation is suppressed across all energies, reaching values as low as $\ 0.1$. When all processes are combined, the most notable case is $\psi_{\rm SM} = e$, which mirrors the features of neutrino spectra from one-step cascades and shows strong suppression along with significant spectral distortions compared to scenarios with only direct DM annihilation. \vspace{-0.1cm}
\item \textbf{Positrons}. The most interesting features are observed for a mediator decaying into $\psi_{\rm SM} =  (e,\,\tau)$. In the former case, the one-step (semi-)annihilation spectra are observed to be the most peculiar one, producing up to a factor (50)40 larger injection spectra around hundreds of GeV compared to the standard DM self-annihilation scenario. 
When mixing all processes, one-step spectral features
can be enhanced when these have comparable cross sections to the annihilation one, producing pronounced spectral features. \vspace{-0.1cm}
\item \textbf{Antiprotons and antinuclei}. Generically, they present similar characteristics, and thus only antiprotons results have been shown. For one-step (semi-)annihilation, the spectra are found to be suppressed by up to an order of magnitude over a wide energy range, depending on the SM final state. 
Specifically, for $\psi_{\rm SM} = b$, which yields a higher antiproton energy output, the ratio $R$ between one-step (semi-)annihilation and direct annihilation spans from $\sim 0.1$ up to $\mathcal{O}(1)$ over the explored energy range. This suggests that AMS-02 antiproton data may be sensitive to these scenarios, given that standard DM annihilation channels have been shown to probe thermal DM up to the TeV scale. When multiple processes contribute, any sizable direct annihilation component is found to dominate the main features of the resulting signal.
\end{itemize}

Although most of this work is devoted to a model-independent analysis, we also discussed concrete particle physics realizations of the scenarios under consideration. A minimal extension of the SM in terms of new degrees of freedom is sufficient to achieve this goal: it is enough to introduce two scalar fields, a complex one $S$ and a real one $\phi$, accounting for the DM particle and the mediator, respectively. This setup is presented in \Sec{sec:modelscalars}. Focusing on regions of parameter space in which DM interactions are dominated by cubic terms in the scalar potential, we showed that all processes in \Fig{fig:diagrams} are present and feature non-vanishing $s-$wave contributions. We also identified an unambiguous connection between the nature of the DM-stabilizing symmetry and the resulting ID phenomenology. We briefly discussed alternative realizations in \Sec{sec:modelsalternative}.

Our results set the stage for several future developments, a few of which we outline here. From a phenomenological perspective, a natural next step is to implement our general injection spectra as inputs for global analyses of ID data, with the goal of extracting constraints on the relative contributions of the different annihilation channels. Because these modifications differ significantly from the standard annihilation scenarios studied in previous work, existing results cannot be recast easily. Dedicated analyses ---covering searches, exclusion limits, and projections --- are therefore required to properly assess the sensitivity to these processes for each messenger, which we leave for future work. In this respect, the framework developed here is primarily intended as a source of spectral templates, upon which future ID studies and dedicated comparisons with observational data can be built. We note that our results are produced taking into account the full dependence of injection spectra on both intermediate and final states masses. This represents an improvement with respect to previous works, and allows to test all separations of scales between the different particles masses involved within
the kinematic region, without the need to specify our results to be valid only for the large hierarchy regime. Additionally, the general framework presented here can be extended to describe dark sectors with additional layers of complexity, such as multi-step semi-annihilation cascades and multiple mediator states.  From a more top-down viewpoint, the illustrative Lagrangian involving only two new scalar fields deserves a more thorough investigation. While this construction has the advantage of minimality, it represents only one possible realization; several alternative setups with a limited number of additional degrees of freedom can reproduce the framework introduced in this work and merit dedicated study.


\acknowledgments
The authors thank Tracy R. Slatyer for useful discussion. F.D. and T.S. are supported by Istituto Nazionale di Fisica Nucleare (INFN) through the Theoretical Astroparticle Physics (TAsP) project, and in part by the Italian MUR Departments of Excellence grant 2023-2027 ``Quantum Frontiers''. The work of T.S. is supported in part by the Italian Ministry of University and Research (MUR) through the PRIN 2022 project n. 20228WHTYC (CUP:I53C24002320006 and C53C24000760006). 
S. M. acknowledges the support of the French Agence Nationale de la Recherche (ANR) under the grant ANR-24-CPJ1-0121-01.

\appendix

\section{Thermal averages}
\label{app:theaver}

We provide a general expression for the thermal average of a binary collision, $\chi_a \, \chi_b \rightarrow \chi_c \, \chi_d$, where each particle $\chi_i$ carries four-momentum $P_i^\mu = (E_i, \vec{p}_i)$ with $i = \{a,b,c,d\}$. We denote the magnitude of the spatial momentum as $p_i \equiv |\vec{p}_i|$, and energy and momentum are related through the dispersion relation $E_i^2 = p_i^2 + m_i^2$, where $m_i$ is the mass of the particle $\chi_i$.

The relevant quantity governing interaction rates is not simply the scattering cross section, but rather its product with the so-called M{\o}ller velocity, defined as
\begin{equation}
\vmol \equiv 
\frac{\sqrt{\left(p_a \cdot p_b\right)^2 - m_a^2 m_b^2}}{E_a E_b} \, .
\label{eq:Mvel}
\end{equation}
This quantity characterizes the effective rate at which the two particles approach each other along their line of impact, neglecting transverse motion. In the center-of-mass frame, it reduces to the conventional relative speed between the incoming particles $\chi_a$ and $\chi_b$. The formal definition for the thermal average of this product is
\be
\algn{
\langle \sigma_{\chi_a \, \chi_b \rightarrow \chi_c \, \chi_d}  \vmol \rangle \equiv & \, \frac{\mathcal{S}_{ab} \mathcal{S}_{cd}}{n_a^{\rm eq} n_b^{\rm eq}}  
\int g_a\frac{d^3 p_a}{2 E_a (2 \pi)^3} g_b\frac{d^3 p_b}{2 E_b (2 \pi)^3} f^{\rm eq}_a(p_a) f^{\rm eq}_b(p_b)\; \obar{|\mathcal{M}_{\chi_a \, \chi_b \rightarrow \chi_c \, \chi_d}|^2}  \, \times \\  & \, \qquad \qquad  
  (2\pi)^4 \, \delta^{(4)}(P_\text{fin} - P_\text{init})  \, g_c\frac{d^3 p_c}{2 E_c (2 \pi)^3} g_d\frac{d^3 p_d}{2 E_d (2 \pi)^3} \ ,}
  \label{eq:thavdef}
\ee
Before discussing the integration over the phase space, we notice two overall factors in front of the integral. The statistical factor $\mathcal{S}_{ij}$ is defined as follows 
\be
\mathcal{S}_{ij} \equiv \left( 1 - \frac{1}{2} \delta_{ij} \right) \ ,
\label{eq:Sabcd}
\ee
and accounts for identical particles in the initial and/or final states. The overall normalization of the thermal average contains also equilibrium number densities $n_i^{\rm eq}$ in the denominator. Quantum degeneracy effects can be neglected in the early universe since $f_i^{\rm eq}(E_i) \ll 1$. Consistently, we treat all the particles involved in the processes under investigation with Maxwell-Boltzmann statistics, according to which $f_i^{\rm eq}(E_i) = \exp[ - E_i / T]$, where $T$ is the temperature of the primordial thermal bath. The zero-th moment of the distribution function leads to the expression for the equilibrium number density
\be
n_i^{\rm eq}(T) = g_i \int \frac{d^3 p_i}{(2 \pi)^3} f_i^{\rm eq}(E_i) = g_i \, \frac{m_i^2\,T}{2\pi^2} \,K_2[m_i/T] \ ,
\label{eq:nSeq}
\ee
where $K_2$ is a modified Bessel function. Finally, we notice that the integral over the phase space sums over all possible final states and averages over all incoming initial states with a weight determined by the phase space distributions. We denote by $g_i$ the number of internal degrees of freedom (e.g.~spin, colors) of the particle $\chi_i$. Furthermore, the above expression is valid under the convention according to which the squared matrix element is averaged over both initial and final states. When only the $s$-wave contribution to the process is retained, as done throughout this work, the integration over the phase space reduces to a straightforward evaluation.

\section{Kinematics of one-step cascades}
\label{app:kinematics}

In this appendix, we derive general expressions for the injection spectra arising from one-step cascades. In contrast to standard DM annihilations, these processes involve an additional intermediate stage, namely the production of metastable mediators $\phi$, which promptly decay in flight into pairs of SM particles. The particles produced in the mediator decay do not correspond directly to the observable signal; rather, they subsequently generate stable SM particles detectable in ID searches. We therefore distinguish between \textit{primary} and \textit{final} SM particles and, consistently with the notation adopted in \Eq{eq:reaction_chain}, denote them by $\psi_{\rm SM}$ and $X$, respectively. The tabulated injection spectra for standard DM annihilations~\cite{Cirelli:2010xx,Arina:2023eic,Arina:2025ner} can then be used to reconstruct the spectra in the mediator rest frame, so that the problem reduces to performing the appropriate Lorentz boost to the desired frame.

In the mediator rest frame, the primary particles produced in the decay $\phi\to \psi_{\rm SM} \overline{\psi}_{\rm SM}$ are monochromatic, each carrying an energy equal to half of the mediator mass. CRs are then generated through the showering of these primary particles, and the resulting differential energy distribution coincides with that obtained from standard DM annihilations with an effective mass $m_\phi/2$. The remaining step is to boost this result to the GF, which up to negligible corrections due to the finite DM kinetic energy coincides with the center-of-mass frame of the DM (semi-)annihilation process.

\paragraph{Boosting back to the Galactic frame.}

Performing the desired boost requires knowledge of the angular distribution of CRs in the mediator rest frame. As explained in the main text, we assume that the decay products are isotropically distributed in this frame, and that this property is inherited by the final states $X$. In other words, we start from the distribution 
\be\label{eq:isotropy-condition}
\frac{dN_X}{dE_X^\prime \, d\cos\theta^\prime} = \frac{1}{2}\, \frac{dN_X}{dE_X^\prime}\,.
\ee

Here, all primed quantities refer to the mediator rest frame. The variable $\theta^\prime$ denotes the emission angle of the CR particle $X$ with respect to the direction of motion of $\phi$. In this notation, $dN_X / d E_X^\prime$ corresponds to the (known) differential energy distribution for standard DM annihilations evaluated for a DM mass equal to $m_\phi/2$.

The energy of $X$ particles in the GF can be related to energy and momenta in the mediator rest frame via the Lorentz boost relation 
\be\label{eq:boosted_energy}
E^\textup{(GF)}_X=\gamma_L(E_X^\prime+\beta_L\,p_X^\prime \cos\theta^\prime)\,\,,
\ee
where the Lorentz boost parameters are expressed in terms of the mediator energy $E_\phi$ and spatial momentum $p_\phi$ as
\be\label{eq:boost}
\gamma_L=\frac{E_\phi}{m_\phi}\,,\qquad\qquad\qquad \beta_L=\frac{p_\phi}{E_\phi} = 
\left(1 - \frac{m_\phi^2}{E^2_\phi} \right)^{1/2} \ .
\ee
The injection spectrum boosted to the GF is then given by
\begin{equation} \label{eq:one-step-gen}
\begin{split}
\frac{dN_X}{dE_X} = & \, \ell\times\int d\cos\theta^\prime dE'_X \frac{dN_X}{dE'_X \, d\cos\theta^\prime}\,\, \delta\left(E_X-E^\textup{(GF)}_X \right) = \\ &
\ell \times\frac{1}{2 \gamma_L\beta_L}\int_{E_X^{'\,min}}^{E_X^{'\,max}}\frac{dN_X}{dE_X'}\frac{dE_X'}{\sqrt{E_X'^{\,2}-m_X^2}} \ .
\end{split}
\end{equation}
The overall factor of $\ell$ corresponds to the number of intermediate mediators, each of which decays and contributes to the visible signal, thereby determining the multiplicity of final states. In the second line, the delta function, combined with \Eq{eq:boosted_energy}, allows the angular integral to be performed analytically. The integration limits are set by the kinematic thresholds 
\begin{subequations}\label{eq:generic one-step energy int extremes}
    \begin{align}
    E_X'^{\,min} & = \max\left\{m_X, \gamma_L \left(E_X - \beta_L\sqrt{E_X^2 - m_X^2} \right)\right\}\,\,,\\
    E_X'^{\,max}&=\min\left\{\frac{m_\phi}{2},\gamma_L \left(E_X + \beta_L\sqrt{E_X^2 - m_X^2} \right) \right\}\,\,.
\end{align}
\end{subequations}

\paragraph{Explicit results for one-step (semi-)annihilations.}

The general result given in \Eq{eq:one-step-gen} is valid for any one-step process once the Lorentz boost factor is specified. We now specialize it to the microscopic processes analyzed in this work, namely one-step annihilations and one-step semi-annihilations. To this end, it is convenient to introduce dimensionless variables: the mass ratios $\epsilon_i$, which encode the mass hierarchies, and the energy fractions $x_i$, which quantify the energy carried by the final states, defined as
\begin{align}\label{eq:one-step variables}
\epsilon_{\phi} \equiv \frac{2m_X}{m_\phi} \ , \qquad \qquad 
\epsdm \equiv \frac{m_\phi}{\mdm} \ , \qquad \qquad
x_\phi \equiv\frac{2E_X^\prime}{m_\phi} \ , \qquad \qquad 
\xdm \equiv \frac{E_X}{\mdm} \ .
\end{align} 
The mass ratios are normalized to unity at the kinematic threshold, with allowed ranges $\epsphi\in[0,1)$ and $\epsdm\in[0,1)$. The energy fraction $x_\phi$ is likewise normalized to unity at threshold and spans the range $x_\phi\in[\epsphi,1)$. The range of $\xdm$ depends on the specific microscopic process responsible for the signal, as discussed below with explicit examples. An additional parameter $\epsilon_{\psi}=2m_{\psi}/m_\phi$ can be introduced to characterize the hierarchy between the primary SM states and mediator masses. This parameter controls the amount of hadronization or final state radiation associated with the available kinetic energy of the decay products of $\phi$. It affects the shape of zero-step spectra at production, but does not play any role in the boosting procedure.

The variables $x_i$ are defined as total energy fractions, since the numerator accounts for the total energy of the final state in either frame. If one instead wishes to express the results in terms of the kinetic energy fraction $K_X/\mdm$, this can be achieved through the relations 
\be
\xdm\longrightarrow\xdm^K+\frac12 \epsphi \epsdm \ , \qquad \qquad  
x_\phi\longrightarrow x^K_\phi+\epsphi \ .
\label{eq:xtoxk}
\ee
All numerical results presented in this work are given in terms of kinetic energy distributions of CR particles. The difference between the two parametrizations is relevant only when the parameters $\epsilon_i$ are not much smaller than unity. We now discuss separately one-step annihilations and one-step semi-annihilations, and present the corresponding results for the injection spectra expressed in terms of the variables $x_i$. The translation to the corresponding spectra differential in the fractional kinetic energy $x_i^K$ is straightforward using the relations in \Eq{eq:xtoxk}.

\begin{itemize}

    \item \textbf{One-step annihilations.} We begin with one-step annihilations, $SS^\star\to \phi\phi$, where the mediators in the final state each carry energy $E_\phi=\mdm$. The derivation of the differential energy spectrum for $X$ particles follows straightforwardly by identifying the corresponding boost parameters through \Eq{eq:boost} and expressing the general result in \Eq{eq:one-step-gen} in terms of the dimensionless variables introduced in \Eq{eq:one-step variables}. Performing the computation, we obtain the final expression for the one-step annihilation injection spectrum~\cite{Mardon:2009rc,Elor:2015tva}
    \be\label{eq:one_step_ann_x}
    \frac{dN^{\scriptscriptstyle SS^\star \to \phi\phi}_X}{d\xdm} =\frac{2}{\sqrt{1-\epsdm^2}}\int_{x_\phi^{\rm a,\, min}}^{x_\phi^{\rm a,\, max}} \frac{dN^{\scriptscriptstyle SS^\star \to \phi\phi}_X}{dx_\phi}\frac{dx_\phi}{\sqrt{x_\phi^2-\epsphi^2}}\,\,,
    \ee
    with integration limits given by
    \begin{subequations}\label{eq:one-step-ann_x_int_extremes}
    \begin{align}
    x_\phi^{\rm a,\, min}&=\max\left\{\epsphi,\,\frac{2\xdm-\sqrt{(1-\epsdm^2)(4\xdm^2-\epsdm^2\epsphi^2)}}{\epsdm^2}\right\}\,\,,\\
    x_\phi^{\rm a,\,max}&=\min\left\{1,\,\frac{2\xdm+\sqrt{(1-\epsdm^2)(4\xdm^2-\epsdm^2\epsphi^2)}}{\epsdm^2}\right\}\,\,.
    \end{align}
    \end{subequations}
    These expressions can be compared with the generalized formula for an $n$-step cascade annihilation process provided in appendix~B of Ref.~\cite{Elor:2015tva}, which is valid for massless final states. We find full agreement up to finite $\epsphi$ corrections, which are retained here. The impact of finite-mass effects will be discussed at the end of this appendix.

    \item \textbf{One-step semi-annihilations.} The presence of a stable DM particle in the final state does not lead to a direct signal in ID searches. However, it modifies the kinematics of the process, with a consequent impact on the predicted signal~\cite{DEramo:2010keq,DEramo:2012fou,Queiroz_2019,Beauchesne:2024tuc}. The calculation proceeds similarly, as the starting point is again an on-shell mediator $\phi$ decaying into visible states. Independently of the specific decay channel, the resulting injection spectrum has half the multiplicity compared to the corresponding one-step annihilation scenario. The final states are in general non-degenerate in mass and must satisfy the hierarchy $m_\phi \leq \mdm$. The mediator energy is given by
    \be\label{eq:semi-ann kinematics}
    E_\phi=\frac{3}{4}\mdm\left(1+\frac{1}{3}\frac{m_\phi^2}{\mdm^2}\right) \ .
    \ee
    In this case, the Lorentz boost is reduced with respect to the one-step annihilation scenario due to the energy carried away by the stable DM particle in the final state. As a result, for fixed mass spectrum and primary channel, the CR injection spectra are shifted towards lower energies and the peak is correspondingly suppressed compared to the one-step annihilation case.

    We again determine the boost parameters from \Eq{eq:boost} and express the injection spectrum in \Eq{eq:one-step-gen} in terms of the dimensionless variables introduced in \Eq{eq:one-step variables}. As in the previous case, we retain the full parametric dependence on the masses of the primary and intermediate states. The modified kinematics leads to
    \be\label{eq:one_step_semi_x}
    \frac{dN^{\scriptscriptstyle SS\to S^\star\phi}_X}{d\xdm}  = \frac{4}{3}\frac{1}{\zeta}\int_{x_\phi^{\rm s,\,min}}^{x_\phi^{\rm s,\,max}} \frac{dN^{\scriptscriptstyle SS\to S^\star\phi}_X}{dx_\phi}\frac{dx_\phi}{\sqrt{x_\phi^2-\epsphi^2}} \, .
    \ee
    Up to an overall factor of $1/2$, accounting for the presence of a single mediator in the final state, the difference with respect to the previous case manifests itself through the appearance of the dimensionless momentum
    \be
    \zeta=\frac{4}{3}\frac{p_\phi}{\mdm}= \sqrt{1 - \frac{10 \, \epsdm^2}{9} + \frac{\epsdm^4}{9}}\ .
    \ee
    The integration limits are now given by 
    \begin{subequations}\label{eq:one-step semiannihilation x int extremes}
    \begin{align}
    x_\phi^{\rm s,\,min}&=\max\left\{\epsphi,\,\frac{8 \xdm \left(\epsdm^2+3\right)-3 \zeta  \sqrt{64 \xdm^2-\epsphi ^2 \left(\left(\epsdm^2+3\right)^2-9 \zeta ^2\right)}}{\left(\epsdm^2+3\right)^2-9 \zeta ^2}\right\}\,\,,\\
    x_\phi^{\rm s,\,max}&=\min\left\{1,\,\frac{8 \xdm \left(\epsdm^2+3\right)+3 \zeta  \sqrt{64 \xdm^2-\epsphi ^2 \left(\left(\epsdm^2+3\right)^2-9 \zeta ^2\right)}}{\left(\epsdm^2+3\right)^2-9 \zeta ^2}\right\}\,\,.
    \end{align}
    \end{subequations}

\end{itemize}

\paragraph{On finite mass corrections.}

The key results of this appendix are \Eqs{eq:one_step_ann_x}{eq:one_step_semi_x}, which provide the injection spectra for one-step cascade annihilations and semi-annihilations, respectively. The former has already been derived in the well-studied context of one-step annihilations. In particular, we have verified that it agrees with the result presented in appendix~B of Ref.~\cite{Elor:2015tva} in the limit of massless final states. We devote the final part of this appendix to discussing the qualitative and quantitative impact of finite particle masses on the CR injection spectra produced by one-step cascade processes. In particular, following the notation introduced in \Eq{eq:one-step variables}, we quantify finite $\epsphi$ and $\epsdm$ corrections. These corrections are expected to be relevant when considering massive CR species, such as antiprotons or antinuclei, or when the mediator is not significantly lighter than the DM particle. We focus on ID signals sourced by one-step annihilation processes, as this scenario has been extensively studied in the literature and provides both numerical and analytical results for comparison. Similar considerations apply to semi-annihilating DM, since the discussion is primarily driven by kinematics.

We first clarify which mass effects are relevant for our discussion. The masses of the primary states (i.e., the particles denoted by $\psi_{\rm SM}$) are already taken into account numerically in the tabulated spectra at production, as they determine the amount of hadronization or EW showering for a given mediator mass. On the other hand, the masses of both the mediator $\phi$ and the final states $X$ can significantly affect the resulting spectrum in one-step processes. This is evident from their explicit appearance in the Lorentz boost relation and in the integration limits (see \Eqs{eq:one-step-gen}{eq:generic one-step energy int extremes}).

In the presence of a large mass hierarchy between $\phi$ and $X$, the expressions in \Eqs{eq:one_step_ann_x}{eq:one-step-ann_x_int_extremes} can be evaluated in the limit $\epsphi\to0$. This procedure correctly reproduces eqs.~(B.3) and (B.4) of Ref.~\cite{Elor:2015tva}, where the $\gamma$-ray spectrum is computed (indeed, $\epsphi=0$ for massless CR). Imposing in addition a hierarchy between DM and the mediator, quantified by the condition $\epsdm\ll1$, one recovers eq.~(5) of the same reference for $\gamma$ rays produced in one-step annihilations. The impact of $\mathcal{O}(\epsdm^2)$ corrections is also discussed in Ref.~\cite{Elor:2015tva}: increasing $\epsdm$ corresponds to the production of less energetic mediators, until in the limit $\epsdm\to1$ they are produced at rest and no Lorentz boost occurs, so that the CR spectrum reduces to that of a decaying particle at rest. Varying $\epsdm\in(0,1)$ therefore interpolates between these two limiting regimes (see figure~10 therein).

\begin{figure}[!t]
    \centering
    \includegraphics[width=0.9\linewidth]{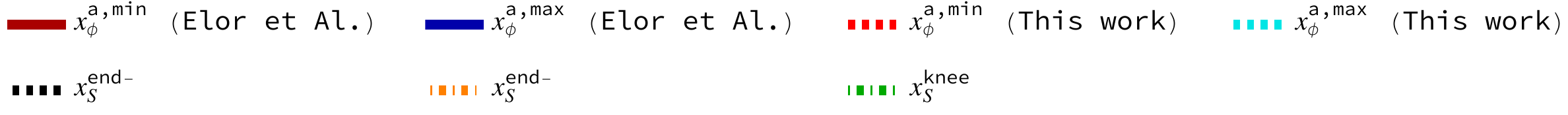}\\\medskip
    \subfloat[\label{fig:extremes_1}]{\includegraphics[width=0.46\linewidth]{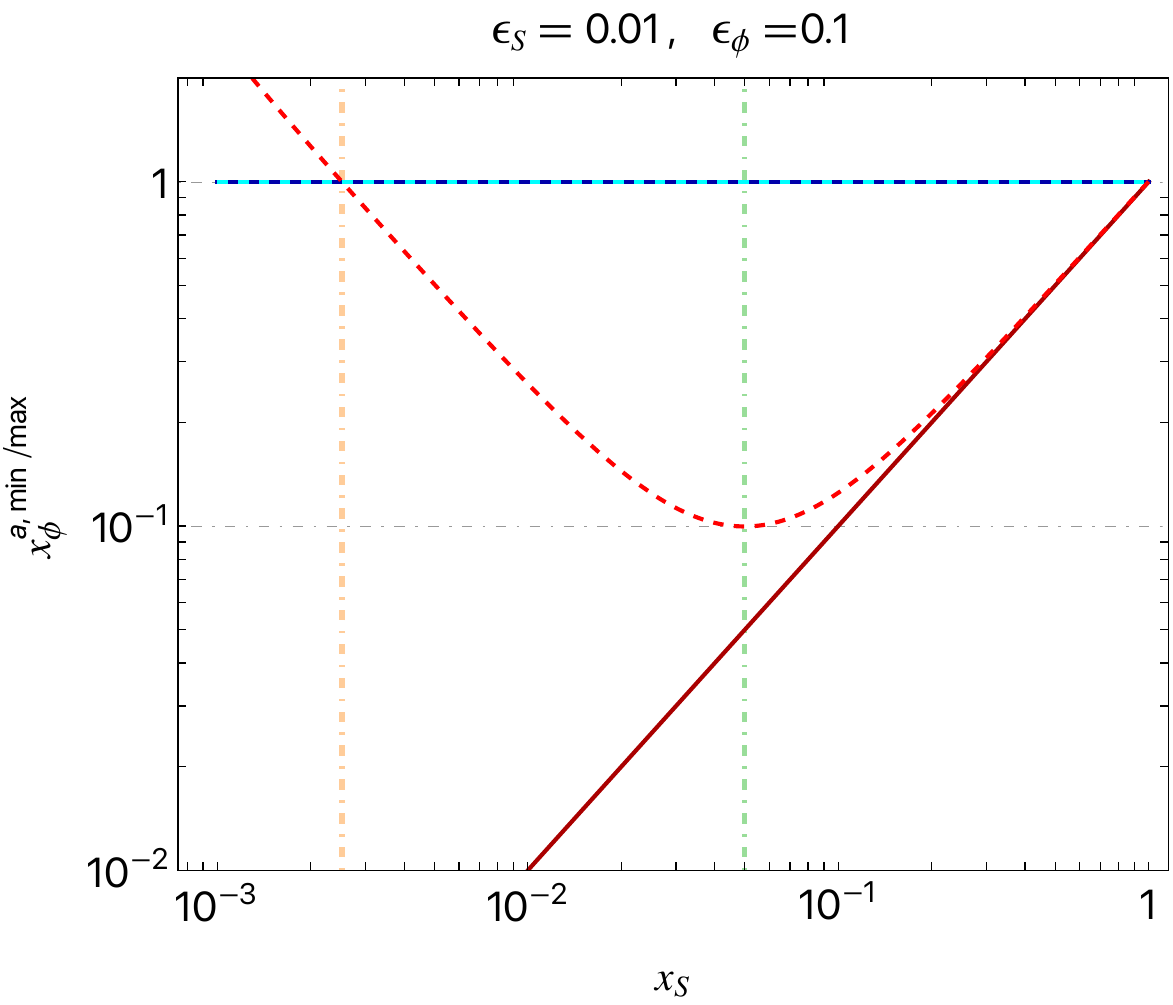}}\qquad
    \subfloat[\label{fig:extremes_2}]{\includegraphics[width=0.46\linewidth]{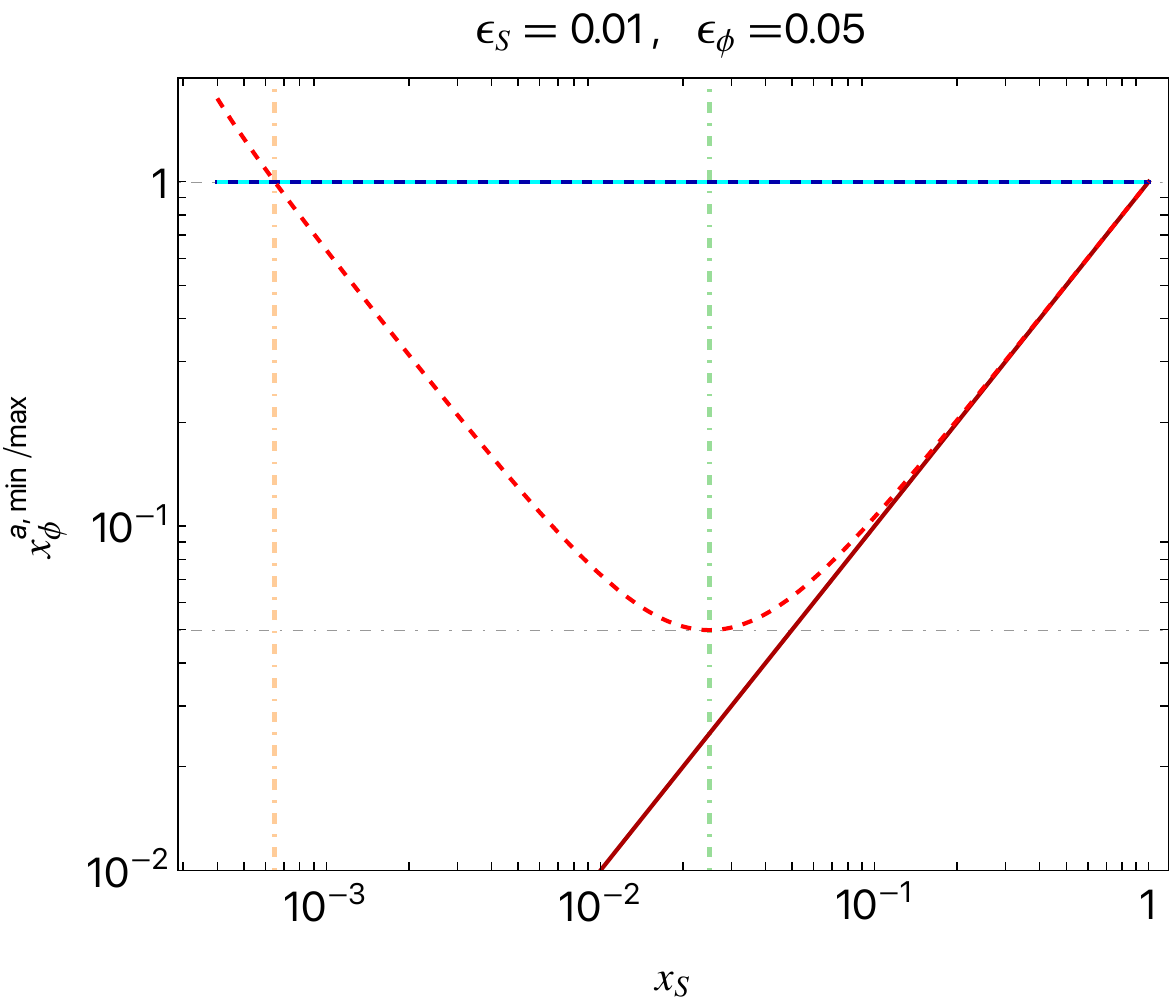}}\\
    \subfloat[\label{fig:extremes_3}]{\includegraphics[width=0.46\linewidth]{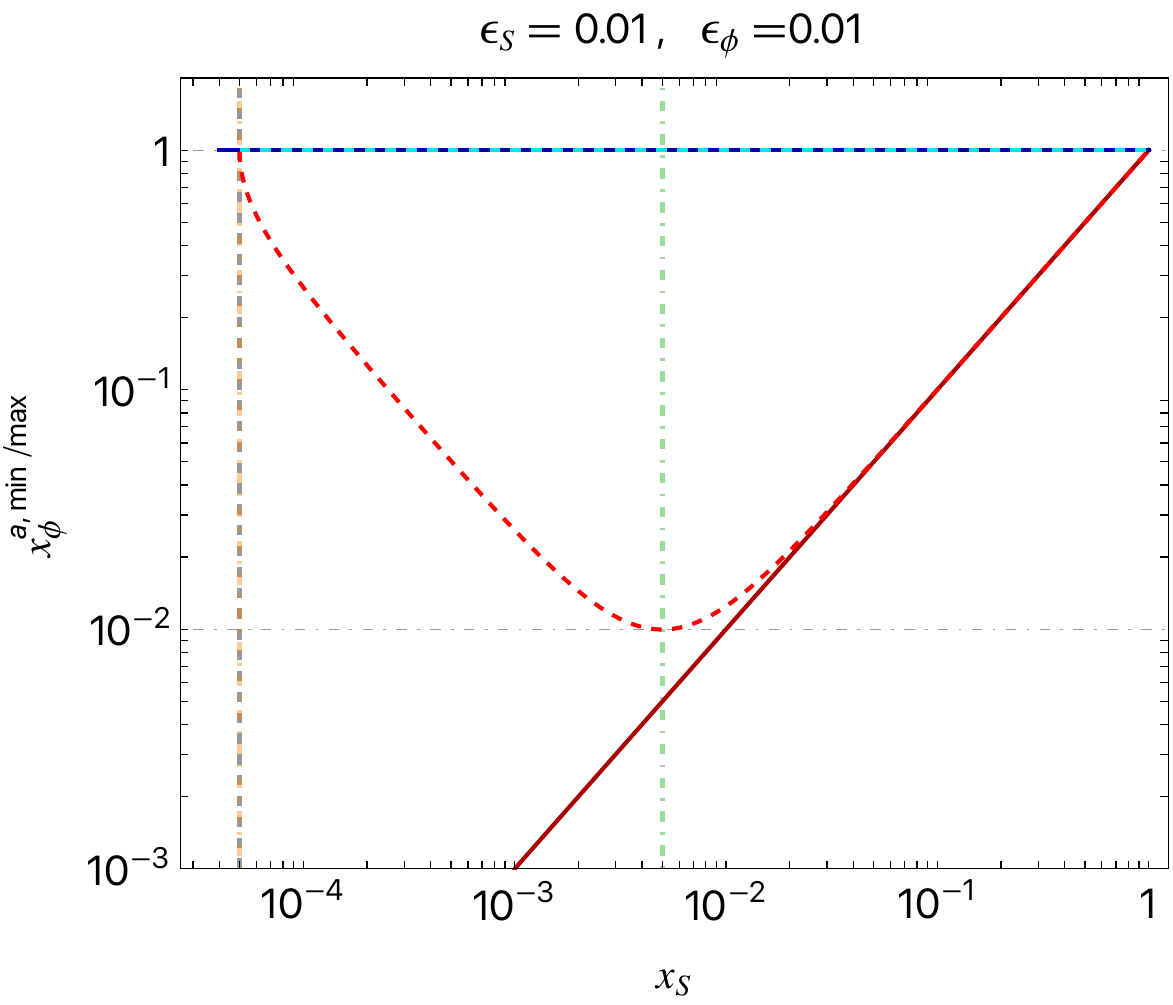}}\qquad
    \subfloat[\label{fig:extremes_4}]{\includegraphics[width=0.46\linewidth]{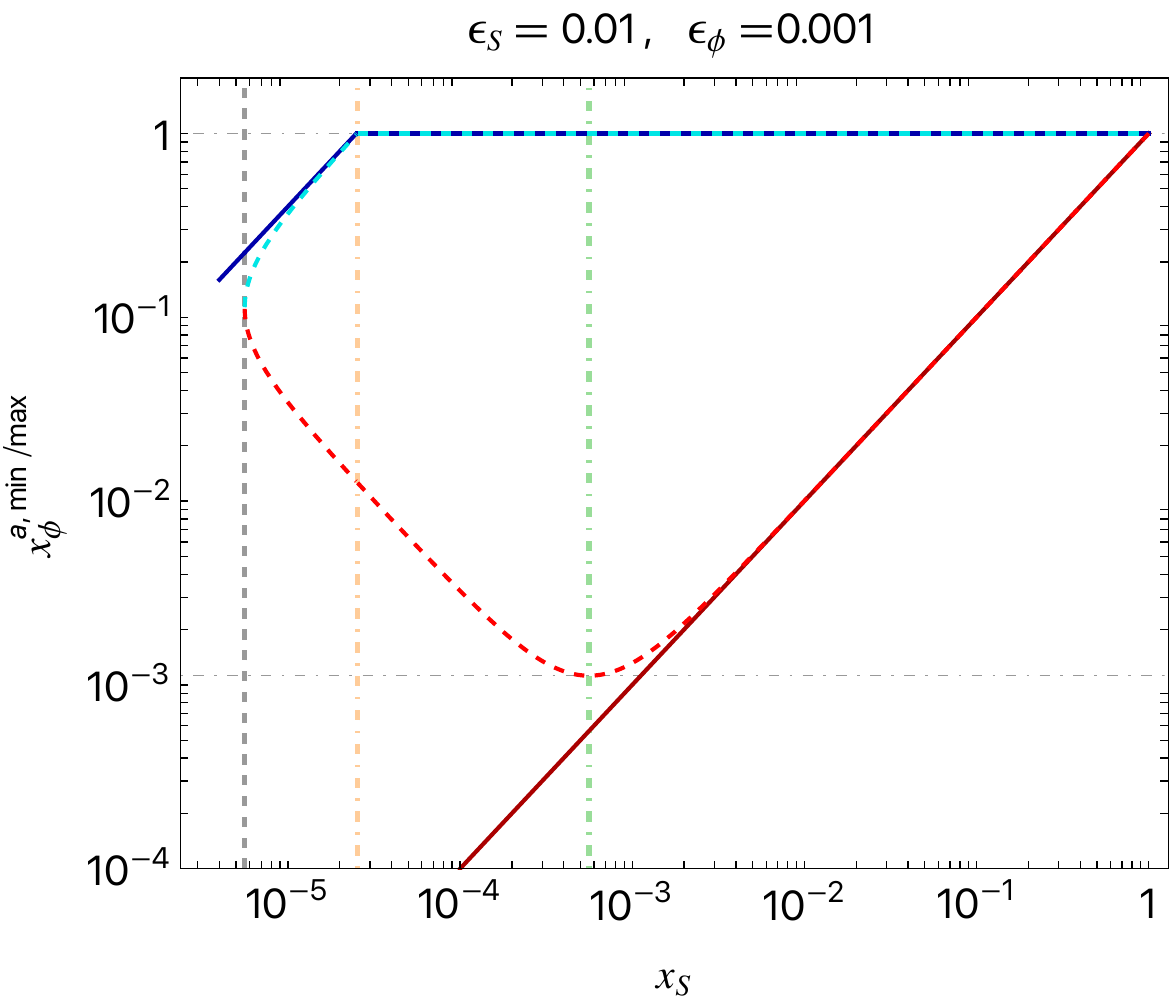}}
    \caption{Upper and lower integration limits of the one-step annihilation injection spectrum for antiprotons ($X=\overline{p}$) as a function of $\xdm$. We fix $\epsdm=0.01$ and vary $\epsphi=(0.1,\,0.05,\,0.01,\,0.001)$ across the four panels. Solid lines correspond to the results of Ref.~\cite{Elor:2015tva}, obtained in the large-hierarchy limit $(\epsphi\ll1,\,\epsdm\ll1)$, while dashed lines are computed using the full expressions in \Eq{eq:one-step-ann_x_int_extremes}. In each panel, we also indicate the positions of the endpoint and the knee, as defined in \Eqs{eq:endpoint}{eq:knee}.
    }
    \label{fig:cfr_extremes}
\end{figure}
\begin{figure}[!t]
    \centering
    \includegraphics[width=0.7\linewidth]{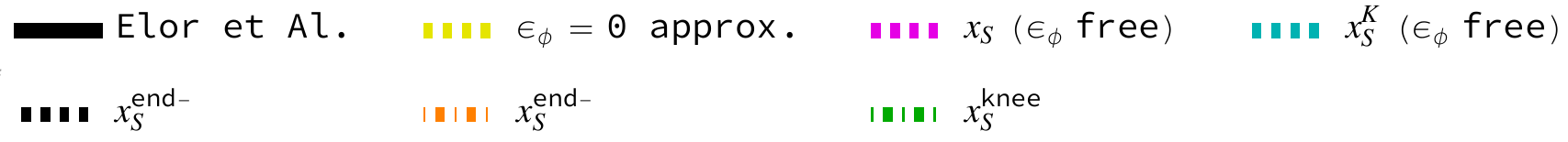}\\\medskip
    \subfloat[\label{fig:Elor_cfr_1}]{\includegraphics[width=0.45\linewidth]{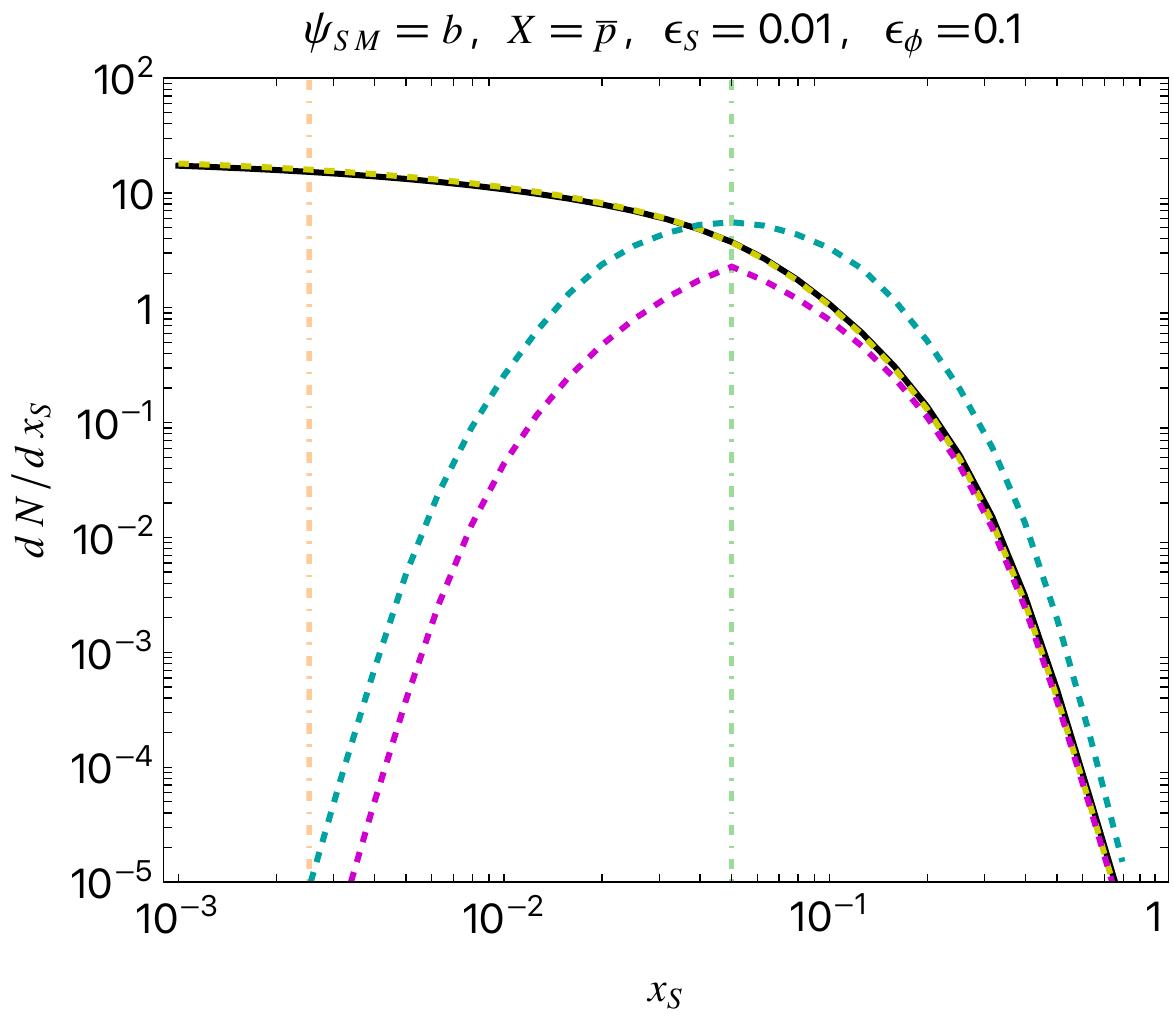}}\quad
    \subfloat[\label{fig:Elor_cfr_2}]{\includegraphics[width=0.45\linewidth]{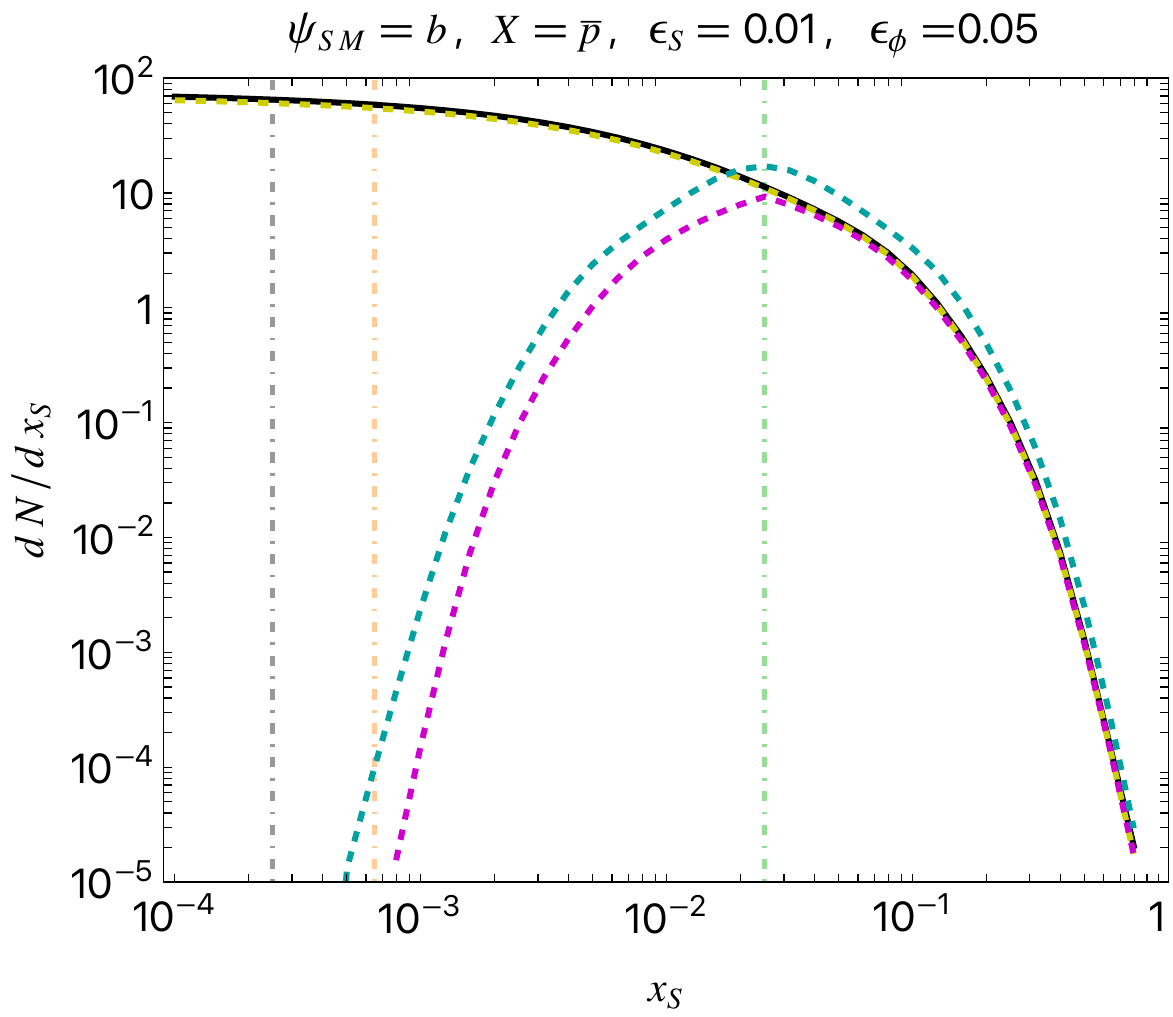}}\\
    \subfloat[\label{fig:Elor_cfr_3}]{\includegraphics[width=0.45\linewidth]{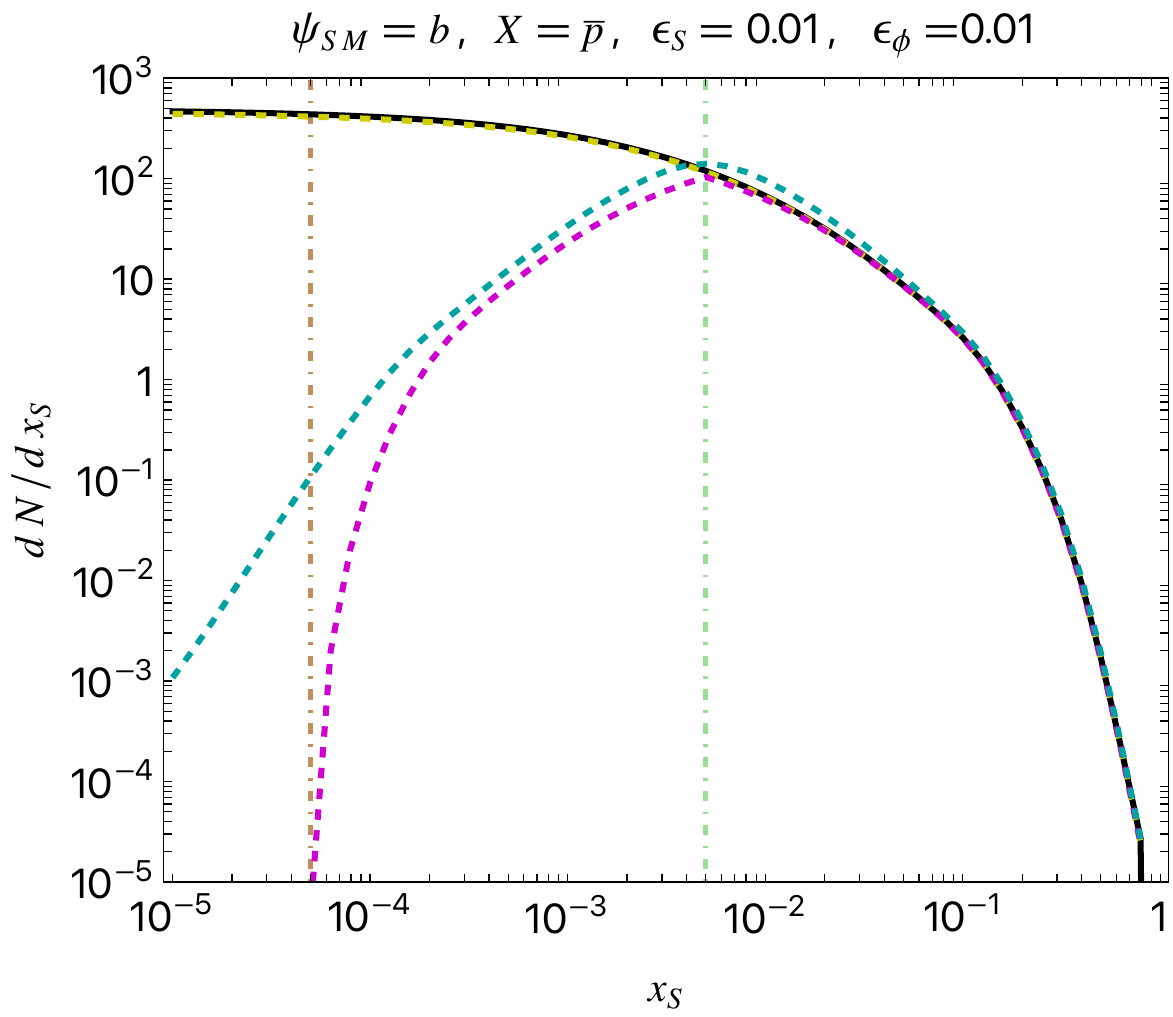}}\quad
    \subfloat[\label{fig:Elor_cfr_4}]{\includegraphics[width=0.45\linewidth]{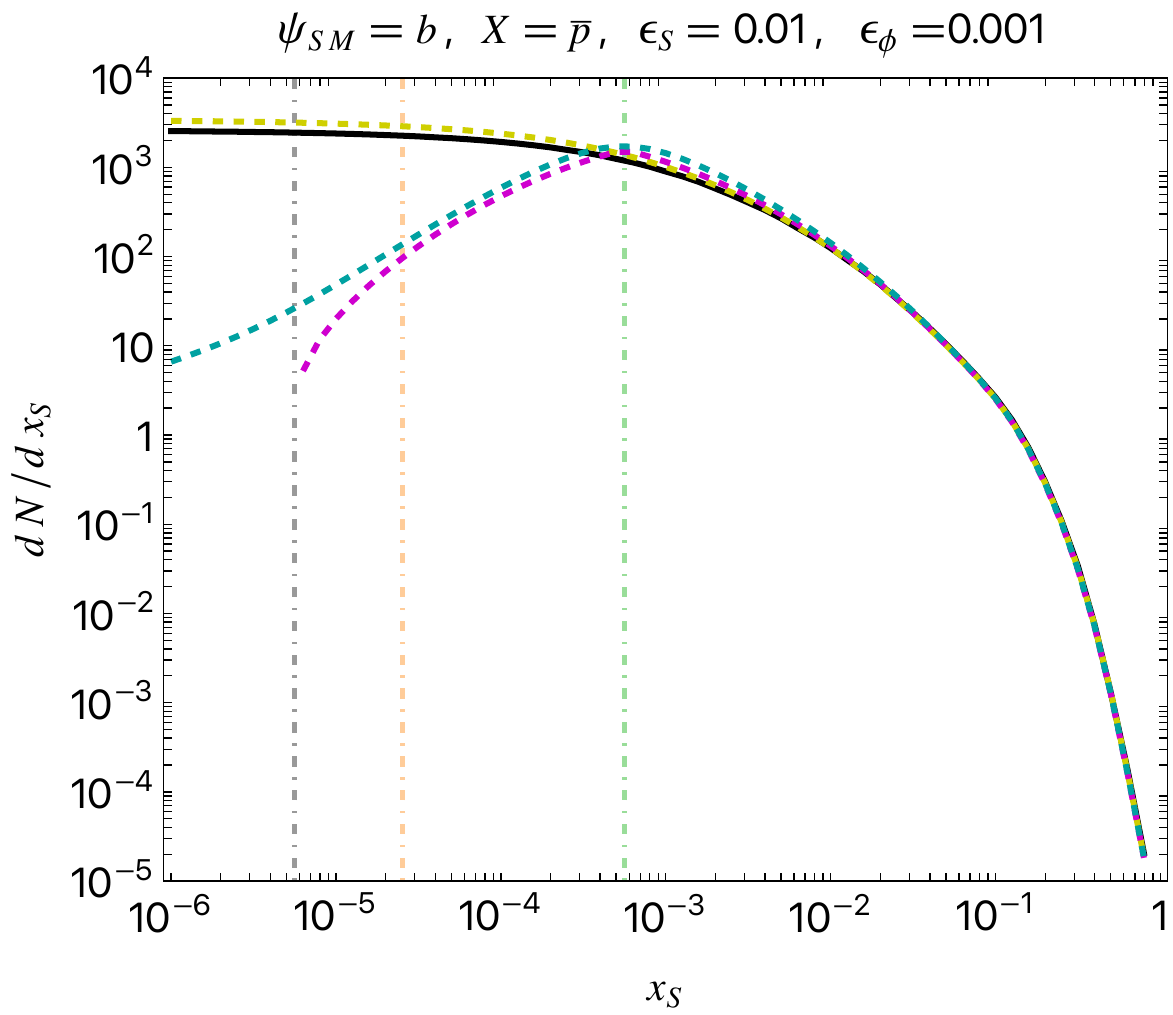}} 
    \caption{Comparison of antiproton injection spectra from one-step DM annihilations computed using different approximations. We fix $\psi_{\rm SM}=b$ as the primary production channel, set $\epsdm=0.01$, and vary $\epsphi=(0.1,\,0.05,\,0.01,\,0.001)$. Solid \textit{black} lines correspond to the results of Ref.~\cite{Elor:2015bho}, obtained within the large-hierarchy approximation $(\epsphi\ll1,\,\epsdm\ll1)$. These results coincide with the dashed \textit{yellow} lines, obtained by setting $\epsphi=0$ in \Eq{eq:one_step_ann_x}. Dashed \textit{magenta} and \textit{cyan} lines correspond to spectra computed with the full expression, and expressed in terms of total and kinetic energy fractions, respectively. Dot-dashed vertical lines indicate the positions of the knee-like feature and of the low-energy endpoints already identified in \Fig{fig:cfr_extremes}(for spectra differential in the total energy fraction).}
    \label{fig:cfr_elor_epsphi}
\end{figure}

The situation becomes more interesting and subtle when considering the injection spectra of massive CR species, such as antiprotons or antinuclei, since both $\epsdm$ and $\epsphi$ come into play. The role of $\epsdm$ has been clarified in the previous paragraph, so we now focus on the impact of the final-state mass, encoded in $\epsphi$. If one were to compute the injection spectra in terms of the \textit{total} CR energy, i.e. kinetic energy plus the mass of $X$, the large-hierarchy approximation would fail to accurately describe the low-energy tail of the spectrum. In this variable, the spectral distribution exhibits two characteristic features: a sharp cutoff at low energies and a pronounced \textit{knee-like} structure (as shown later in \Fig{fig:cfr_elor_epsphi}). This behavior originates from the fact that the energy of the observable particle cannot be smaller than its mass $m_X$, as well as from the support of the spectrum enforced by the Dirac delta function in the integrand of \Eq{eq:one-step-gen}.

We can make these statements more quantitative. Starting from the $\epsphi\to0$ limit, the integration window given in \Eq{eq:one-step-ann_x_int_extremes} is approximated by
\begin{subequations}\label{eq:elor_x_extremes}
    \begin{align}
    \lim_{\epsphi \rightarrow 0} x_\phi^{\rm a,\, min} &=\frac{2\xdm}{\epsdm^2}\left(1-\sqrt{1-\epsdm^2}\,\right) \ ,\\
    \lim_{\epsphi \rightarrow 0} x_\phi^{\rm a,\,max} &=\min\left\{1,\frac{2\xdm}{\epsdm^2}\left(1+\sqrt{1-\epsdm^2}\,\right)\right\} \ .
\end{align}
\end{subequations}
In this limit, the function $x_\phi^{\rm a,\, min}$ depends linearly on $\xdm$, while $x_\phi^{\rm a,\, max}$ is typically saturated by $1$, unless the second term inside the minimum function becomes smaller than $1$.

The situation changes significantly once finite $\epsphi$ corrections are included. Retaining only $\mathcal{O}(\epsphi^2)$ terms, both linear branches in \Eq{eq:elor_x_extremes} acquire corrections proportional to $1/\xdm$, with opposite sign. While these corrections have a limited impact on the upper integration boundary, they play a crucial role in shaping the lower one. In particular, the correction to the lower limit enters with a positive sign, implying that as $\xdm$ varies, the integration interval first increases, reaches a maximum, and then decreases again, eventually closing when the lower and upper bounds coincide. At this point, the spectrum develops an endpoint determined by the condition
\be
\xdm^{\rm end-}=
\begin{cases}
 \frac{1}{2} \left(1-\sqrt{(1-\epsdm^2)(1-\epsphi^2)}\right)\qquad\,\text{for } \ x_\phi^{\mathrm{a,max}}= \frac{2 \xdm+\sqrt{(1-\epsdm^2)(4 \xdm^2-\epsdm^2 \epsphi^2)}}{\epsdm^2}\\
 \frac{\epsdm\,\epsphi}{2}\hspace{12em}\text{for } \  x_\phi^{\mathrm{a,max}}=1
\end{cases} \ ,
\label{eq:endpoint}
\ee
Furthermore, the spectrum develops a knee-like feature at the value of $\xdm$ for which $x_\phi^{\rm a,\, min}$ is minimized, and the integration region is correspondingly maximized. This occurs at
\be
\xdm^{\rm knee}=\frac{\epsphi}{2} \ .
\label{eq:knee}
\ee

We illustrate the impact of the integration region in the four panels of \Fig{fig:cfr_extremes}. We fix $\epsdm=0.01$ to ensure a clear hierarchy between the DM particle and the mediator, and explore how the integration limits depend on $\epsphi$. In particular, for $\epsphi=(0.1,\,0.05,\,0.01,\,0.001)$, we show the dependence of $x_\phi^{\rm a,\, min}$ and $x_\phi^{\rm a,\, max}$ on the variable $\xdm$. All the features discussed in the previous paragraph are clearly visible. In the limit $\epsphi\to0$, corresponding to the integration limits in \Eq{eq:elor_x_extremes}, the integration region reduces to a triangular domain bounded by two straight lines (shown by the solid red and blue curves). Finite $\epsphi$ corrections are displayed by the dashed lines, corresponding to the full expressions in \Eq{eq:one-step-ann_x_int_extremes}. The endpoint and the knee of the spectrum are also indicated in the figure.

The above discussion shows that it is more convenient to parametrize the differential distribution of messenger particles in terms of their kinetic energy (fraction), which can be achieved through the change of variables in \Eq{eq:xtoxk}. In the limit $\epsphi=0$, the two parametrizations coincide. The behavior of the integration extremes as a function of $x_S^K$, including finite $\epsphi$ corrections, follows similar lines, since this change of variables amounts to a constant shift. In this parametrization, the endpoint at $\xdm=m_X/\mdm$ is no longer present, as CR particles can in principle have arbitrarily small kinetic energy. However, the existence of an endpoint at small kinetic energy is not entirely excluded a priori, due to the Dirac delta function, which still constrains the support of the spectrum through the Lorentz boost. Moreover, the spectra of massive CR species continue to exhibit a maximum, while the sharp knee-like feature is smoothed out by the introduction of the shifted variable.

We provide a concrete example to highlight the features discussed so far. Ref.~\cite{Elor:2015bho} presents a collection of CR injection spectra from multi-step cascade annihilations of DM particles computed in the large-hierarchy approximation, with $\epsilon_{\psi}\in[0.01,0.5]$, starting from PPPC spectra at production. In particular, it includes antiproton injection spectra for a subset of primary $\psi_{\rm SM}$ channels. This provides a direct comparison between existing results for massive CR and those obtained using the full expression in \Eq{eq:one_step_ann_x}, which is valid for arbitrary mass hierarchies. The comparison is shown in \Fig{fig:cfr_elor_epsphi}, where the characteristic features of massive CR spectra from one-step annihilations are clearly visible, together with the deviations from the results available in the literature. Our results extend those of Ref.~\cite{Elor:2015bho}, as we consistently account for finite particle masses without assuming any hierarchy among the states at any stage of the cascade.

\bibliographystyle{JHEP}
\bibliography{OneStepSemi}

\end{document}